\pdfoutput=1
\documentclass[aps,prd,preprint,superscriptaddress,showpacs,nobibnotes,nofootinbib,a4paper]{revtex4-1}

\usepackage{graphicx}
\usepackage{amsmath}
\usepackage{bm}
\usepackage{amssymb,enumerate}
\usepackage[caption=false]{subfig}
\usepackage{setspace} 
\usepackage{hyperref}

\def\p{\partial}
\def\del{\partial}
\def\sech{\mathop{\mathrm{sech}}\nolimits}
\def\arctan{\mathop{\mathrm{arctan}}\nolimits}
\def\Li{\mathop{\mathrm{Li}}\nolimits}
\def\=:{=\hspace{-.7em}\raisebox{1.1ex}{.}\hspace{.1em}\raisebox{-0.2ex}{.} }
\newcommand{\beq}{\begin{eqnarray}}
\newcommand{\eeq}{\end{eqnarray}}
\newcommand{\non}{\nonumber\\}

\newcommand{\Tr}{{\rm Tr}}

\def\gd{{\mathop{\rm gd}}}
\begin{document}

\title{Skyrmions confined as beads on a vortex ring}

\author{Sven Bjarke Gudnason}
\affiliation{{\it Institute of Modern Physics, Chinese Academy of
    Sciences, Lanzhou 730000, China}} 
\author{Muneto  Nitta}
\affiliation{{\it Department of Physics, and Research and
    Education Center for Natural Sciences, Keio University, Hiyoshi
    4-1-1, Yokohama, Kanagawa 223-8521, Japan}}

\date{\today}
\begin{abstract}
A very simple, quadratic potential is used to construct vortex strings
in a generalized Skyrme model and an additional quadratic potential is
used to embed sine-Gordon-type halfkinks onto the string worldline,
yielding half-Skyrmions on a string.
The strings are furthermore compactified onto a circle and the
halfkinks are forced to appear in pairs; in particular $2B$ halfkinks
(half-Skyrmions) will appear as beads on a ring with $B$ being the
number of times the host vortex is twisted and also the baryon number
(Skyrmion number) from the bulk point of view.
Finally, we construct an effective field theory on the torus,
describing the kinks living on the vortex rings. 
\end{abstract}

\maketitle

\tableofcontents

\newpage
\section{Introduction}
 
Among various topological solitons, Skyrmions are the ones having the
longest history \cite{Skyrme:1962vh,Skyrme:1961vq}.
Nevertheless they are still under active consideration, partly
because they have been proposed to be identified with baryons of
(large-$N_c$) QCD and provide a first-principles angle to nuclear
physics; for some recent results see
e.g.~\cite{Gudnason:2016mms,Haberichter:2015qca,Karliner:2015qoa,Adam:2015lra,Foster:2015cpa,Adam:2015rna,Lau:2014baa}. 

In a series of papers, we have studied a number of different
incarnations of
Skyrmions \cite{Nitta:2012wi,Nitta:2012rq,Gudnason:2014hsa}; the
simplest form is just the Skyrmion \cite{Skyrme:1962vh,Skyrme:1961vq},
whereas the topological charge of the Skyrmion -- the baryon number --
can be absorbed into a host soliton, yielding a daughter soliton
living in the world volume of its host or mother soliton.
The simplest incarnation is taking a domain wall and embedding either
a baby-Skyrmion or a lump in its world volume \cite{Gudnason:2014nba};
this gives a Skyrmion from the bulk point of view, but of course
with infinite energy due to the infinite (2+1)-dimensional world
volume of the domain wall.

Another class of incarnations of Skyrmions is to construct a vortex
string with a U(1) modulus and subsequently twist said modulus which
will form kinks on the worldline of the string; each kink will 
correspond to either a half or a full baryon charge, depending on
whether the kink winds $\pi$ or $2\pi$ which in turn is determined by
the type of potential in the model.
In Ref.~\cite{Gudnason:2014jga} we explicitly constructed a vortex
string in compactified form; that is, compactified onto a circle.
The potential used in Ref.~\cite{Gudnason:2014jga} was inspired by a
limit of a potential used in Bose-Einstein condensates (BEC) 
\cite{Ruostekoski:2001fc,Battye:2001ec}
(see also \cite{Kawakami:2012zw,Nitta:2012hy})
and it
breaks a U(2) subgroup of the O(4) symmetry of the Skyrme model to
U(1)$\times$U(1); one of them is used for constructing the vortex and
the other will be the mentioned U(1) modulus living on the string
worldline.
In Ref.~\cite{Gudnason:2014hsa} we constructed straight vortex strings
using the same BEC potential and embedded kinks and halfkinks on their 
worldlines.
The last possibility in this direction is to compactify the strings
and embed kinks on them; in this case the number of kinks is forced to
be integer as the baryon number is quantized for finite energy
configurations; this implies that when the vortex ring possesses
halfkinks, it must have an even number of halfkinks.
This case of a vortex ring with embedded kinks on its worldline has
not been explicitly constructed before; this will be a new result in
the present paper. 
A lower dimensional analog, however, has been studied in the
baby-Skyrme model in 2+1 dimensions \cite{Kobayashi:2013ju}
by compactifying a domain wall worldline, with baby-Skyrmions 
on it \cite{Nitta:2012xq}, to a circle. 

The last incarnation of Skyrmions is a half-Skyrmion inside a
monopole \cite{Gudnason:2015nxa}. 
The unit-charged Skyrmion is split into a set of monopole and
anti-monopole both of which carry a half baryon number. 
If a half-Skyrmion is separated from its other half, then it will have
a divergent total energy because it is a global monopole.
This also has a lower dimensional analogue in the baby-Skyrme
model \cite{Kobayashi:2013aja}, in which case half a baby-Skyrmion
lives inside a global vortex. 

Besides the four-dimensional incarnations of Skyrmions, in five 
dimensions it is possible to create a string with Skyrme charge that
ends on a domain wall \cite{Gudnason:2014uha}; interestingly, this is 
possible in the O(4) model, which is just the standard Skyrme model
(but generalized to 4+1 dimensions). 

In this paper, we show that there exists a simpler potential than the
BEC-type potential used in
Refs.~\cite{Gudnason:2014jga,Gudnason:2014hsa} admitting a vortex
string and vortex rings.
Using a notation with two complex fields $\phi_{1,2}\in\mathbb{C}$,
the BEC-type potential is of the form $|\phi_1|^2(1-|\phi_1|^2)$,
whereas the simplest possibility for forming a vortex string is the
potential we study in this paper, i.e.~of the form $(1-|\phi_1|^2)$.
An interesting side mark about this potential is that it is induced at
the classical level by introducing an isospin-breaking chemical
potential of the form $\p_0 U\to \p_0 U-i[\mu\sigma^3,U]$.
However, this chemical potential -- at the classical level -- also
induces other terms; in particular, the Skyrme term induces a
noncanonical kinetic term which effectively can drive the coefficient
of the kinetic term negative for large enough $\mu$. 
Although such chemical potential at the classical level, has been
discussed in the
literature \cite{Loewe:2004ic,Loewe:2005kq,Loewe:2006hv,Ponciano:2007jm},
Ref.~\cite{Cohen:2008tf} pointed 
out that due 
to a subtlety in the large-$N_c$ counting of $\mu$, one should not
include the effects of the chemical potential at the classical level
if one wishes to study the proton or neutron, because the large-$N_c$
nature of the model will pick out the largest spin state of the
nucleon; hence not the proton or neutron.
Therefore, if one does not include the effect of the chemical
potential at the classical level, there is no presence of the
potential that we use to construct vortex strings.
We also do not consider the other terms that would be induced from
said chemical potential.
We simply take the potential and prove that it can be used to
construct strings in the Skyrme model.

In this paper, as in
Refs.~\cite{Gudnason:2014hsa,Gudnason:2014jga,Gudnason:2013qba,Gudnason:2014gla,Gudnason:2015nxa},
we compare the Skyrme term to the sixth-order derivative term, which
is composed by squaring the baryon current; this term is inspired by
the BPS-Skyrme model \cite{Adam:2010fg,Adam:2010ds,Adam:2015ele}.
The BPS-Skyrme model was motivated by the long-standing problem of the
large binding energies in the standard Skyrme model; in the BPS-Skyrme
model, which is a particular submodel of the Skyrme model, the BPS
bound can be saturated -- unlike \cite{Manton:1986pz} the one in the
standard Skyrme model \cite{Faddeev:1976pg} -- and so classically the
binding energies vanish. 

Finally, in Ref.~\cite{Gudnason:2014gla} we constructed a framework of
effective field theories for solitons living on host solitons of
generic shapes. In Ref.~\cite{Gudnason:2014gla} we applied it to
straight vortices with the BEC potential.
In this paper, we will use the same framework for the vortices in the
new potential and use it to construct kinks directly in the effective
field theory approach. Finally, as a new result, we derive the
effective field theory for sine-Gordon halfkinks living on vortex
rings -- that is, vortex strings compactified onto a circle -- and use
it to calculate the kinks and baryon charge density in the effective
theory approach. 

The paper is organized as follows.
In Sec.~\ref{sec:model} we introduce the model, set the notation and
discuss the vacua and symmetries.
In Sec.~\ref{sec:straightvortex} we construct the straight, infinitely
long, vortex string, which we in Sec.~\ref{sec:vortexrings} compactify
onto a circle with one and two twists, yielding a $B=1$ and $B=2$
vortex ring, respectively.
In Sec.~\ref{sec:quadpot} we then, finally, embed kinks onto both the
straight vortex and the vortex ring and we also construct the relevant
leading-order effective field theories for all cases.
We conclude in Sec.~\ref{sec:discussion} with a discussion of our
results.
The appendix \ref{app:stringsplitting} provides evidence for the
two-vortex to split up into two separate Skyrmions.

\section{Skyrme-like model}
\label{sec:model}

We consider a Skyrme-type model in 3+1 dimensions
\beq
\mathcal{L} = \frac{1}{4}\Tr(L_\mu L^\mu)
+ c_4 \mathcal{L}_4 + c_6 \mathcal{L}_6 - V(U),
\label{eq:L}
\eeq
that includes the Skyrme term and a sixth-order term -- made of 
the square of the baryon current -- which we will call the BPS-Skyrme
term \cite{Adam:2010fg,Adam:2010ds,Adam:2015ele}, 
\begin{align}  
\mathcal{L}_4 &= \frac{1}{32} \Tr\left([L_\mu,L_\nu]^2\right),
\label{eq:Skyrme-term}\\
\mathcal{L}_6 &=
\frac{1}{144}\left(\epsilon^{\mu\nu\rho\sigma} 
\Tr\left[L_\nu L_\rho L_\sigma\right]\right)^2,
\label{eq:6th-term}
\end{align}
where we have defined the $\mathfrak{su}(2)$-valued left-invariant
current $L_\mu\equiv U^\dag\p_\mu U$, $U$ is the 2-by-2 nonlinear
sigma-model field obeying the constraint $U^\dag U=\mathbf{1}_2$, the
spacetime indices $\mu,\nu,\rho,\sigma=0,1,2,3$ run over all 3+1
dimensions, the Lagrangian coefficients $c_4\geq 0$ and $c_6\geq 0$
are both positive semi-definite\footnote{Either $c_4$ or $c_6$ has to
be positive in order for the model to possess a stable Skyrmion.}
and finally, we use the mostly-positive metric signature. 

Since (one of) our objective(s) is to study vortices, it will prove
convenient to switch notation from the matrix field $U$ to a complex
vector field $\phi$ as
\beq
\phi \equiv
\begin{pmatrix}
\phi_1 \\
\phi_2
\end{pmatrix}.
\eeq
The two fields are related as follows
\beq
U =
\begin{pmatrix}
\phi & -i\sigma^2 \phi^*
\end{pmatrix}
=
\begin{pmatrix}
\phi_1 & -\phi_2^*\\
\phi_2 & \phi_1^*
\end{pmatrix},
\eeq
and the nonlinear sigma-model constraint translates to
\beq
\det U = |\phi_1|^2 + |\phi_2|^2 = 1.
\label{eq:NLSMconstraint}
\eeq
Rewriting the Lagrangian density \eqref{eq:L} using the field $\phi$,
we get 
\beq
\mathcal{L} 
= -\frac{1}{2}\p_\mu\phi^\dag\p^\mu\phi
+ c_4\mathcal{L}_4 + c_6 \mathcal{L}_6 - V(\phi,\phi^\dag),
\label{eq:masterL_phi}
\eeq
where the Skyrme term and BPS-Skyrme term now read
\begin{align}
\mathcal{L}_4 &=
-\frac{1}{4}(\p_\mu\phi^\dag\p^\mu\phi)^2  
+\frac{1}{16}(\p_\mu\phi^\dag\p_\nu\phi + \p_\nu\phi^\dag\p_\mu\phi)^2,
\\
\mathcal{L}_6 &=
\frac{1}{4}
\left(\epsilon^{\mu\nu\rho\sigma}
  \phi^\dag\p_\nu\phi\p_\rho\phi^\dag\del_\sigma\phi\right)^2.
\end{align}

The Lagrangian density \eqref{eq:L} enjoys manifest
$\mathrm{SU}(2)\times\mathrm{SU}(2)$ 
symmetry when the potential is switched off; this symmetry is however 
spontaneously broken to its diagonal subgroup,
$\mathrm{SU}(2)\times\mathrm{SU}(2)\to\mathrm{SU}(2)$ by the presence
of any finite-energy configuration. 
Turning on a mass term for the pions, e.g.~$V\sim\Tr[\mathbf{1}_2-U]$,
results in the same symmetry breaking, however explicitly.
This symmetry breaking is important because it is the basis of the
existence of the Skyrmion or simply baryon charge.
The Skyrmions are characterized by the degree of the map from space
with infinity identified as a point
($\mathbb{R}^3\cup\{\infty\}\simeq S^3$) to SU(2),
\beq
\pi_3({\rm SU}(2)) \simeq \pi_3(S^3) = \mathbb{Z} \ni B.
\eeq
The integer $B$ is the degree, the topological charge or simply the
baryon number, and can be calculated from a configuration as 
\begin{align}
B &= -\frac{1}{24\pi^2} \int d^3x \;
\epsilon^{ijk}\, \Tr\left(L_i L_j L_k\right) \non
&= -\frac{1}{4\pi^2} \int d^3x \; \epsilon^{ijk} 
  \phi^\dag\del_i\phi\del_j\phi^\dag\del_k\phi.
\label{eq:B}
\end{align}
In order to construct a vortex in our model at hand, we need a special
potential that breaks the symmetry explicitly and further than to
simply SU(2). One such potential -- which was inspired by
Bose-Einstein condensates \cite{Ruostekoski:2001fc,Battye:2001ec} 
-- was considered in
Ref.~\cite{Gudnason:2014hsa}. That potential is fourth order in $\phi$
and of the form $|\phi_1\phi_2|^2=|\phi_1|^2(1-|\phi_1|^2)$.
This potential breaks the symmetry of the model to U(1)$\times$U(1),
explicitly. 
One of the U(1)s are then used to construct the vortex and the other
manifests itself as a U(1) modulus living on the vortex worldsheet. 

In this paper, we show that there is an even simpler potential than
that considered in Ref.~\cite{Gudnason:2014gla,Gudnason:2014hsa},
which allows for vortices in the model and it reads
\beq
V^{\rm vortex} = \frac{1}{2}
m^2\left(1 - \frac{1}{2}\phi^\dag(\mathbf{1}_2 + \tau^3)\phi\right)
= \frac{1}{2} m^2\left(1 - |\phi_1|^2\right).
\label{eq:Vvortex}
\eeq
This is one of the purposes of this paper, namely to construct the
vortex strings in the simplest potential in Skyrme-type models. 

Although not manifest in the formulation of the
Lagrangian \eqref{eq:masterL_phi} in terms of $\phi$, it contains a 
symmetry group \cite{Gudnason:2014hsa} -- which is a subgroup of O(4)
-- in the absence of the potential term 
\beq
\tilde{G} \simeq \mathrm{U}(2),
\eeq
which can be seen as formed by 
\beq
\mathrm{U}(2)\simeq\mathrm{SU}(2)_{\rm L}\times\mathrm{U}(1)_{\rm R},
\eeq
where the latter U(1) group is generated by $\tau^3$ in
SU(2)$_{\rm R}$. 
The potential that allows for vortices will break
\beq
\tilde{G}\to G = \mathrm{U}(1)_0\times\mathrm{U}(1)_3,
\eeq
explicitly; U(1)$_0$ is the tracepart of U(2) and U(1)$_3$ is the U(1) 
subgroup of SU(2). 
Explicitly, each group acts on $\phi$ as
\begin{align}
\mathrm{U}(1)_0:&\qquad \phi\to e^{i\alpha}\phi,\\
\mathrm{U}(1)_3:&\qquad \phi\to e^{i\beta\tau^3}\phi.
\end{align}
Unlike the potential that was considered in
Ref.~\cite{Gudnason:2014hsa} which possessed two distinct vacua with
in turn each of their kind of vortices, the 
potential \eqref{eq:Vvortex} only has the vacuum
\beq
\langle\phi\rangle^{\rm T} = (e^{i\alpha},0),
\eeq
and the unbroken symmetry group $H$ is
\beq
H = \mathrm{U}(1)_{0-3}:&\qquad \phi\to
e^{i\alpha(\mathbf{1}_2-\tau^3)}\phi.
\eeq
We can therefore write the moduli space as
\beq
\mathcal{M}\simeq G/H =
\frac{\mathrm{U}(1)_0\times\mathrm{U}(1)_3}{\mathrm{U}(1)_{0-3}}\simeq 
\mathrm{U}(1)_{0+3},
\label{eq:Mvortex}
\eeq
which in turn gives rise to the nontrivial homotopy group
\beq
\pi_1(\mathcal{M}) = \mathbb{Z} \ni n,
\eeq
which admits vortices and we will denote the vortex winding number by
$n$. 

Although, as mentioned above, the system at hand generally contains
also nonzero baryon charge, the straight vortex in this model does
not \emph{a priori} contain such.
Inside the core of the vortex $|\phi_1|\simeq 0$ and so the vortex
carries a U(1) modulus, $|\phi_2|\simeq 1$. 
In order for the configuration to acquire baryon charge, we need to
twist the modulus; that is, the modulus needs to wind along the
string. A $2\pi$ winding of the U(1) modulus corresponds topologically 
to a single sine-Gordon kink on the vortex worldsheet.

Let us explicitly construct a potential that gives rise to
(sine-Gordon) kinks 
\beq
V^{\rm kink} = -\frac{1}{2} m_2^2 \left(\Re\phi_2\right)^2.
\label{eq:Vkink}
\eeq
Now with the addition of this kink potential, we should discuss the
vacuum manifold again in order to assess the topological charges
present in the theory. 

The total potential is given by 
\beq
V = V^{\rm vortex} + V^{\rm kink}
= \frac{1}{2}m^2\cos^2 v - m_2^2\cos^2 v\cos^2\beta,
\label{eq:Vtotal}
\eeq
where we have parametrized the field configuration as
\beq
\langle\phi\rangle^{\rm T} = (e^{i\alpha}\sin v, e^{i\beta}\cos v),
\label{eq:vacparm}
\eeq 
with $\alpha,\beta,v \in [0,2\pi)$, 
and thus for $m>m_2$, the vacuum solution is
\beq
v = \frac{\pi}{2} + q\pi, \qquad
\label{eq:vac1}
\eeq
where $q\in\mathbb{Z}$ is an integer that we without loss of
generality can set to zero: $q=0$, 
and $\beta$ is undetermined. 
For $m<m_2$ there exists another vacuum which possesses no vortices
and the vacuum solution is $v=q\pi$ and $\beta=q'\pi$ with
$q'\in\mathbb{Z}$. 
For both solutions $V^{\rm vortex} + V_2^{\rm kink}=0$ in the
vacua. 
Since we are interested in vortices, which exist only in the first
case we will choose $m>m_2$ and hence the first vacuum solution.
It is important here to note that the vacuum manifold, with the
addition of the quadratic kink potential, is unchanged with respect to 
the one for only the vortex potential \eqref{eq:Mvortex}.
Thus
\beq
\mathcal{M}_{V^{\rm vortex}+V^{\rm kink}} = \mathcal{M}_{V^{\rm vortex}} = U(1)_{0+3}.
\label{eq:Mvacp2}
\eeq

We can think of the vortex solution in the following way. If we
identify $\alpha=\phi$ to be the winding phase and $\sin v$ to be the
radial profile function of the vortex, then $\beta$ is a U(1)
modulus field living on the vortex string when the kink potential is
turned off.
When we turn on the kink potential $m_2>0$, then although $\beta$ can 
take any value in the vacuum, $\beta=q'\pi$ induces the smallest
vortex mass (and thus smallest vortex tension); the vacua of the
modulus on the string worldsheet are $\beta=q'\pi$, again with
$q'\in\mathbb{Z}$. 

We can now evaluate the topological kink number on the vortex core.
The vortex core is defined by $\phi_1=0$ and so due to the nonlinear
sigma model constraint \eqref{eq:NLSMconstraint}, we have
$|\phi_2|=1$. In the parametrization \eqref{eq:vacparm} this
corresponds to $v=q\pi$ and the vacua for $\beta=q'\pi$ on the string
worldsheet as mentioned just above. 
The topological kink number is thus
\beq
\pi_0(\mathcal{M}^{\rm vortex}) = \pi_0(\mathbb{Z}) = \mathbb{Z} \ni k,
\eeq
where $k$ is the number of sine-Gordon halfkinks.
We define these kinks to be halfkinks because they only wind $\pi$ 
in the U(1) modulus, as compared to $2\pi$, for which we would call
the kinks full kinks or simply kinks.

The baryon number in the bulk of the total system can be calculated as 
\beq
B = \frac{n k}{2},
\eeq
where $n$ is the vortex number and $k$ is the sine-Gordon halfkink
number.

\section{The straight vortex}\label{sec:straightvortex}

We are now ready to construct the vortex in the model introduced in
the last section.
With no potential, a string-like configuration is of course unstable
to decay \cite{Nitta:2007zv}, but the potential \eqref{eq:Vvortex}
considered here admits a stable vortex string. 
We begin with the case where the vacuum is at
$|\phi_1|=1$ and so $v=\pi/2$.
This for the kinkless case: $m_2=0$. 
Now we can choose an appropriate Ansatz for the vortex as
\beq
\phi_1 = \sin f(r) e^{i\phi}, \qquad
\phi_2 = \cos f(r) e^{i\chi}, \label{eq:vortex_ansatz}
\eeq
where $f(r)$ is the profile of the vortex and $(r,\phi)$ are polar
coordinates in the $xy$-plane: $x=r\cos\phi$ and $y=r\sin\phi$.
The constant $\chi$ is the U(1) modulus residing on the vortex string
and in this section we take it to be its vacuum value which is
$\chi=q'\pi$ and so without loss of generality we can set $\chi=0$.
Inserting the above Ansatz into the Lagrangian
density \eqref{eq:masterL_phi}, we get 
\beq
-\mathcal{L} = 
\frac{1}{2}f_r^2
+\frac{1}{2r^2}\sin^2 f
+\frac{c_4}{2r^2}\sin^2(f) f_r^2
+\frac{1}{2} m^2 \cos^2(f),
\label{eq:Lf}
\eeq
which by variation with respect to the vortex profile function $f$,
leads to the equation of motion
\beq
f_{rr} + \frac{1}{r} f_r - \frac{1}{2r^2} \sin 2f
+\frac{c_4}{r^2} \sin^2 f\left(f_{rr} - \frac{1}{r} f_r\right)
+\frac{c_4}{2r^2} \sin(2f) f_r^2
+\frac{1}{2} m^2 \sin 2f = 0, \label{eq:eomf}
\eeq
where $f_r\equiv \p_rf$, etc.
In order to construct a vortex solution, we need to impose the
following boundary conditions on the vortex profile function $f$,
\beq
f(0) = 0, \qquad
f(\infty) = \frac{\pi}{2}.
\label{eq:BCvortex}
\eeq

First notice that the vortex Lagrangian density does not contain any 
term stemming from the sixth-order BPS-Skyrme term present in
Eq.~\eqref{eq:masterL_phi}. This is clear because first when the
vortex string has a nontrivial dependence along the string (the
$z$-direction in our frame), then the baryon charge operator (which is
the square root of the BPS-Skyrme term) can acquire nonvanishing
values.

Around $r=0$ we can expand $f$ in a power series in $r$ and expanding
the equation of motion \eqref{eq:eomf} up to third order, determines
$f$ up to fifth order
\beq
f = f_1 r - \frac{24f_1 + f_1^3 - 2f_1^5}{12 + 48f_1^2}\, r^3
+ \mathcal{O}(r^5),
\eeq
in terms of $f_1$, which is called the shooting parameter and encodes
nonperturbative information about the vortex.
Expanding instead the vortex profile function around spatial infinity
leads to
\beq
f \simeq \gd(m r),
\label{eq:asymptoticf}
\eeq
where $\gd$ is the Gudermannian function defined by
$\gd(z)\equiv 2\arctan(e^z) - \pi/2$.

A comment in store is about the energy. The energy density for static
configurations is simply given by $-\mathcal{L}$, where the Lagrangian
density is that of Eq.~\eqref{eq:Lf}.
The total energy, however, diverges, in more than one (spatial)
direction. Considering first the plane transverse to the vortex (for
concreteness we can choose this plane to be the $xy$-plane), the
vortex tension diverges logarithmically because the vortex is a global 
vortex \footnote{This is in agreement with Derrick's
theorem \cite{Derrick:1964ww}. }.
Next, if the vortex is not compactified on a circle, i.e.~if it is an
infinitely long straight string, then the total energy diverges like
$L\log L$, where $L$ is a cutoff in length scale.

Let us now consider the tension in more detail. As we already
mentioned, the tension (energy per unit length) diverges
logarithmically. Let us consider the contributions from the different 
terms to the asymptotic tension, by plugging the asymptotic form of
the profile function \eqref{eq:asymptoticf} into the energy
($-\mathcal{L}$ in Eq.~\eqref{eq:Lf}) and integrate over the
$xy$-plane (the plane transverse to the vortex). 
The kinetic term has two contributions to the energy
\begin{align}
&\lim_{R\to\infty}\left[\int dr\; r f_r^2
\sim \int dr\; \frac{2m e^{m r}}{1 + e^{2m r}}
= 2r\arctan e^{m r} - \frac{i}{m}\Li_2\left(-i e^{m r}\right)
  + \frac{i}{m}\Li_2\left(i e^{m r}\right) \right]^{r=R}=0, \nonumber\\
&\int^R dr\; \frac{1}{r}\sin^2f
\sim \int^R dr\; \frac{\tanh^2 m r}{r}
\sim \log R,
\end{align}
but only the latter diverges (logarithmically).
The contribution from the Skyrme term, however, is finite
\beq
\lim_{R\to\infty}\left[\int dr\; \frac{1}{r}\sin^2(f) f_r^2
\sim \int dr\; \frac{m}{r}\sech m r \tanh m r\right]^{r=R} = 0,
\nonumber
\eeq
as is that of the potential
\begin{align}
&\lim_{R\to\infty}\left[\int dr\; r m^2 \cos^2 f
\sim \int dr\; \frac{m^2 r e^{2 m r}}{(1 + e^{2 m r})^2}
=\frac{m R(1 + \tanh m R) -\log[1 + e^{2m R}]}{4}
\right]^{r=R} = 0. \nonumber
\end{align}

\begin{figure}[!tp]
\begin{center}
\mbox{\subfloat[]{\includegraphics[width=0.49\linewidth]{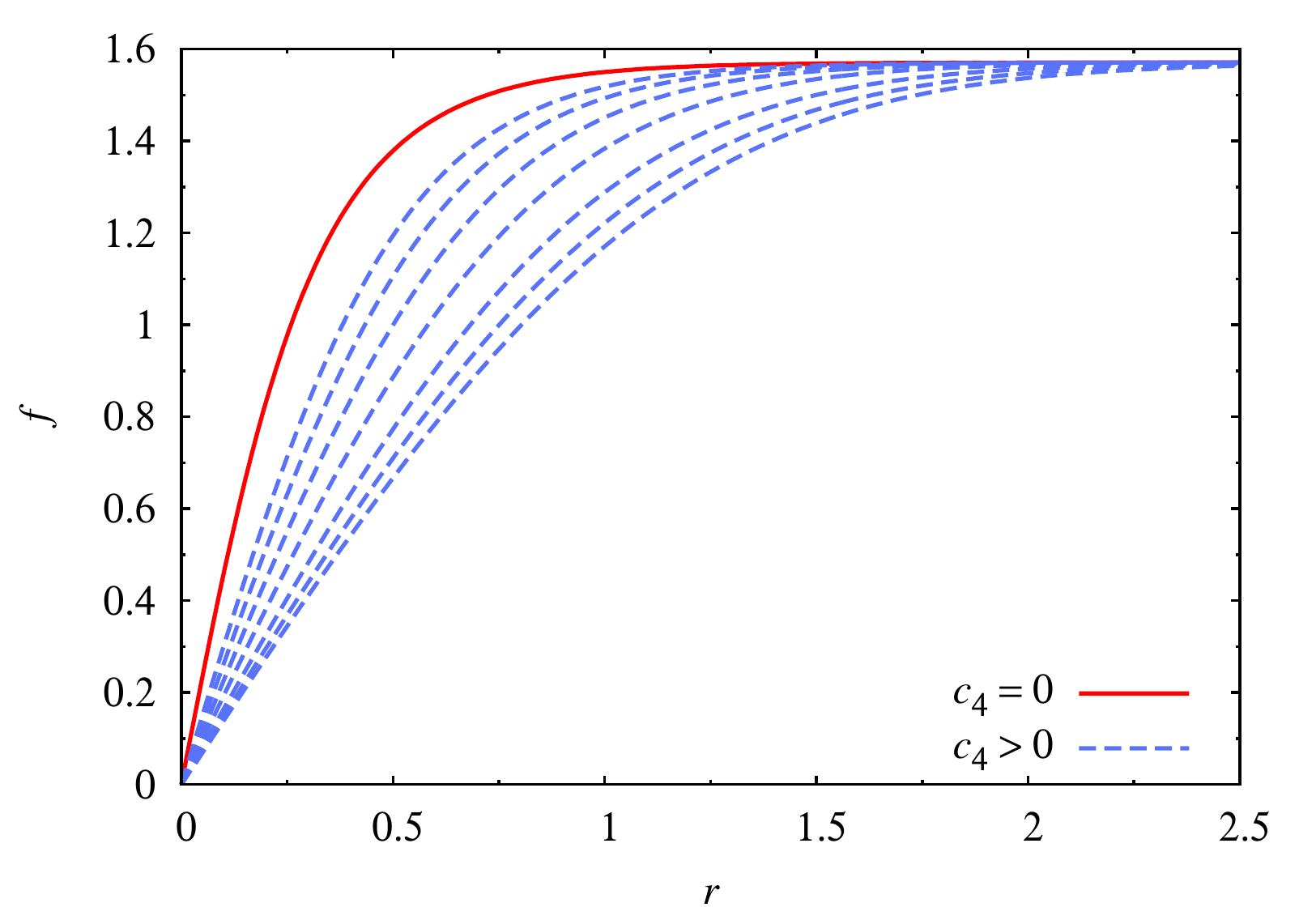}}
\subfloat[]{\includegraphics[width=0.49\linewidth]{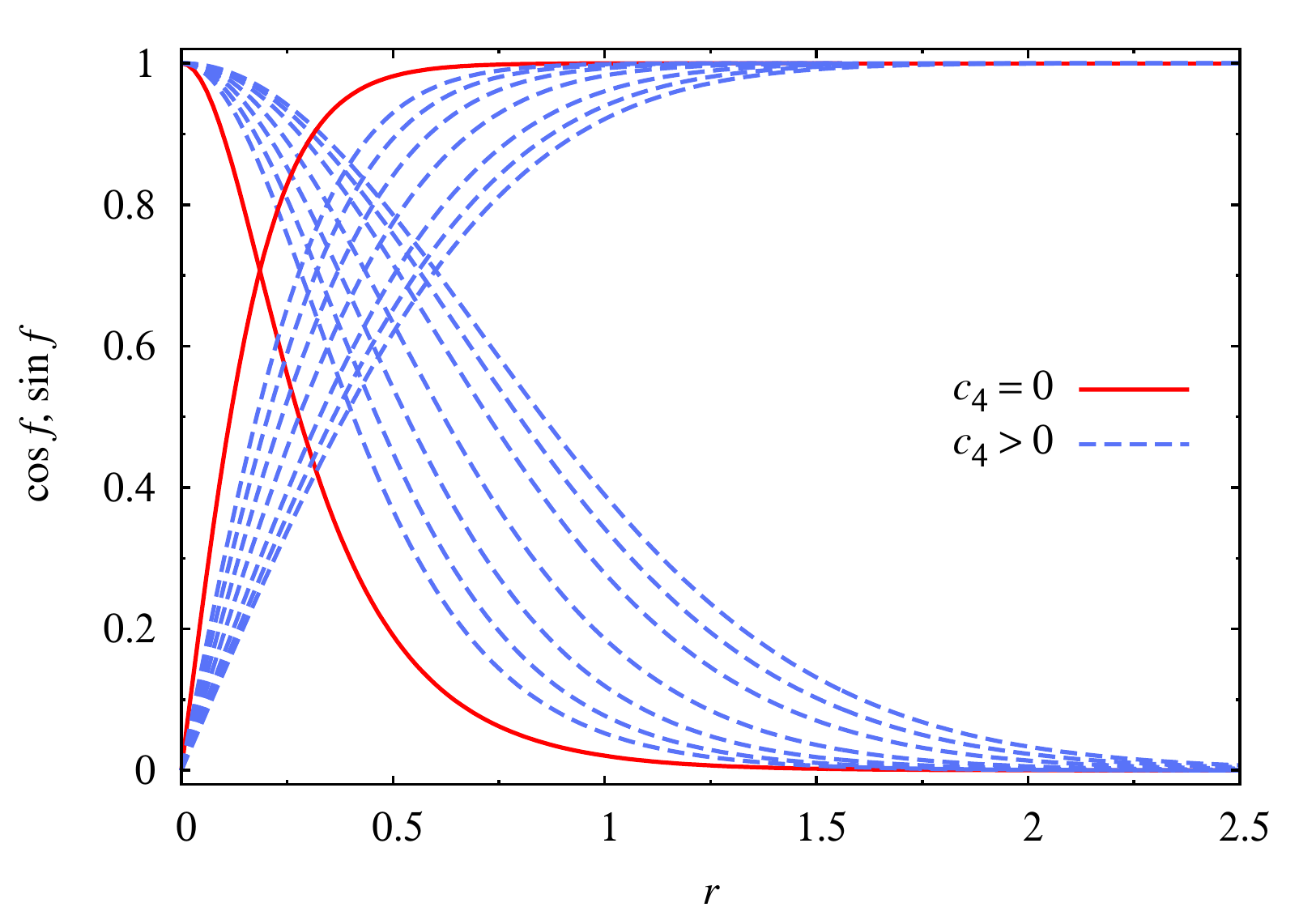}}}
\mbox{\subfloat[]{\includegraphics[width=0.49\linewidth]{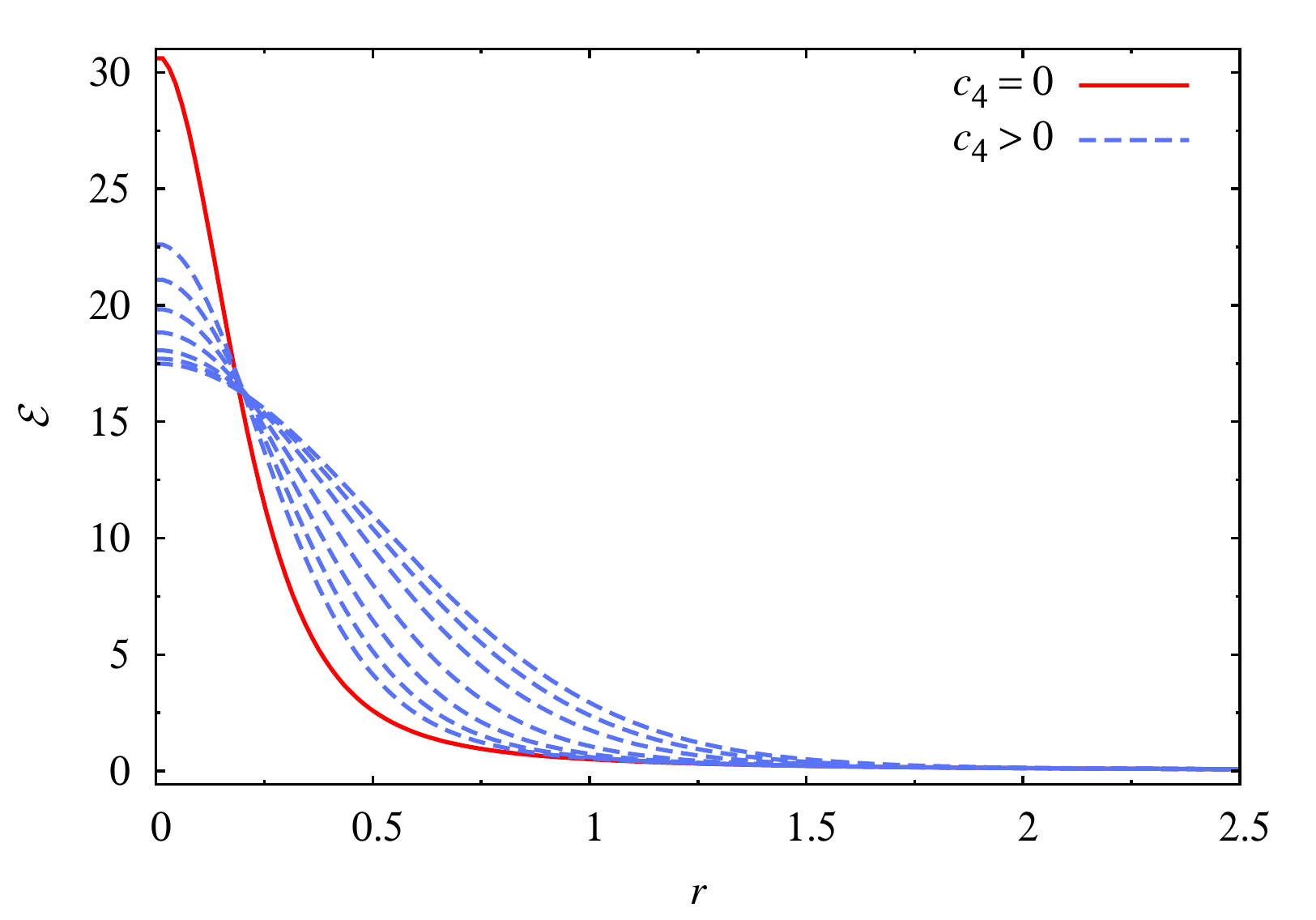}}
\subfloat[]{\includegraphics[width=0.49\linewidth]{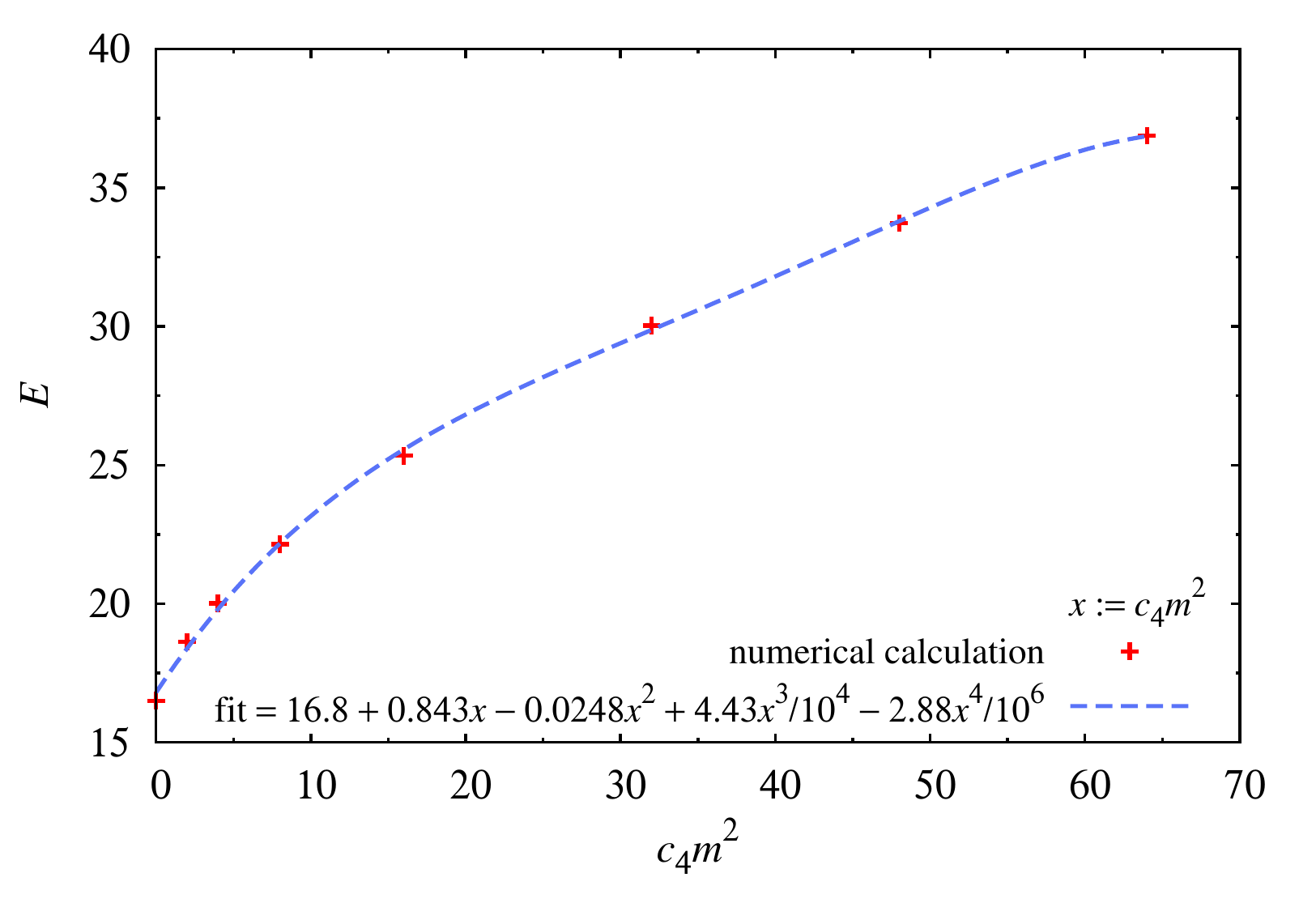}}}
\caption{Vortex solutions for various values of
$c_4=0,0.125,0.25,0.5,1,2,3,4$: (a) profile function $f$, (b)
the condensate $|\phi_1|=\sin f$ and $|\phi_2|=\cos f$, (c) energy
density and (d) total energy. 
For the figure we have fixed the length scales by setting $m=4$. }
\label{fig:skvtx1}
\end{center}
\end{figure}

As the equation of motion \eqref{eq:eomf} cannot be solved
analytically, we turn to numerically methods; namely we use a
fourth-order Runge-Kutta method to find numerical solutions for
various values of $c_4$. The solutions, the corresponding condensate, 
energy densities and integrated energies are shown in
Fig.~\ref{fig:skvtx1}. 

The mass parameter $m$ can be scaled away and physically just
corresponds to setting the length scale. The only free parameter in
the system is then $c_4 m^2$ which turns on the Skyrme term and
results in a wider vortex string. The energy density of the wider
vortex string is likewise wider but also has a lower density at the
vortex core, see Fig.~\ref{fig:skvtx1}(c). The total energy, however, 
increases monotonically as function of $c_4m^2$, as expected, see
Fig.~\ref{fig:skvtx1}(d).

\section{Vortex rings as Skyrmions}\label{sec:vortexrings}

In this section, we study the vortex ring in the simpler vortex
potential \eqref{eq:Vvortex} as compared to that considered in
Refs.~\cite{Gudnason:2014gla,Gudnason:2014jga,Gudnason:2014hsa}.
We will consider this potential in two variants of our model, which we 
shall call the 2+4 model and the 2+6 model
\begin{align}
2+4\;{\rm model}: \quad & c_4=1,\quad c_6=0,\\
2+6\;{\rm model}: \quad & c_4=0,\quad c_6=1.
\end{align}
This choice of coefficients is made such that we can see the
differences between the normal Skyrme term and the sextic term of the 
BPS-Skyrme model.

Since the potential \eqref{eq:Vvortex} breaks spherical symmetry and
the potential \eqref{eq:Vkink} furthermore breaks axial symmetry, we
will perform the numerical calculations of the partial differential
equations (equations of motion) on a cubic lattice using the finite
difference method in conjunction with the relaxation method.
We typically use $121^3$ lattices, but also smaller lattices like
$81^3$ and $101^3$, where the denser grid is not necessary.
We should warn the reader that the configurations shown in the figures
are cropped to better show the features of the plots and thus the size
of the grids used appears to be smaller than what was used for the
calculations.

\subsection{Singly twisted vortex rings as Skyrmions}

\begin{figure}[!tp]
\begin{center}
\mbox{
\subfloat{\includegraphics[width=0.245\linewidth]{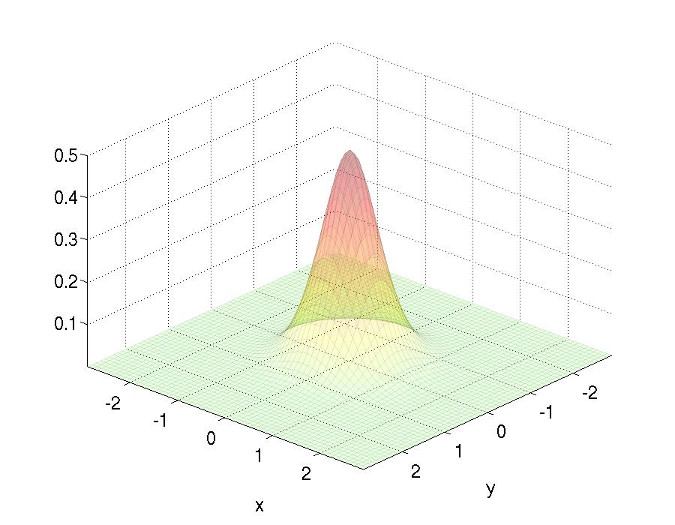}}
\subfloat{\includegraphics[width=0.245\linewidth]{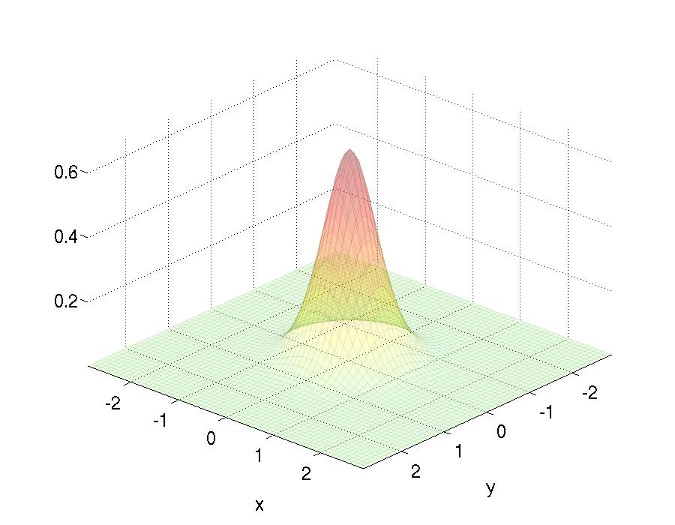}}
\subfloat{\includegraphics[width=0.245\linewidth]{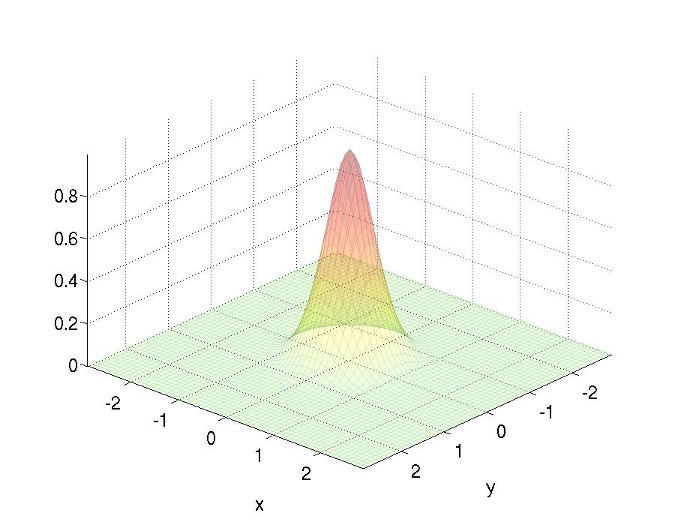}}
\subfloat{\includegraphics[width=0.245\linewidth]{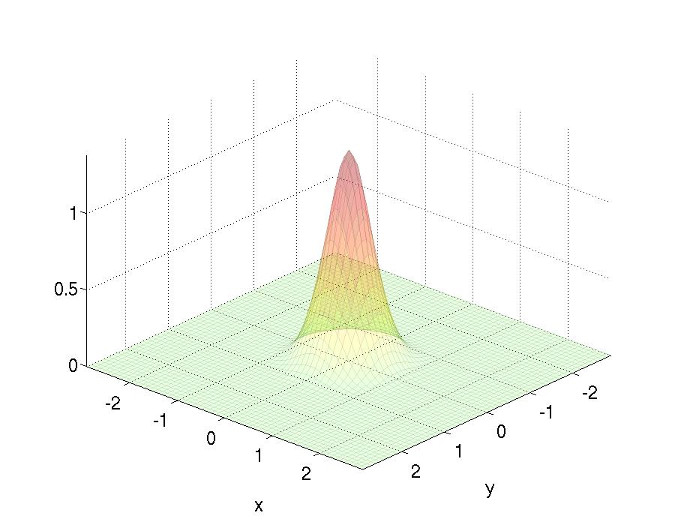}}}
\mbox{
\subfloat{\includegraphics[width=0.245\linewidth]{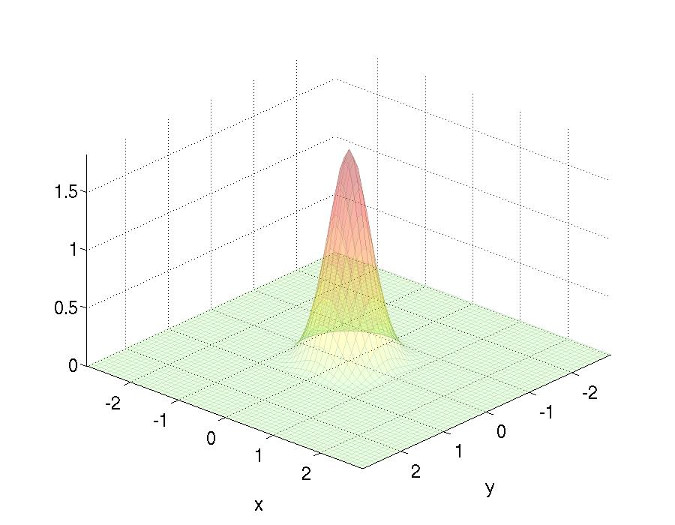}}
\subfloat{\includegraphics[width=0.245\linewidth]{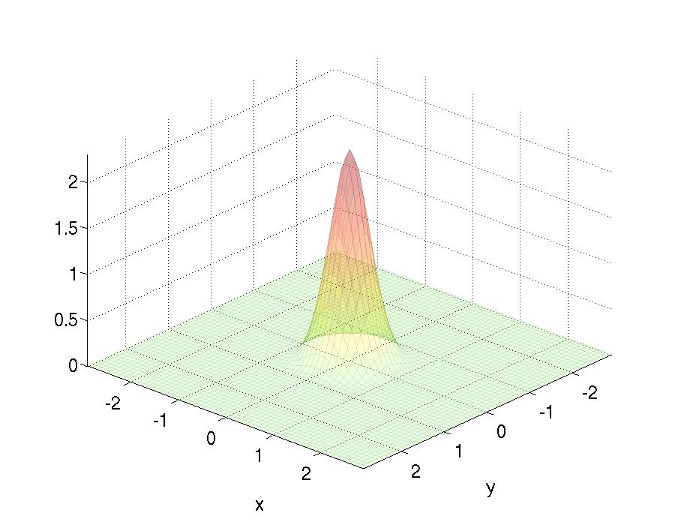}}
\subfloat{\includegraphics[width=0.245\linewidth]{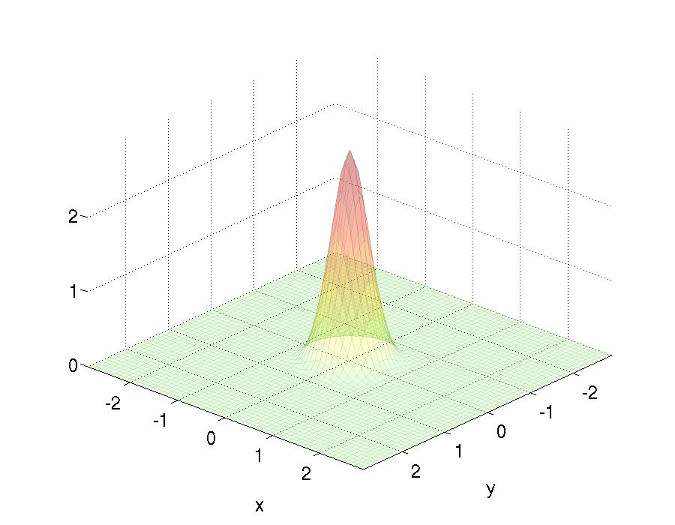}}
\subfloat{\includegraphics[width=0.245\linewidth]{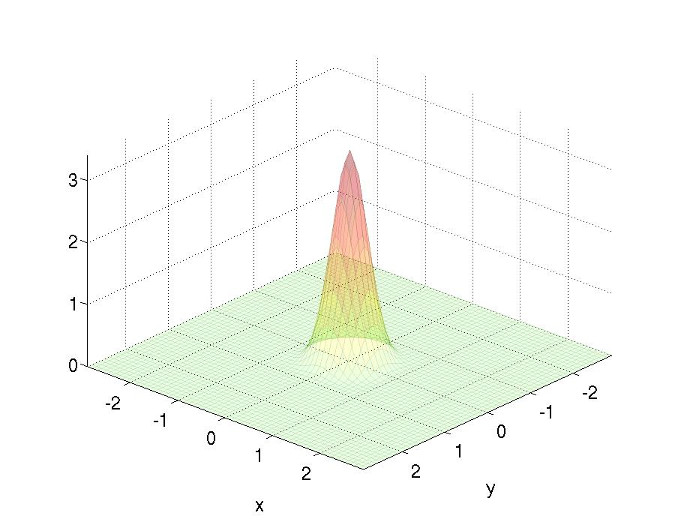}}}
\caption{Single vortex ring in the 2+4 model. The figure is an
$xy$-slice of the baryon charge density at $z=0$ for the vortex
potential mass parameter $m=0,1,2,3,4,5,6,7$ from top-left to
bottom-right 
panel. }
\label{fig:nout81_1_0_ms_baryonslice2}
\vspace*{\floatsep}
\mbox{
\subfloat{\includegraphics[width=0.245\linewidth]{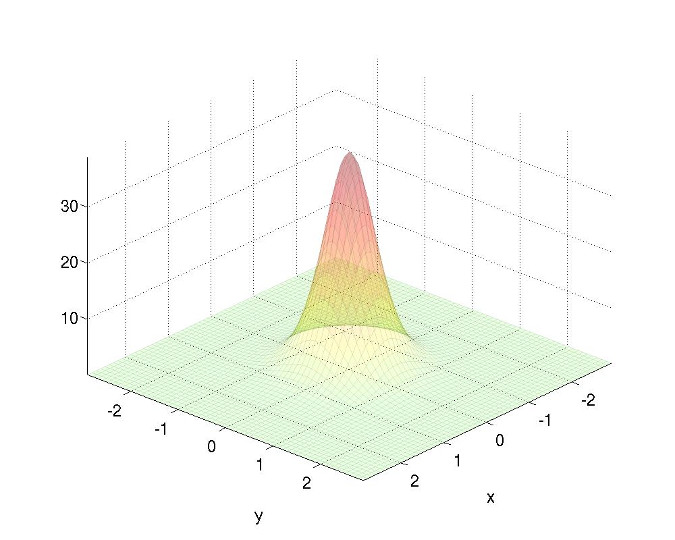}}
\subfloat{\includegraphics[width=0.245\linewidth]{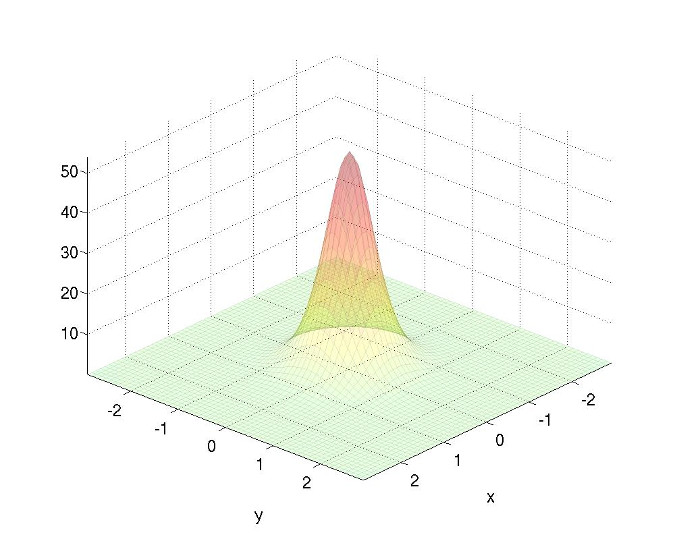}}
\subfloat{\includegraphics[width=0.245\linewidth]{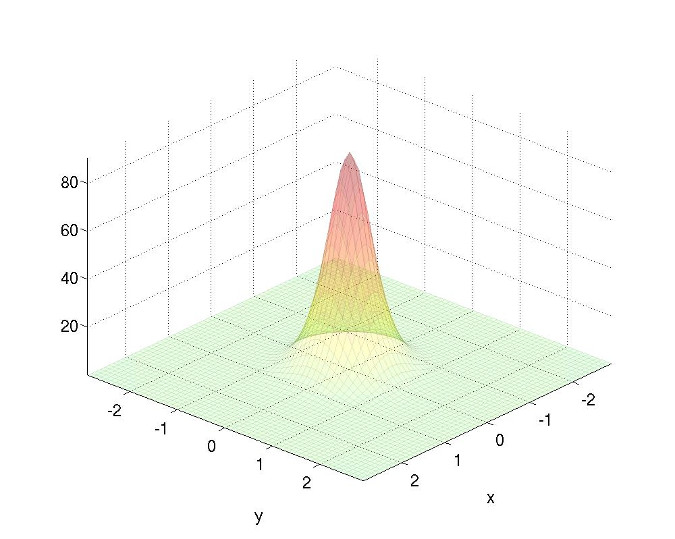}}
\subfloat{\includegraphics[width=0.245\linewidth]{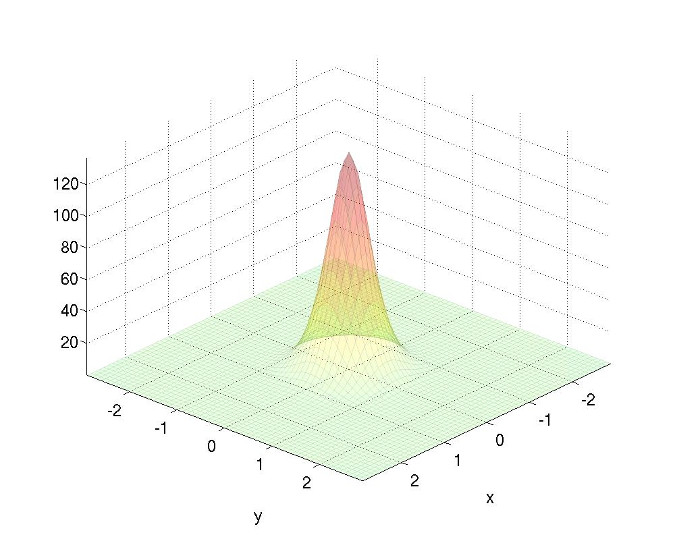}}}
\mbox{
\subfloat{\includegraphics[width=0.245\linewidth]{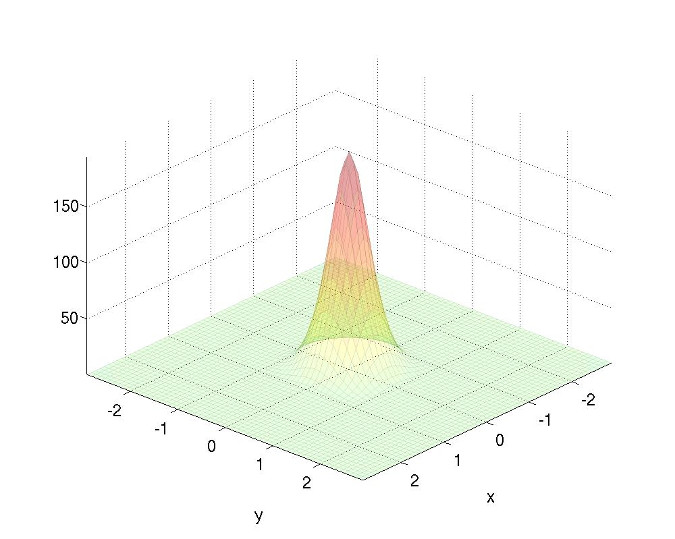}}
\subfloat{\includegraphics[width=0.245\linewidth]{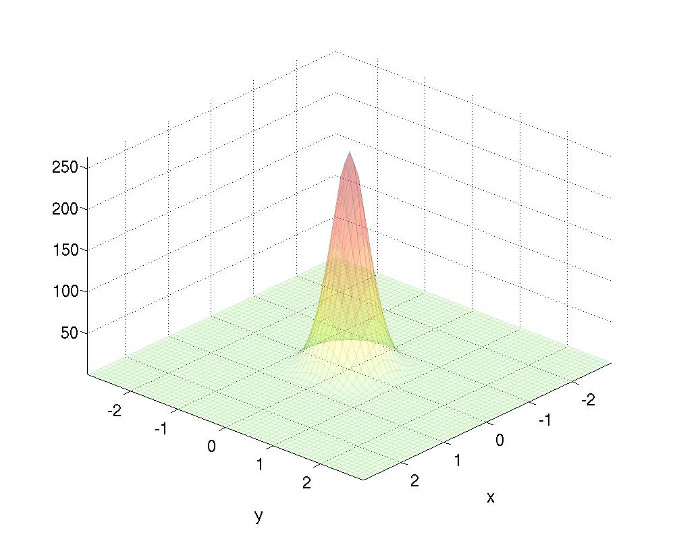}}
\subfloat{\includegraphics[width=0.245\linewidth]{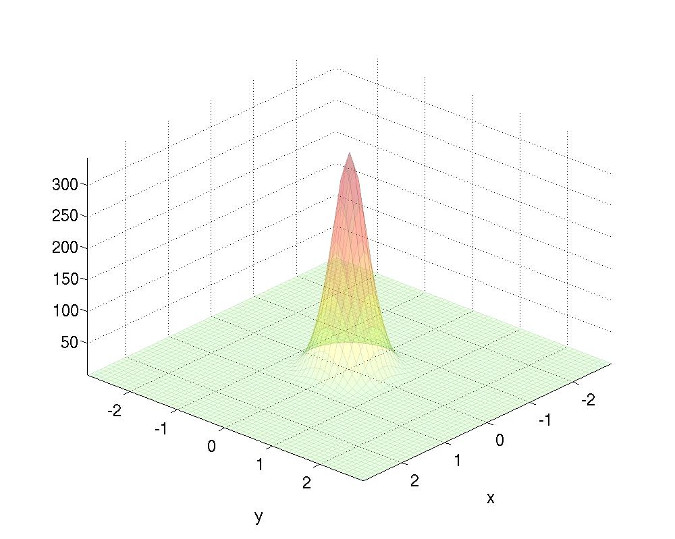}}
\subfloat{\includegraphics[width=0.245\linewidth]{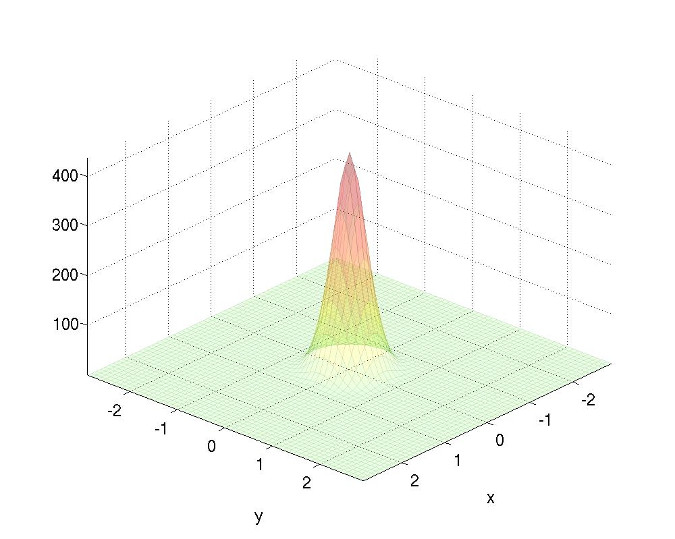}}}
\caption{Single vortex ring in the 2+4 model. The figure is an
$xy$-slice of the energy density at $z=0$ for the vortex potential 
mass parameter $m=0,1,2,3,4,5,6,7$ from top-left to bottom-right
panel. }
\label{fig:nout81_1_0_ms_energyslice2}
\end{center}
\end{figure}

We start by performing numerical calculations of the standard hedgehog
Skyrmion in the 2+4 model, gradually turning on the vortex
potential \eqref{eq:Vvortex}.
Our initial condition for the relaxation method is taken as the
hedgehog Ansatz in the following form
\beq
\phi_{\rm initial}^{\rm T} =
\left(\cos f - i\sin f\cos \theta,e^{i\phi}\sin f\sin\theta\right).
\label{eq:vortexring_ansatz}
\eeq
Figs.~\ref{fig:nout81_1_0_ms_baryonslice2}
and \ref{fig:nout81_1_0_ms_energyslice2} show $xy$-slices at $z=0$ of
the baryon charge densities and energy densities for the single
Skyrmion in the 2+4 model for various values of the vortex potential
mass parameter $m=0,1,\ldots,7$.
As can be seen from the figures, all the Skyrmion configurations are
nearly spherical and so we have shown only the $xy$-slices. 

\begin{figure}[!htp]
\begin{center}
\includegraphics[width=0.5\linewidth]{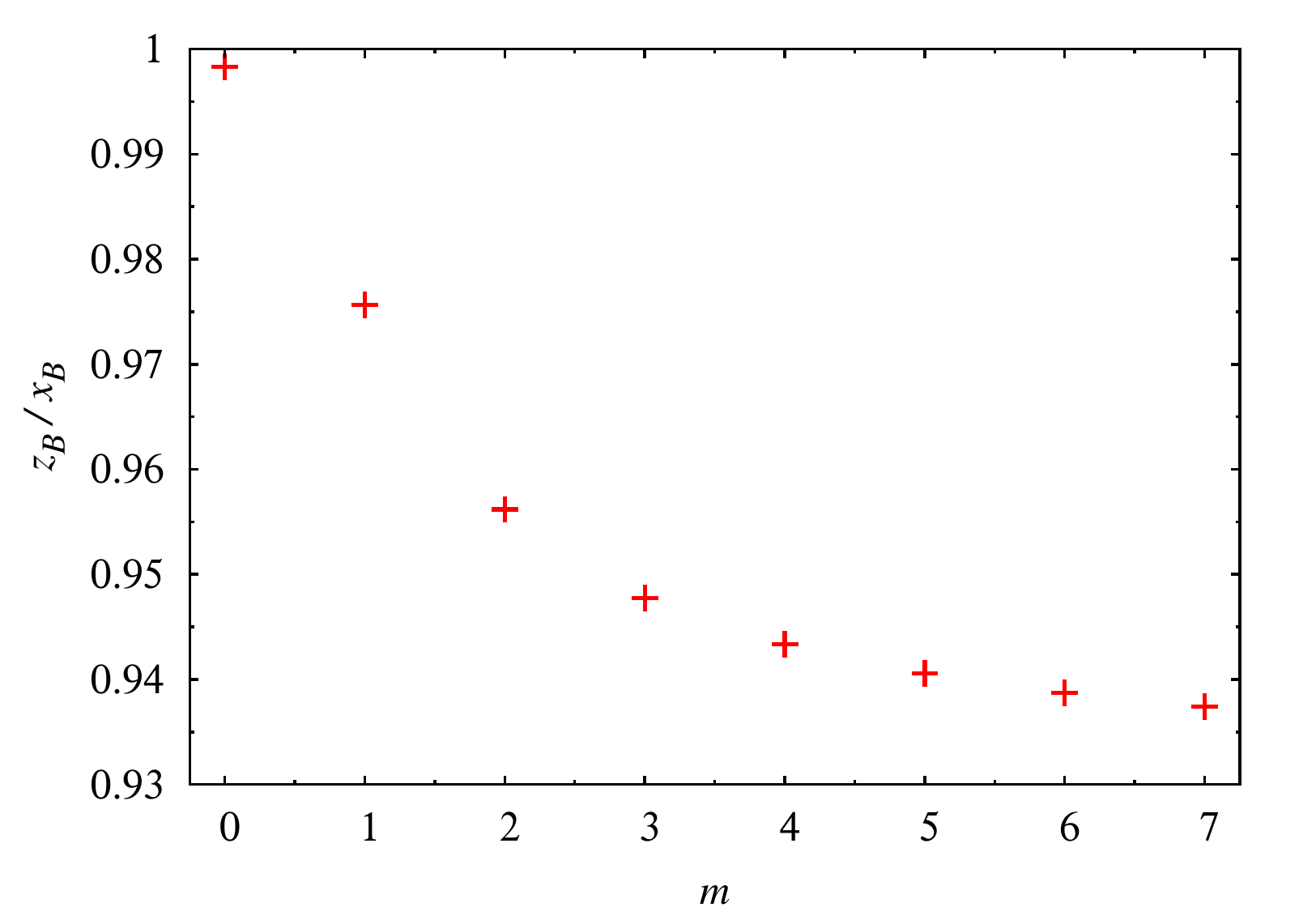}
\caption{The ratio of the square root of the second moment of the
baryon charge density in the $z$ and in the $x$ direction, $z_B/x_B$,
for the single vortex ring in the 2+4 model as function of the vortex
potential mass parameter $m$. }
\label{fig:ring240}
\end{center}
\end{figure}

In fact, the only effect of the vortex potential on the Skyrmion solutions
other than shrinking them and increasing their energies, is that their
spherical symmetry gets broken; that is, they become slightly squashed
spheres. To measure the squashing of the Skyrmions as function of the
vortex potential mass parameter $m$, we define
\beq
(x_B^i)^2 \equiv \frac{1}{B}\int d^3x\; (x^i)^2 \mathcal{B},
\label{eq:xiBsq}
\eeq
$i$ not summed over and
\beq
\mathcal{B} = -\frac{1}{4\pi^2}\epsilon^{ijk}
\phi^\dag\p_i\phi\p_j\phi^\dag\p_k\phi^\dag,
\eeq
is the baryon charge density. 
Then the size in one spatial direction is taken as
$x_B^i=\sqrt{(x_B^i)^2}$.
In Fig.~\ref{fig:ring240} we show the ratio of the size in the
$z$-direction to the $x$-direction (for these configurations
$x_B=y_B$). 

\begin{figure}[!tp]
\begin{center}
\mbox{
\subfloat{\includegraphics[width=0.245\linewidth]{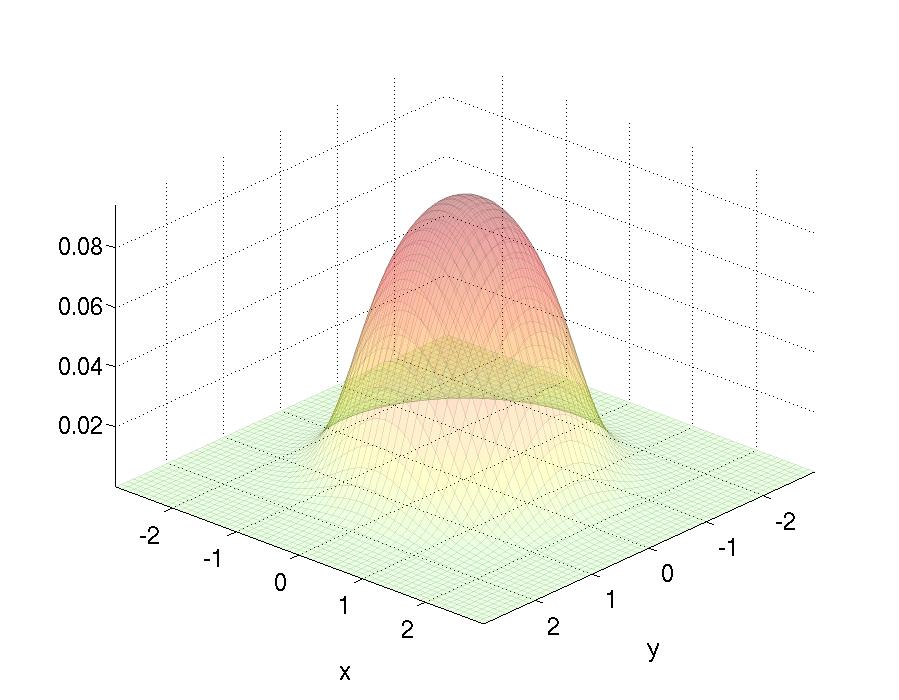}}
\subfloat{\includegraphics[width=0.245\linewidth]{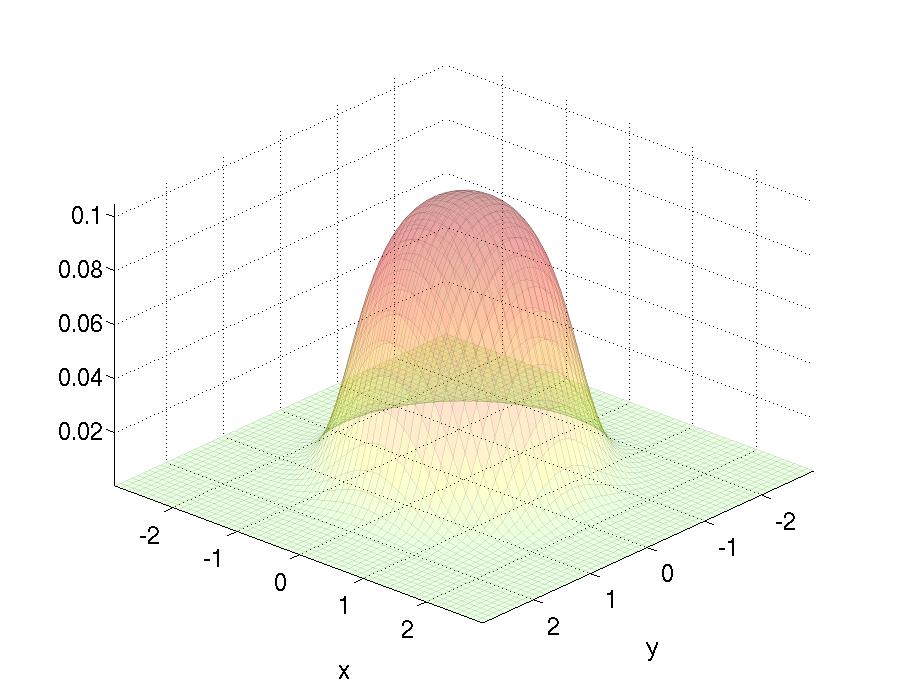}}
\subfloat{\includegraphics[width=0.245\linewidth]{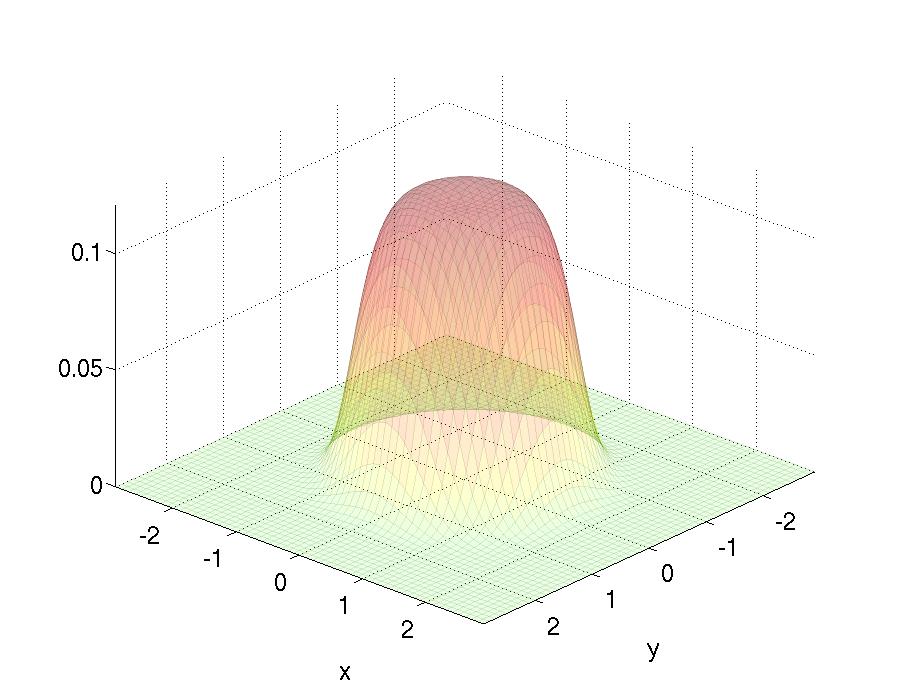}}
\subfloat{\includegraphics[width=0.245\linewidth]{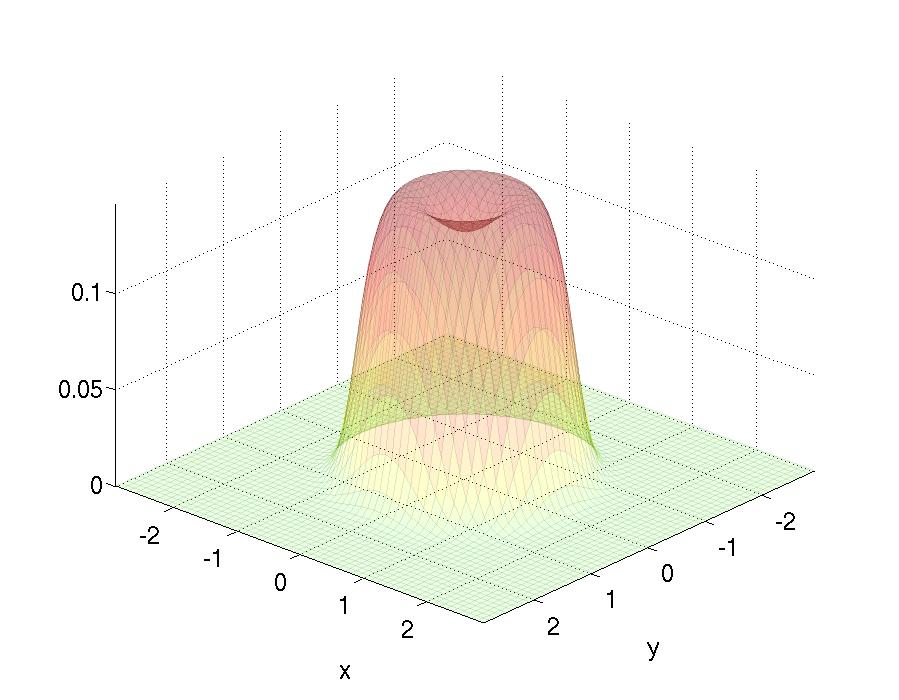}}}
\mbox{
\subfloat{\includegraphics[width=0.245\linewidth]{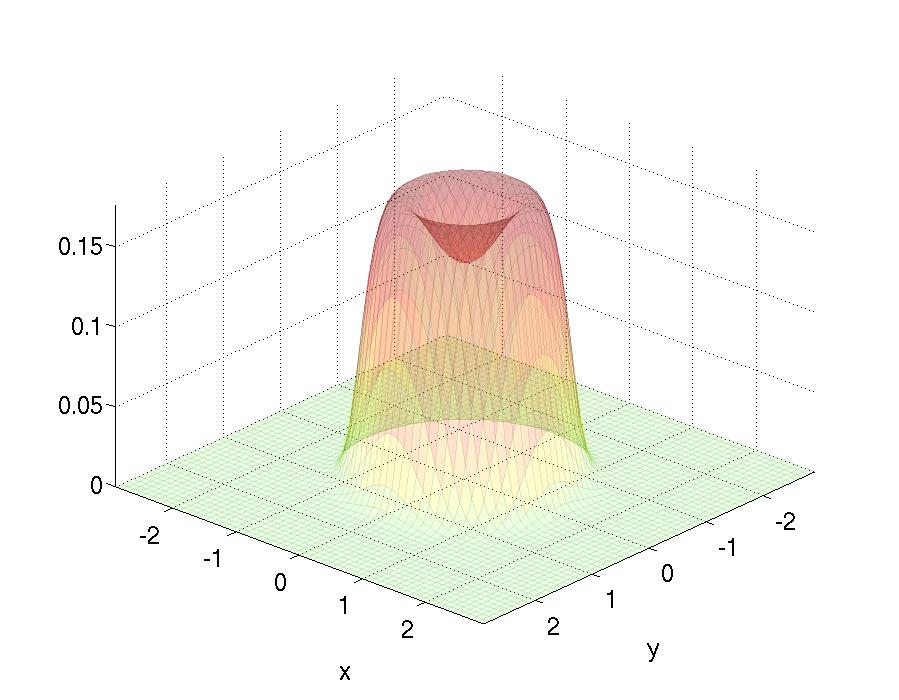}}
\subfloat{\includegraphics[width=0.245\linewidth]{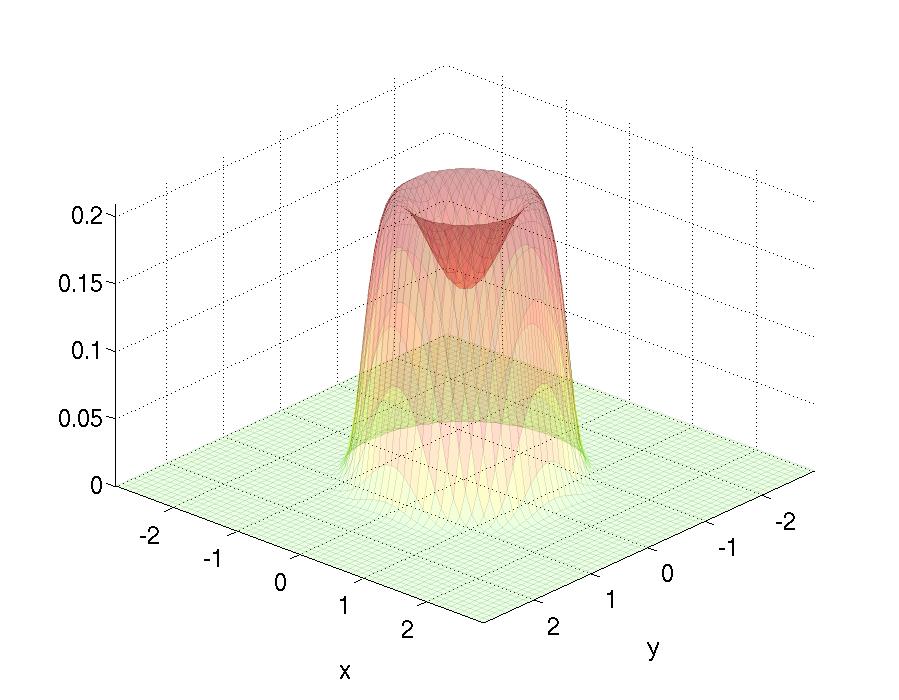}}
\subfloat{\includegraphics[width=0.245\linewidth]{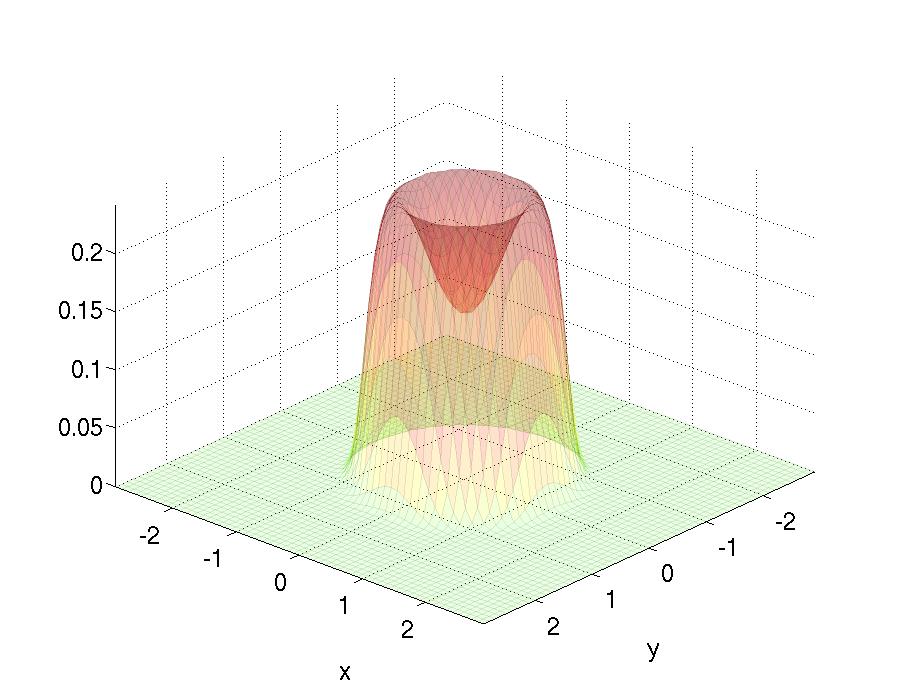}}
\subfloat{\includegraphics[width=0.245\linewidth]{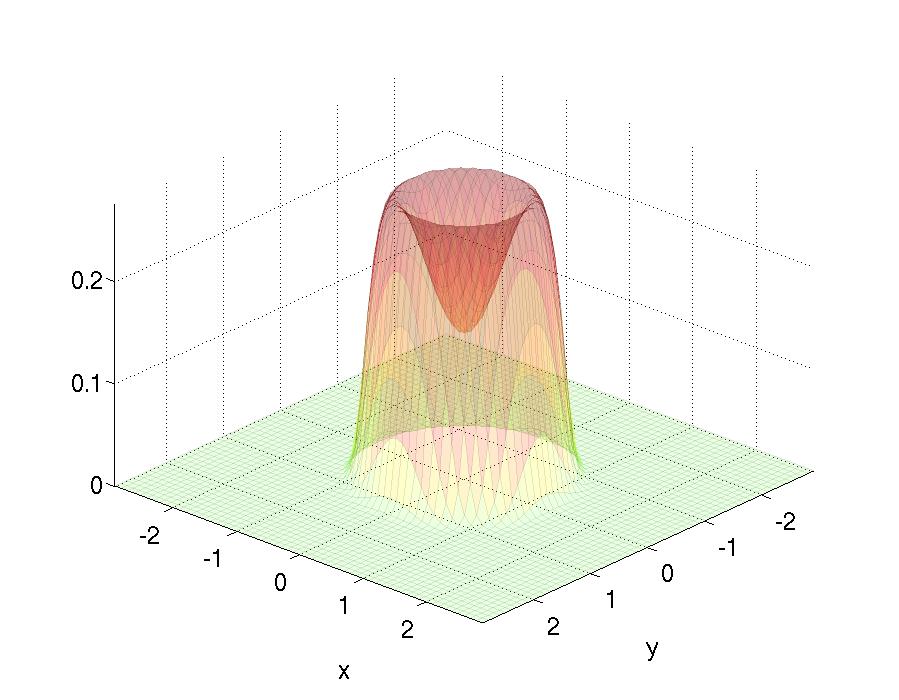}}}
\caption{Single vortex ring in the 2+6 model. The figure is an
$xy$-slice of the baryon charge density at $z=0$ for the vortex
potential mass parameter $m=0,1,2,3,4,5,6,7$ from top-left to
bottom-right panel. }
\label{fig:nout81_0_1_ms_baryonslice2}
\vspace*{\floatsep}
\mbox{
\subfloat{\includegraphics[width=0.245\linewidth]{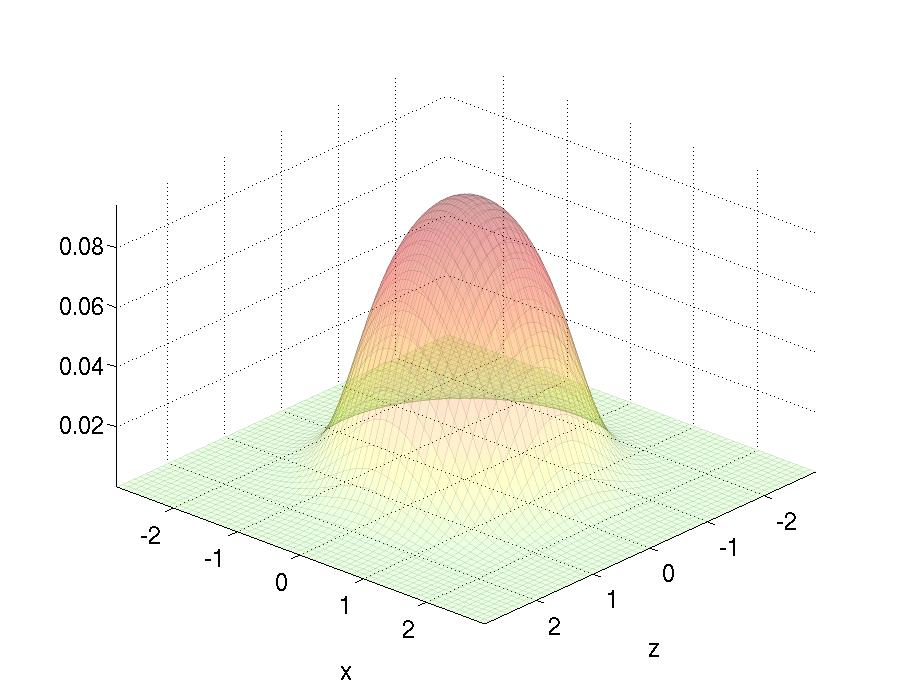}}
\subfloat{\includegraphics[width=0.245\linewidth]{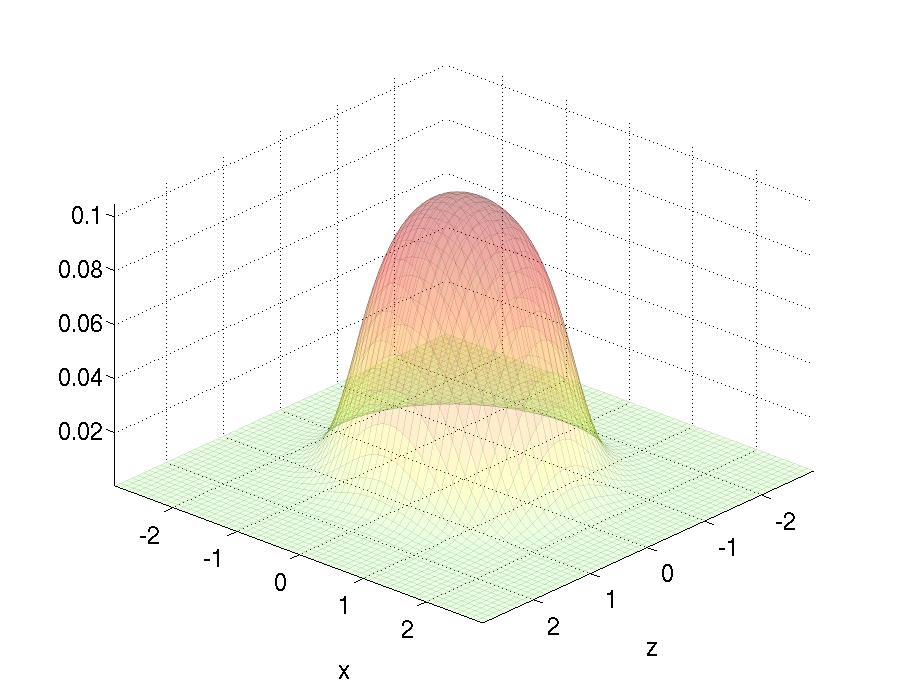}}
\subfloat{\includegraphics[width=0.245\linewidth]{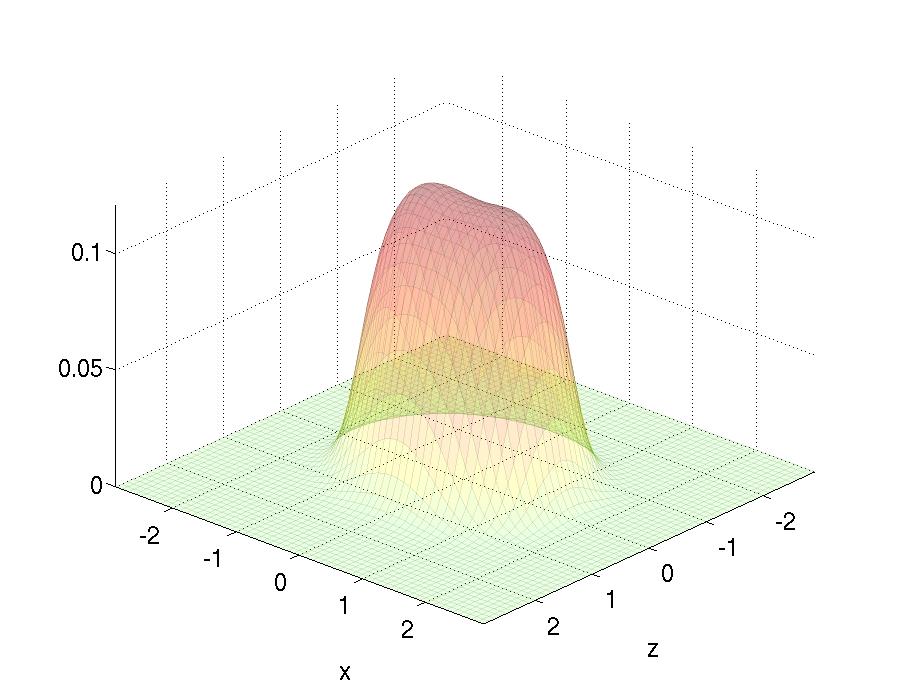}}
\subfloat{\includegraphics[width=0.245\linewidth]{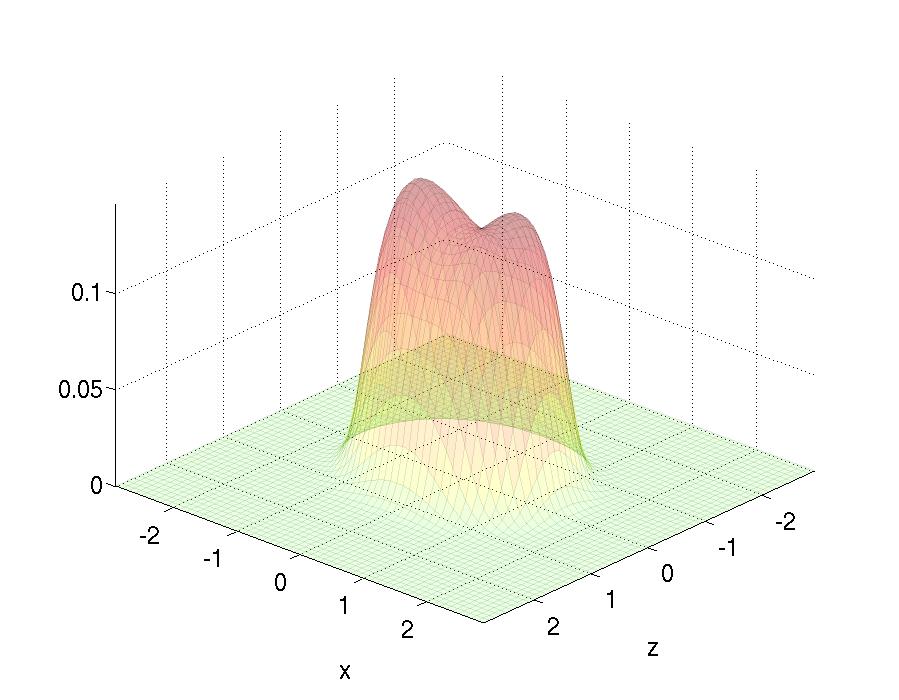}}}
\mbox{
\subfloat{\includegraphics[width=0.245\linewidth]{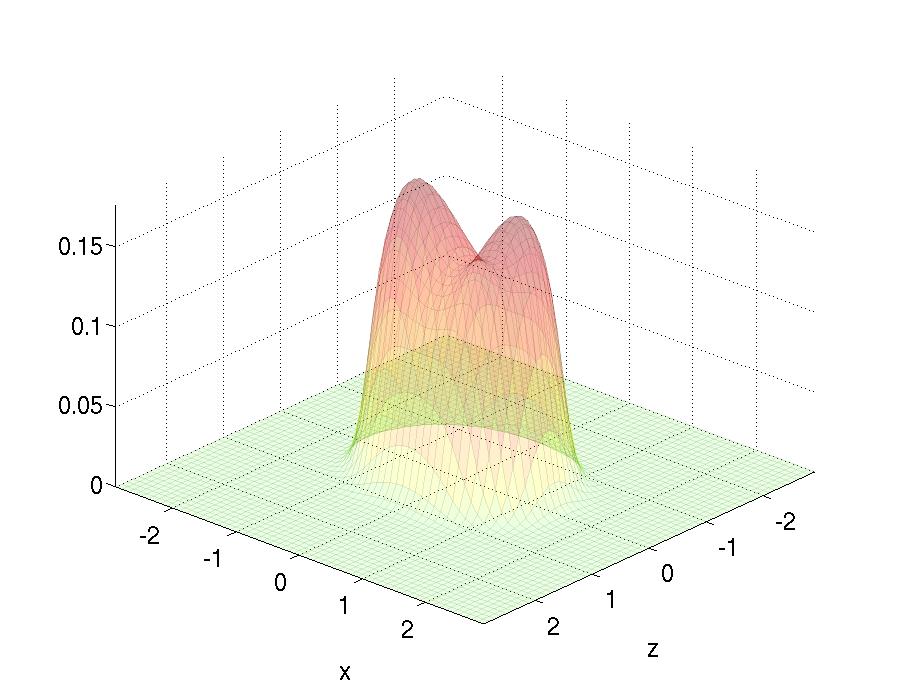}}
\subfloat{\includegraphics[width=0.245\linewidth]{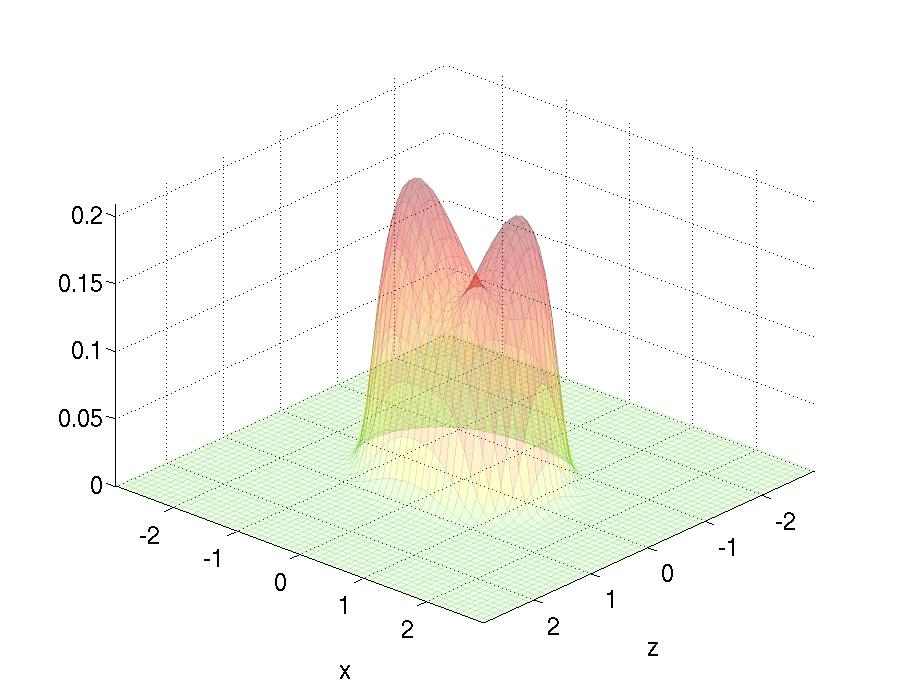}}
\subfloat{\includegraphics[width=0.245\linewidth]{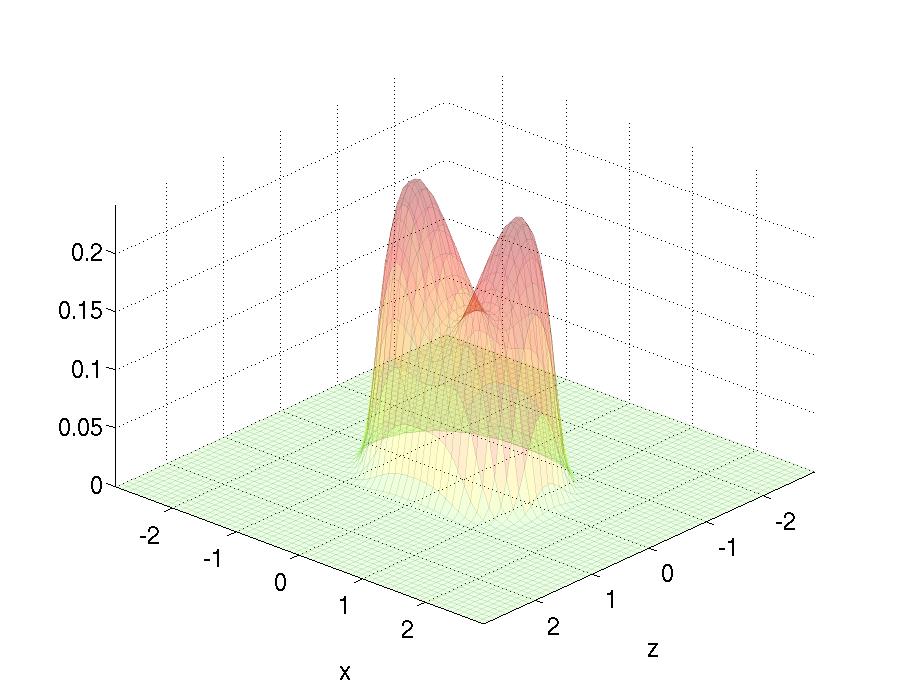}}
\subfloat{\includegraphics[width=0.245\linewidth]{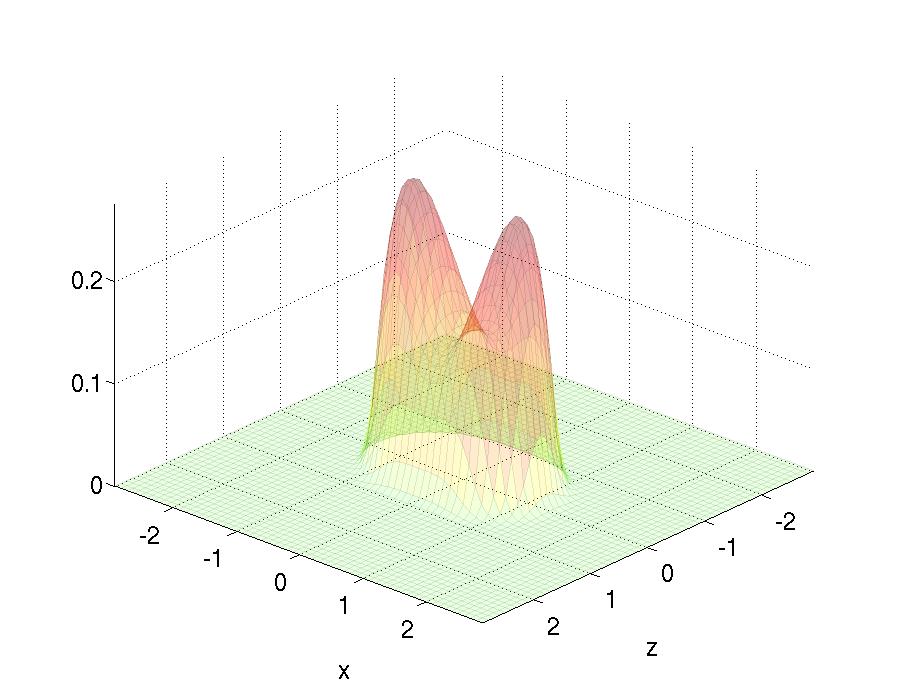}}}
\caption{Single vortex ring in the 2+6 model. The figure is an
$xz$-slice of the baryon charge density at $y=0$ for the vortex
potential mass parameter $m=0,1,2,3,4,5,6,7$ from top-left to
bottom-right panel.}
\label{fig:nout81_0_1_ms_baryonslice}
\end{center}
\end{figure}

\begin{figure}[!tp]
\begin{center}
\mbox{
\subfloat{\includegraphics[width=0.245\linewidth]{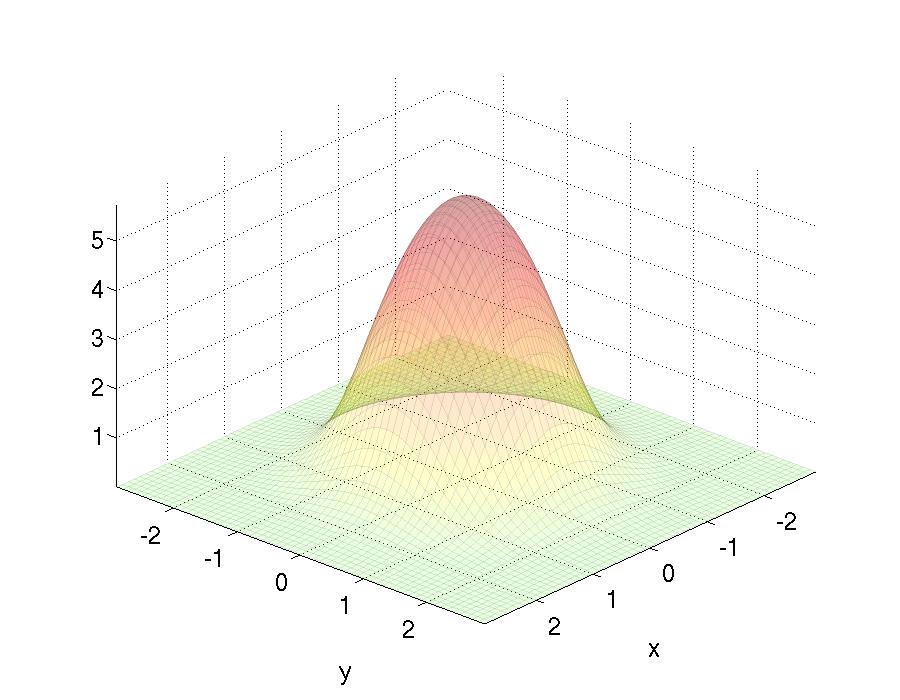}}
\subfloat{\includegraphics[width=0.245\linewidth]{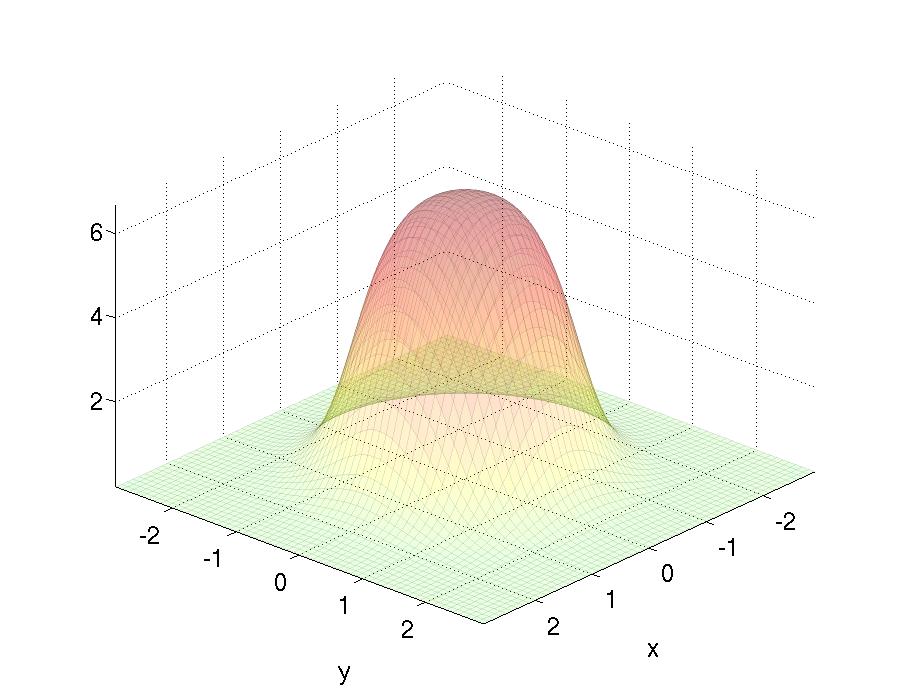}}
\subfloat{\includegraphics[width=0.245\linewidth]{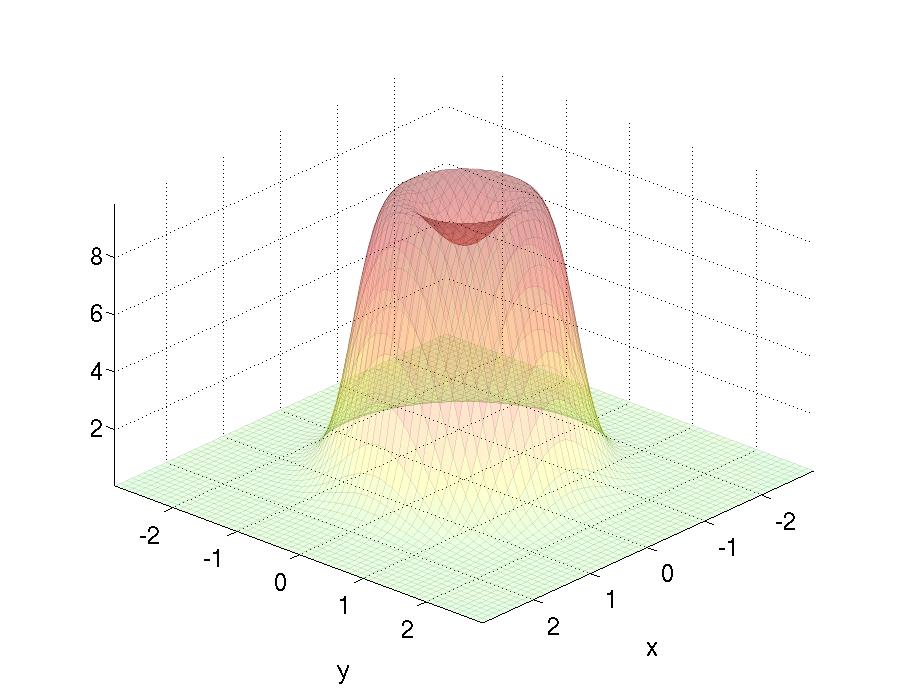}}
\subfloat{\includegraphics[width=0.245\linewidth]{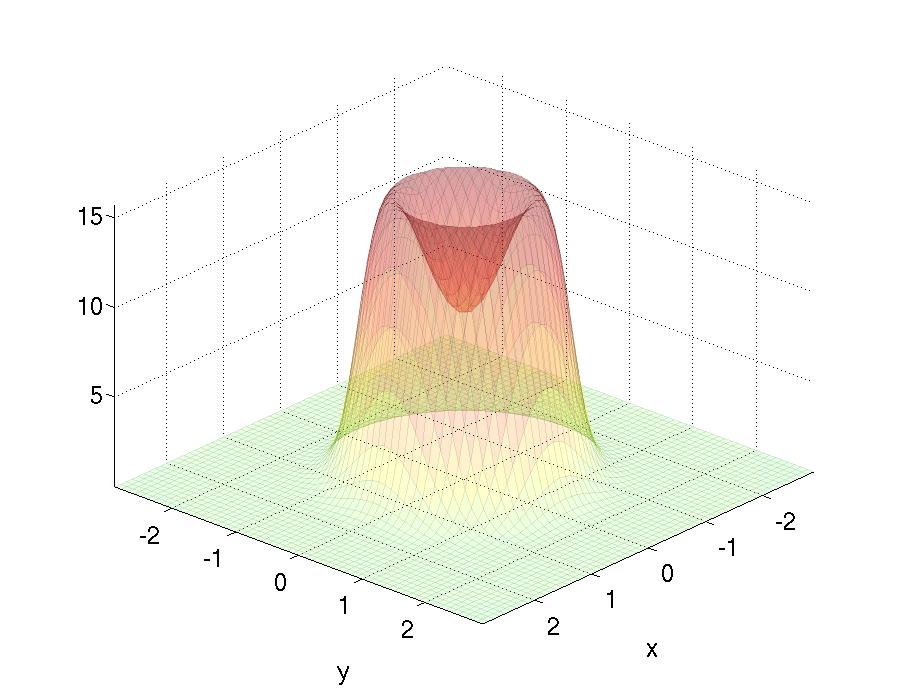}}}
\mbox{
\subfloat{\includegraphics[width=0.245\linewidth]{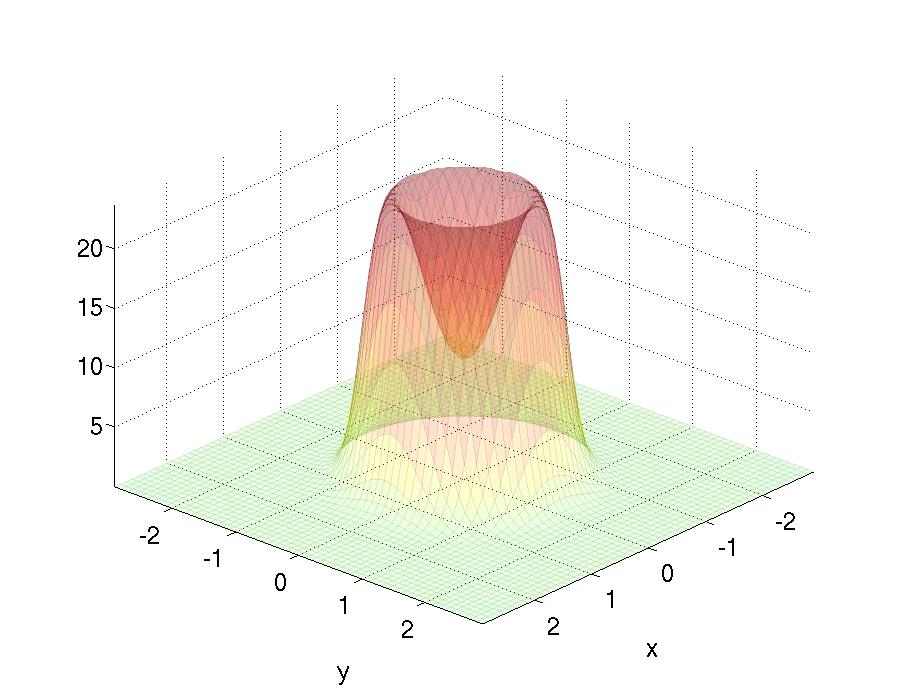}}
\subfloat{\includegraphics[width=0.245\linewidth]{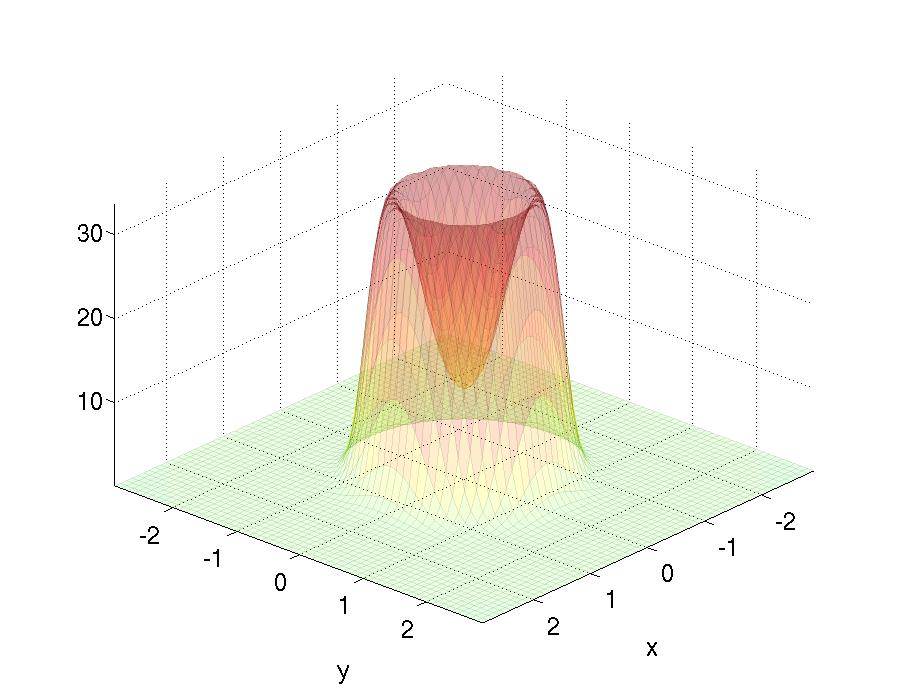}}
\subfloat{\includegraphics[width=0.245\linewidth]{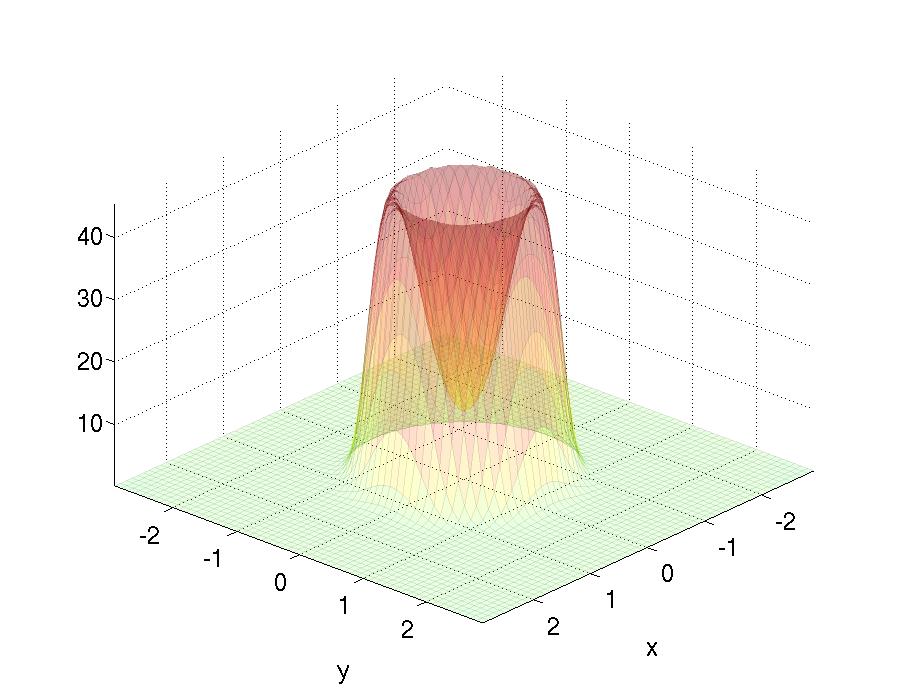}}
\subfloat{\includegraphics[width=0.245\linewidth]{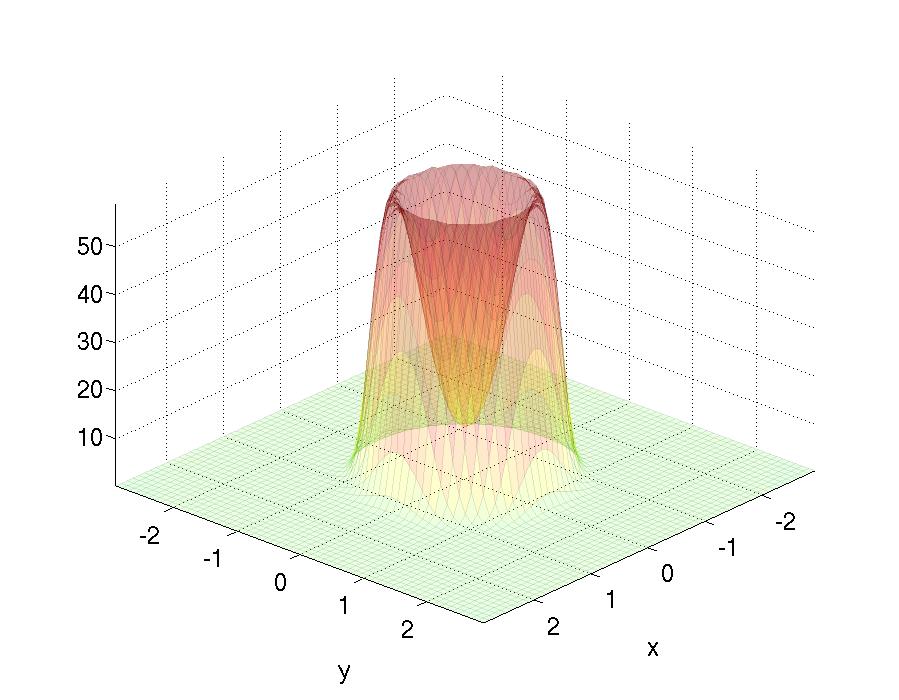}}}
\caption{Single vortex ring in the 2+6 model. The figure is an
$xy$-slice of the energy density at $z=0$ for the vortex potential 
mass parameter $m=0,1,2,3,4,5,6,7$ from top-left to bottom-right
panel. }
\label{fig:nout81_0_1_ms_energyslice2}
\vspace*{\floatsep}
\mbox{
\subfloat{\includegraphics[width=0.245\linewidth]{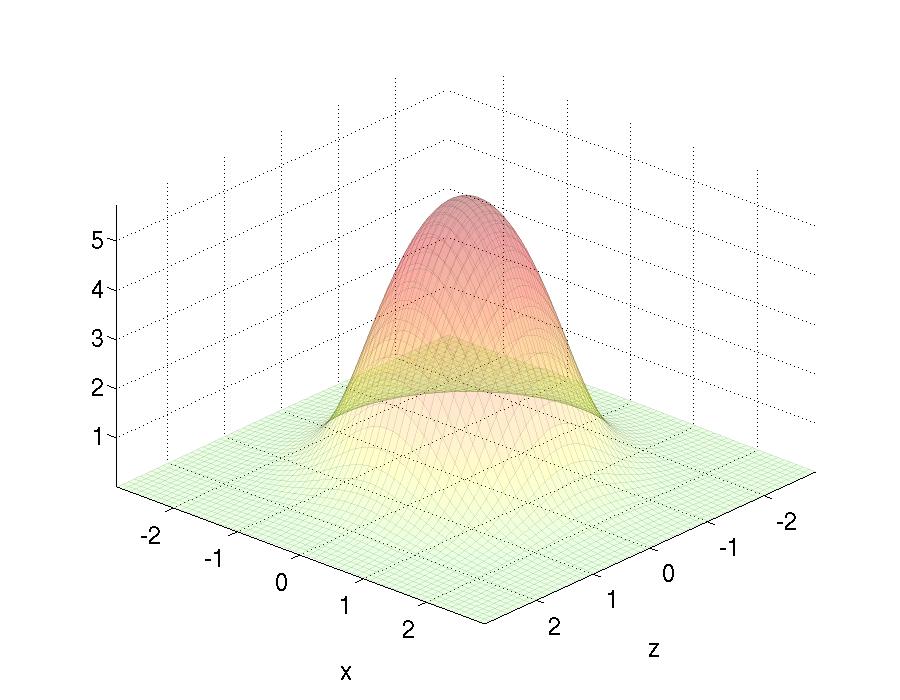}}
\subfloat{\includegraphics[width=0.245\linewidth]{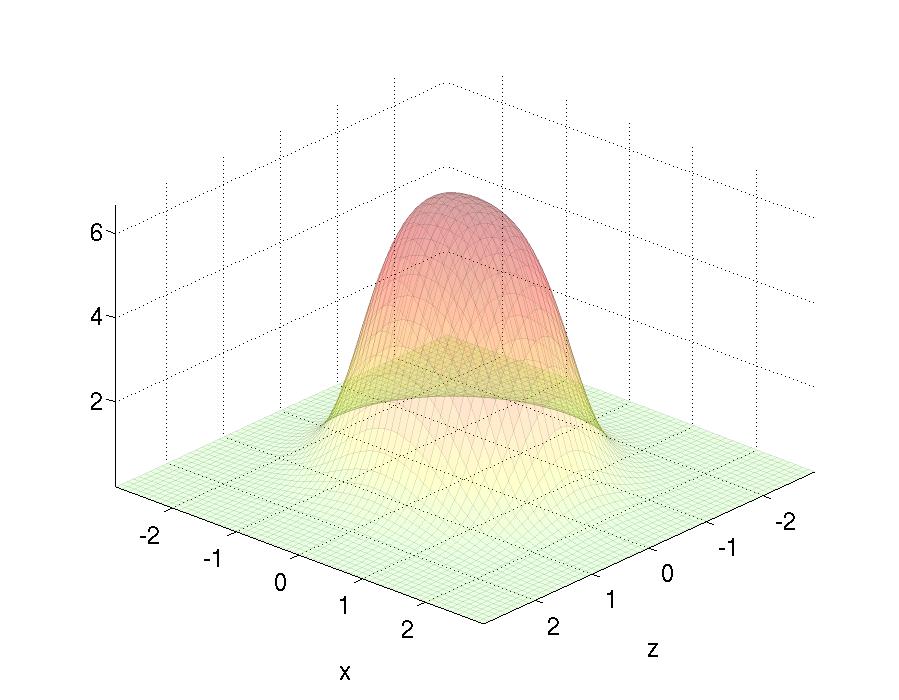}}
\subfloat{\includegraphics[width=0.245\linewidth]{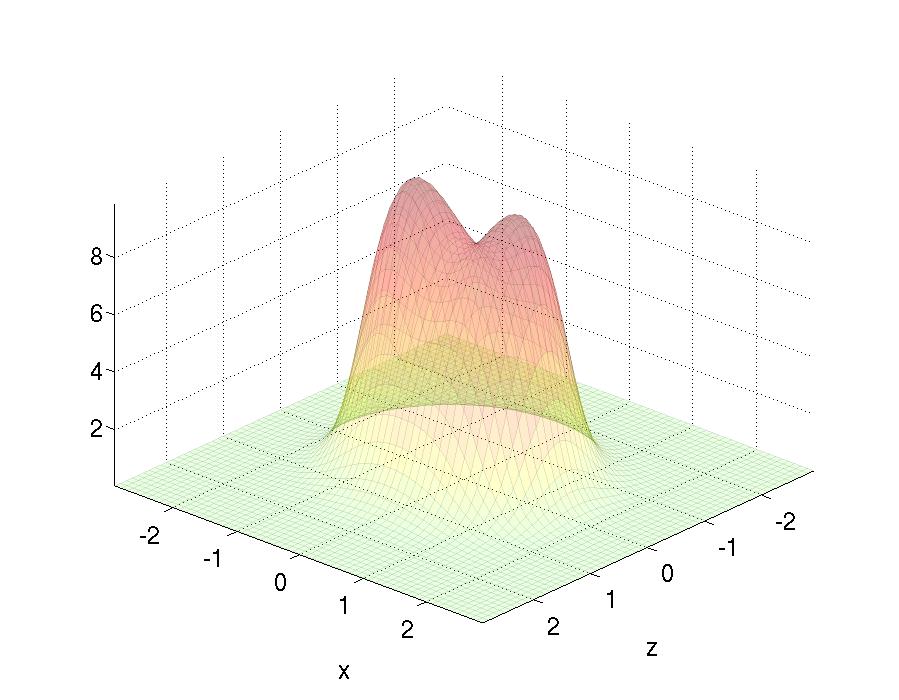}}
\subfloat{\includegraphics[width=0.245\linewidth]{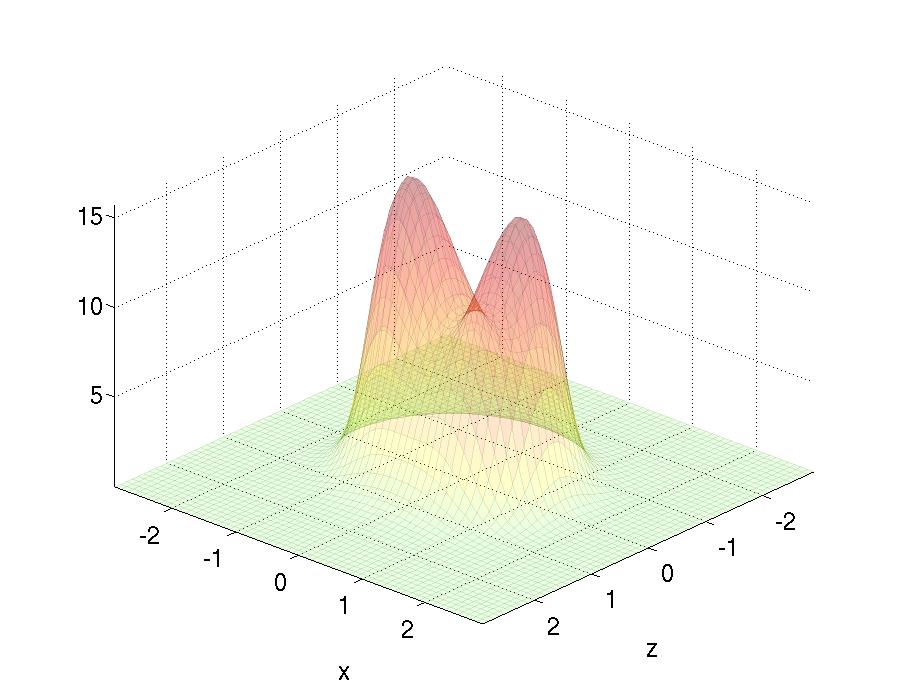}}}
\mbox{
\subfloat{\includegraphics[width=0.245\linewidth]{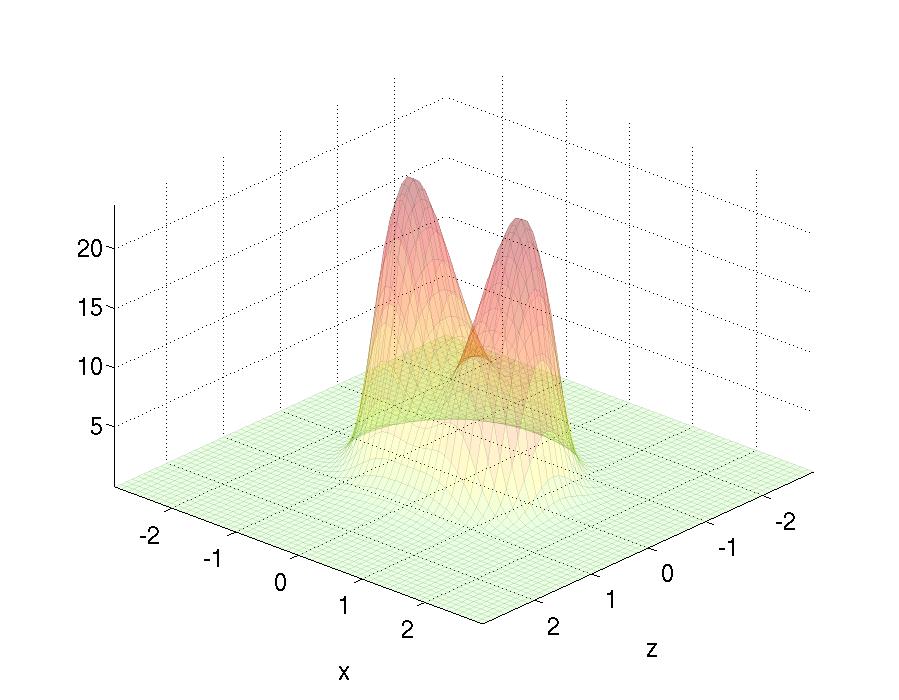}}
\subfloat{\includegraphics[width=0.245\linewidth]{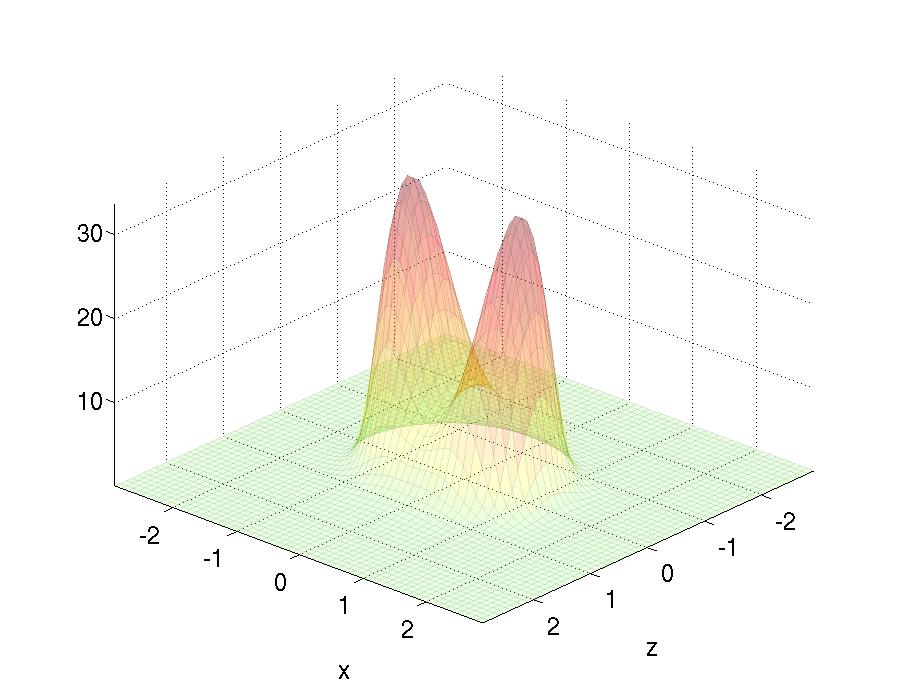}}
\subfloat{\includegraphics[width=0.245\linewidth]{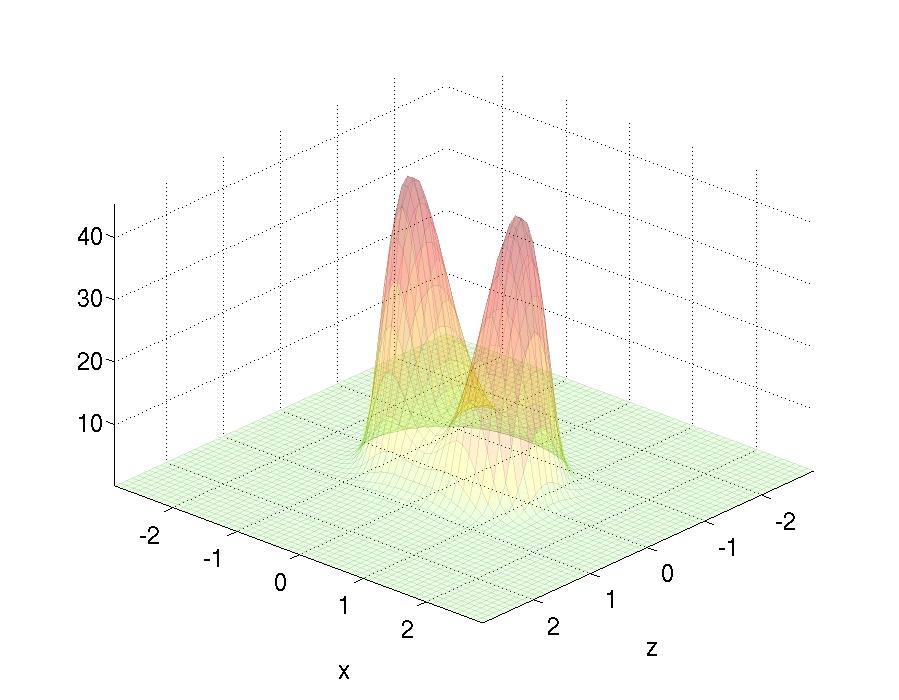}}
\subfloat{\includegraphics[width=0.245\linewidth]{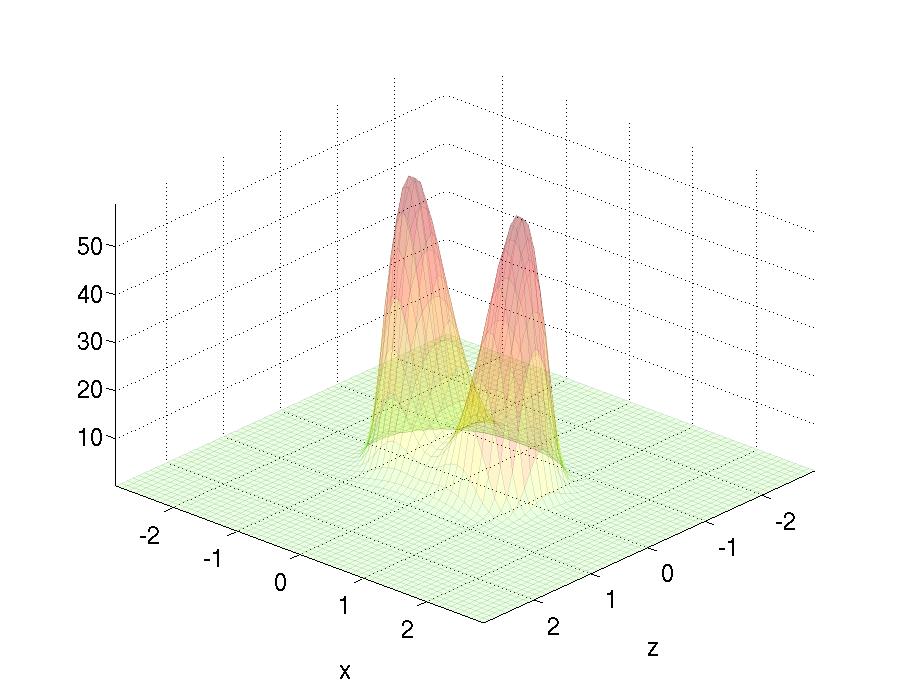}}}
\caption{Single vortex ring in the 2+6 model. The figure is an
$xz$-slice of the energy density at $y=0$ for the vortex potential 
mass parameter $m=0,1,2,3,4,5,6,7$ from top-left to bottom-right
panel.}
\label{fig:nout81_0_1_ms_energyslice}
\end{center}
\end{figure}

Next, we will turn to the 2+6 model, in which the vortex potential has
a quite different effect on the Skyrmion.
Figs.~\ref{fig:nout81_0_1_ms_baryonslice2}
and \ref{fig:nout81_0_1_ms_baryonslice} show $xy$-slices at $z=0$ and
$xz$-slices at $y=0$, respectively, of the baryon charge densities for
the single Skyrmion in the 2+6 model for various values of the vortex
potential mass parameter $m=0,1,\ldots,7$, while 
Figs.~\ref{fig:nout81_0_1_ms_energyslice2}
and \ref{fig:nout81_0_1_ms_energyslice} show $xy$-slices at $z=0$ and
$xz$-slices at $y=0$, respectively, of the energy densities.

We can see from Figs.~\ref{fig:nout81_0_1_ms_baryonslice2}
and \ref{fig:nout81_0_1_ms_baryonslice} that the vortex potential
turns (even) the single Skyrmion into a torus-like object.
The dip in the baryon charge density at the center of the
configuration starts to happen for a critical value of the vortex
potential mass parameter $m_{\rm critical}$ between 2 and 3.
In the last figure depicted in
Fig.~\ref{fig:nout81_0_1_ms_baryonslice2}, the baryon charge density
at the center of the Skyrmion is about 1/2 of the maximum value. 

We observe also -- analogously to what was seen in
Ref.~\cite{Gudnason:2014hsa} -- that the energy density is far more
torus-like than the corresponding baryon charge density.
The critical value of the vortex potential mass parameter $m$ for
which a dip appears in the center of the Skyrmion is also lower than
that for the baryon charge density and is between 1 and 2.

\begin{figure}[!htp]
\begin{center}
\includegraphics[width=0.5\linewidth]{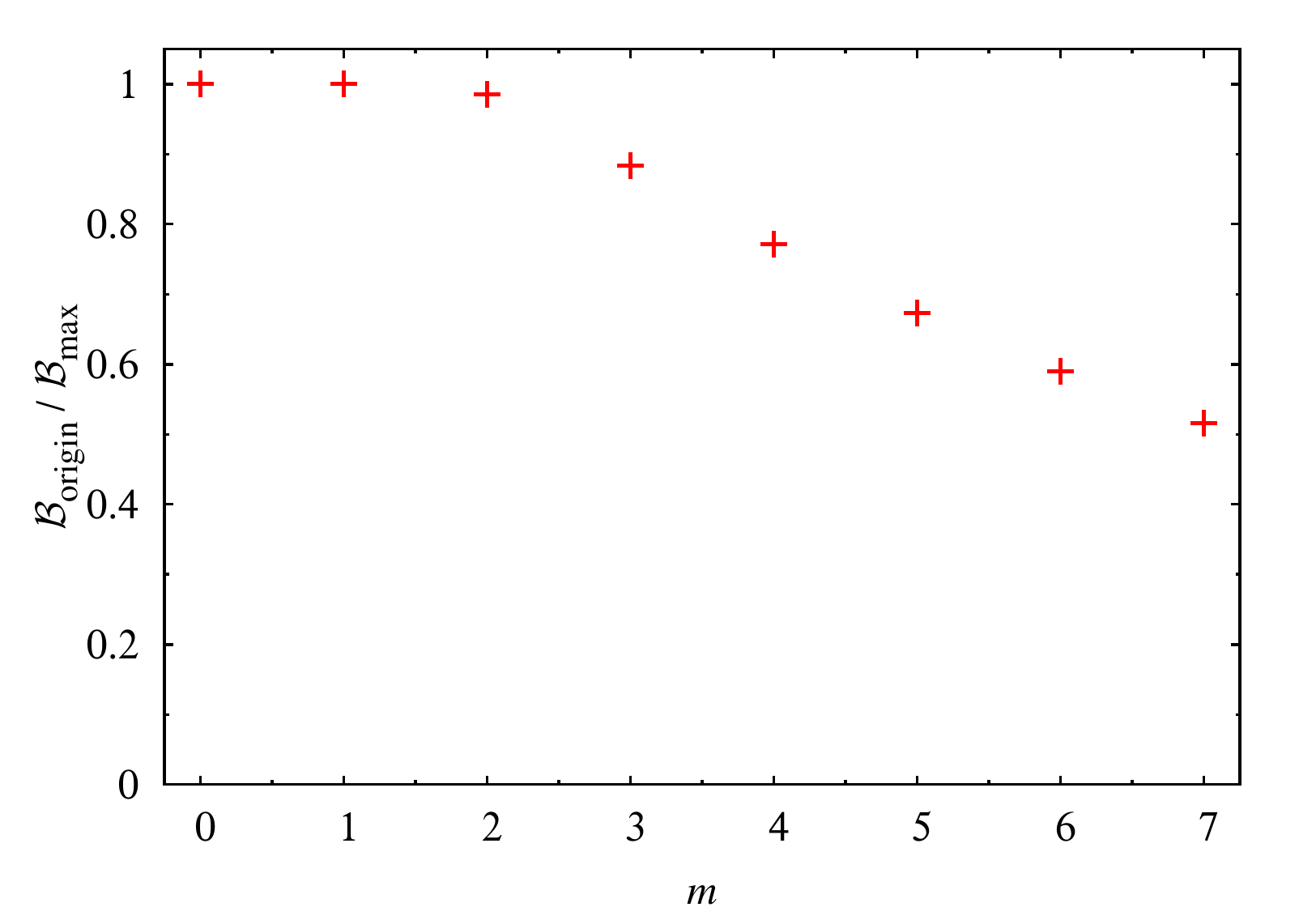}
\caption{The ratio of the baryon charge density at the origin and the
maximum baryon charge density,
$\mathcal{B}_{\rm origin}/\mathcal{B}_{\rm max}$, for the single vortex
ring in the 2+6 model as function of the vortex potential mass
parameter $m$. }
\label{fig:ring260}
\end{center}
\end{figure}

Finally, we show in Fig.~\ref{fig:ring260} the ratio between the
baryon charge density at the center (of mass) of the configuration to
the maximum value, $\mathcal{B}(0)/\mathop{\rm max}(\mathcal{B}(x))$.
It can be seen from the figure that the critical value for the dip to
form is slightly smaller than, but about $m\lesssim 2$.

\subsection{Doubly twisted vortex rings as $B=2$ Skyrmions}\label{sec:B2rings}

In this subsection we consider the vortex rings of baryon charge
$B=2$. As well known, the normal Skyrmion of charge $B=2$ already
takes the shape of a torus; therefore we consider this case for
completeness to study the effect of the vortex potential in the $B=2$
sector and it will serve as a basis for when we later in the paper
want to add kinks to the $B=2$ vortex ring.
Let us clarify that although $B=2$, the vortex has only vortex charge
$n=1$ but it is twisted up twice, in our language.

As the initial condition for the $B=2$ Skyrmion, we use the following
Ansatz
\beq
\phi_{\rm initial}^{\rm T} =
\left(\cos f - i\sin f\cos\theta,
e^{i2\phi}\sin f\sin\theta\right),
\eeq
for an appropriate profile function $f$.
This is of course just the standard axially symmetric generalization
to $B=2$ of the hedgehog. 

\begin{figure}[!tp]
\begin{center}
\mbox{
\subfloat{\includegraphics[width=0.245\linewidth]{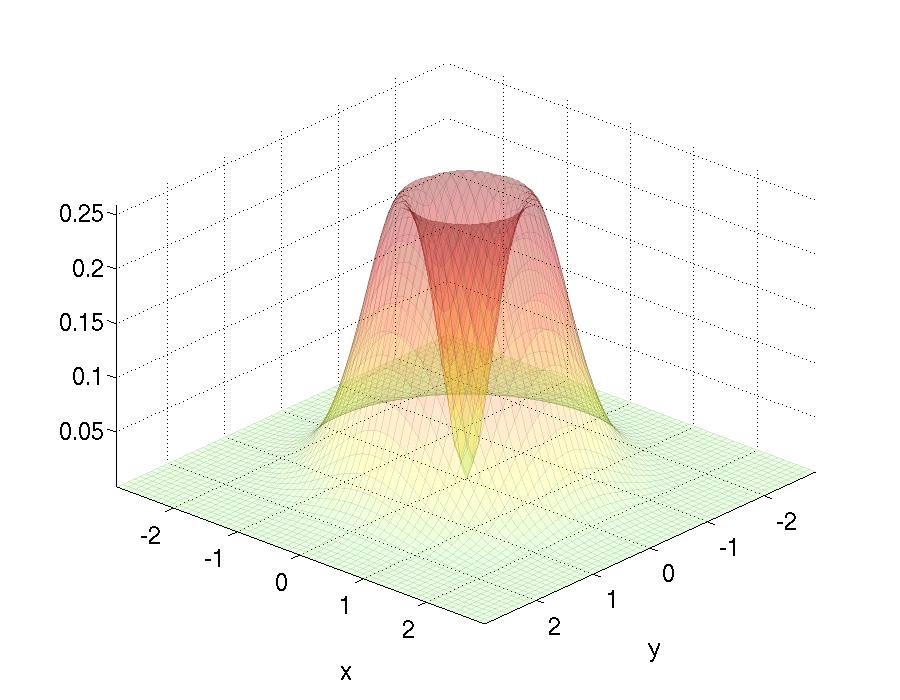}}
\subfloat{\includegraphics[width=0.245\linewidth]{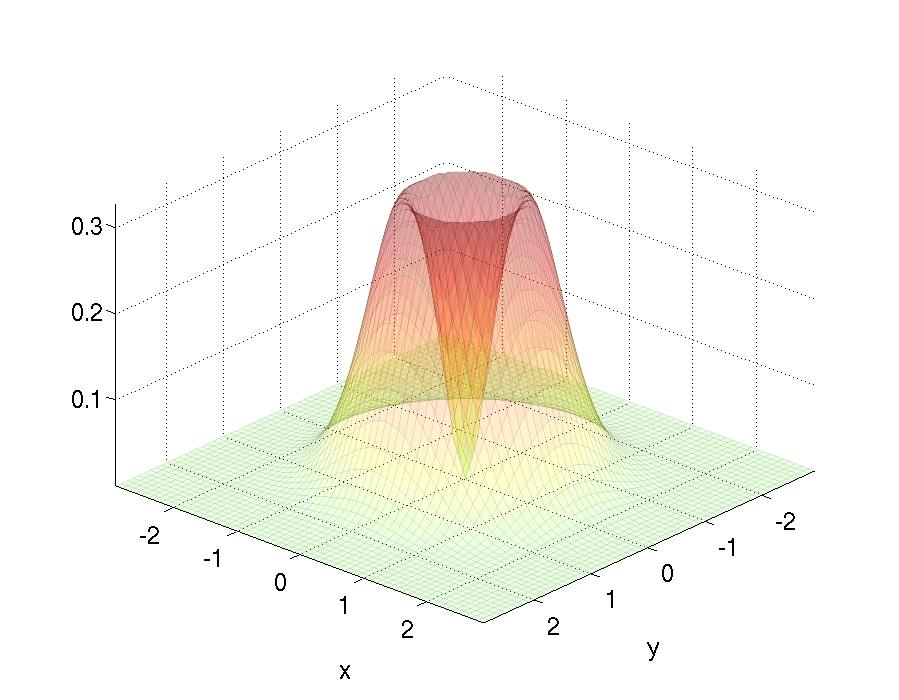}}
\subfloat{\includegraphics[width=0.245\linewidth]{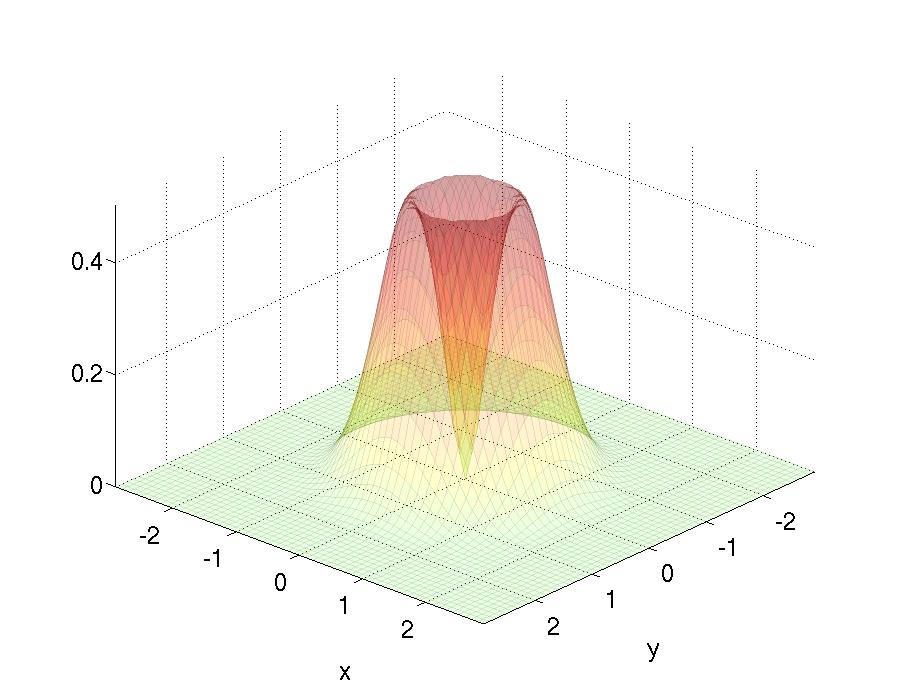}}
\subfloat{\includegraphics[width=0.245\linewidth]{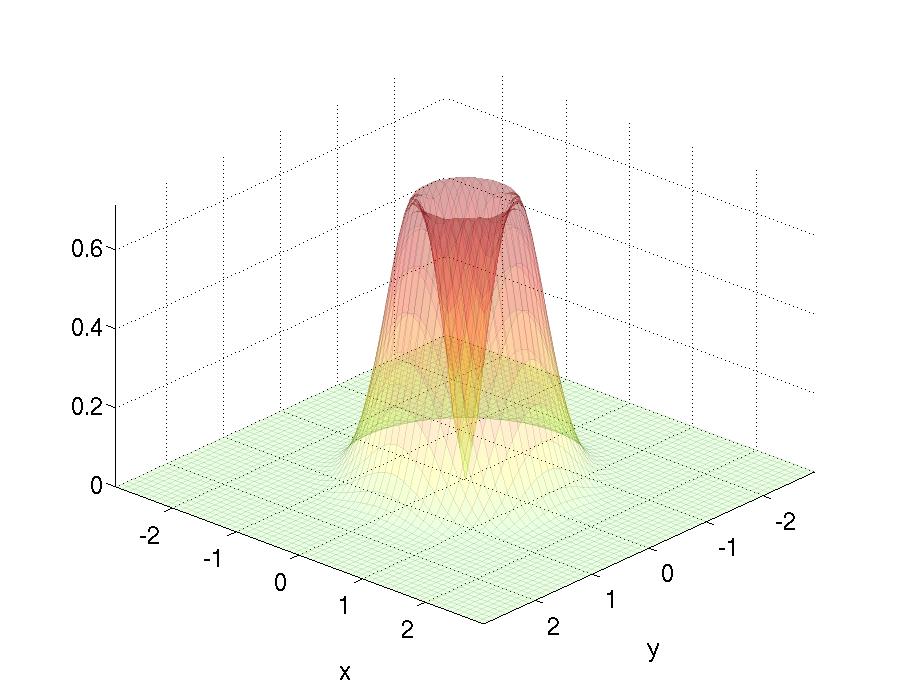}}}
\mbox{
\subfloat{\includegraphics[width=0.245\linewidth]{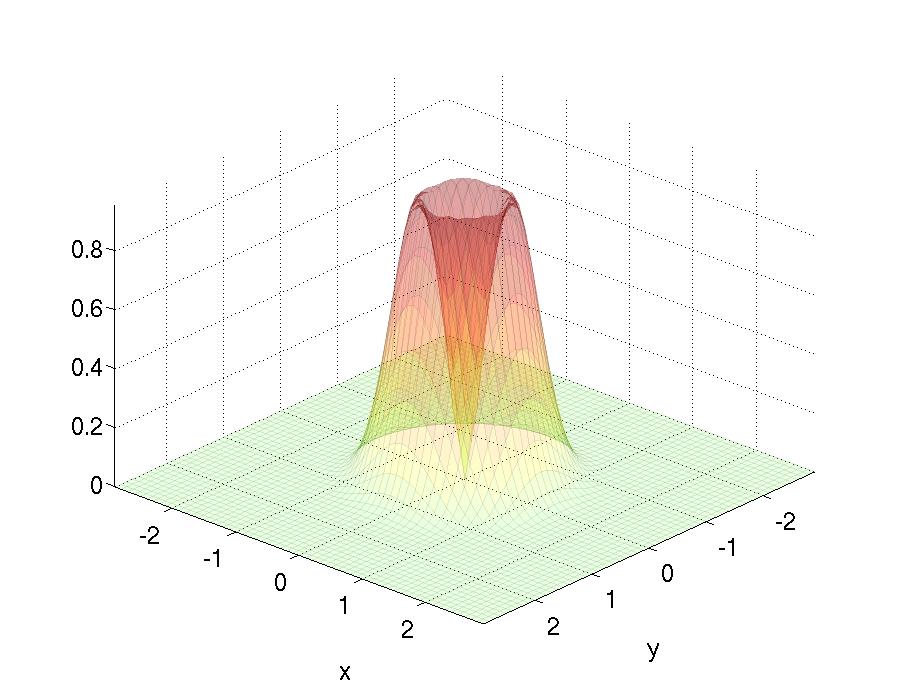}}
\subfloat{\includegraphics[width=0.245\linewidth]{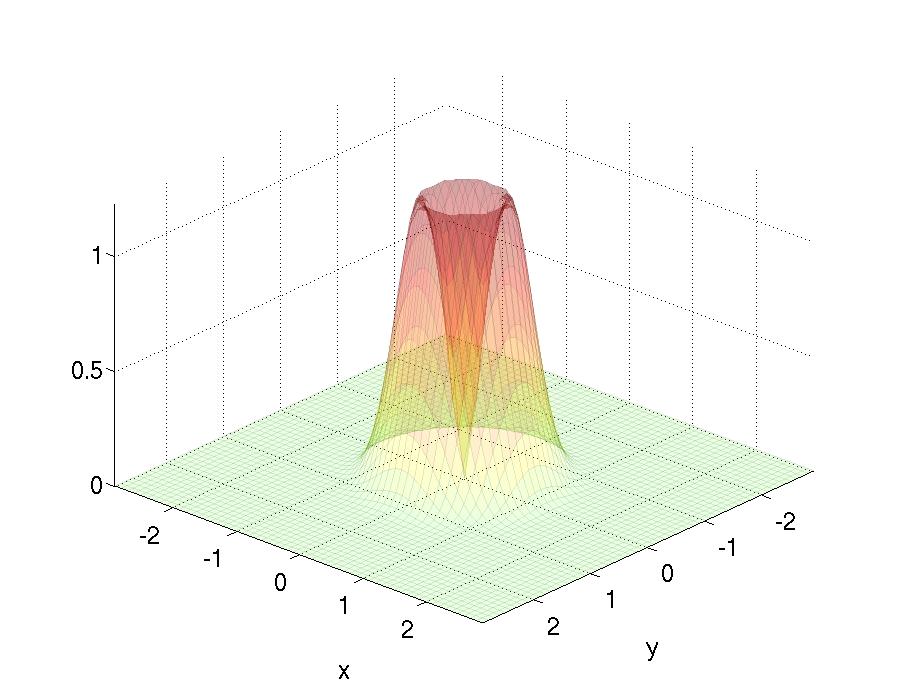}}
\subfloat{\includegraphics[width=0.245\linewidth]{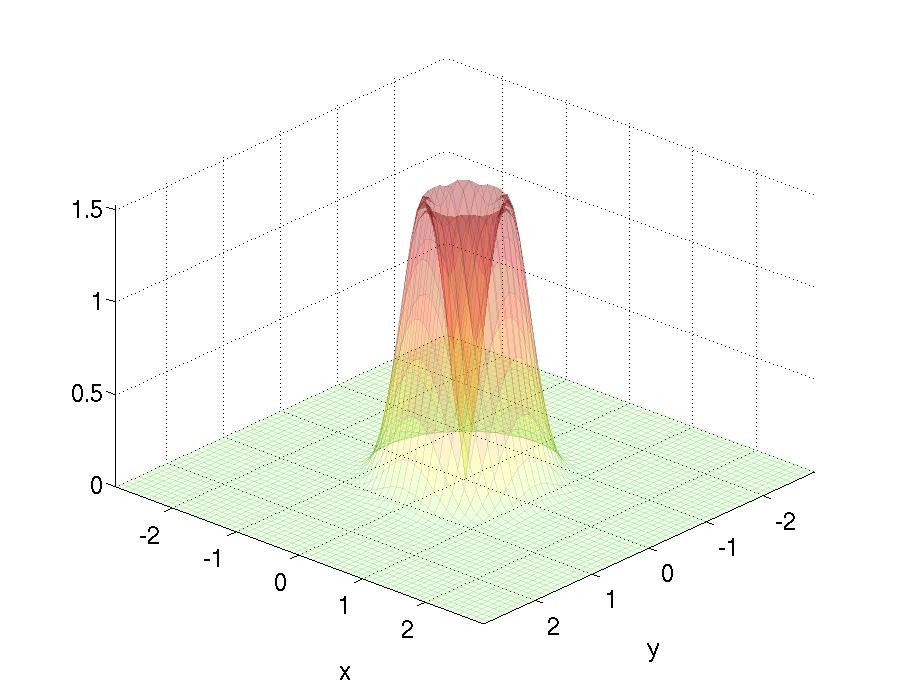}}
\subfloat{\includegraphics[width=0.245\linewidth]{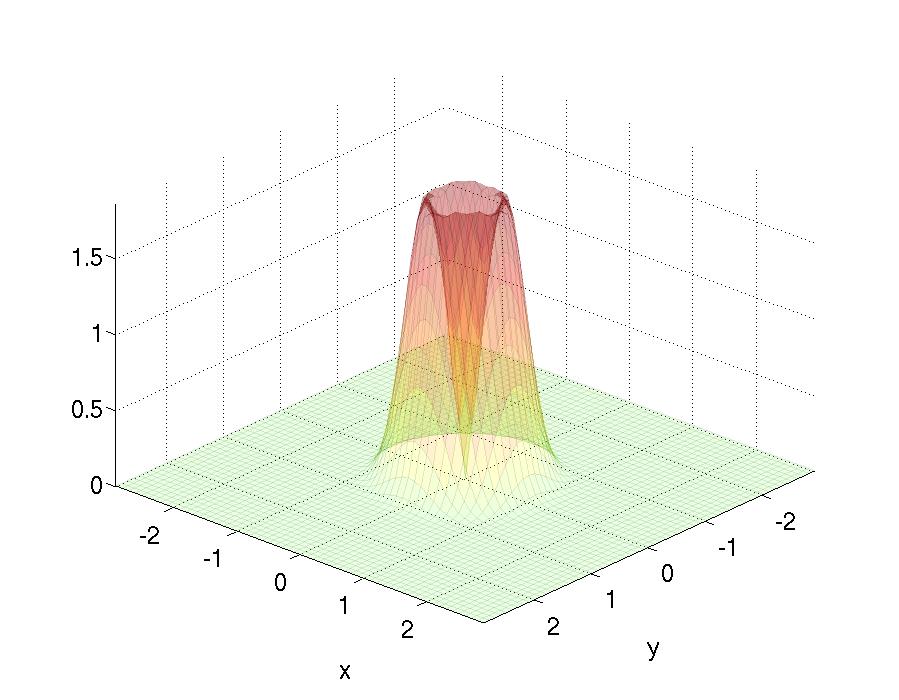}}}
\caption{A $B=2$ vortex ring in the 2+4 model. The figure is an
$xy$-slice of the baryon charge density at $z=0$ for the vortex
potential mass parameter $m=0,1,2,3,4,5,6,7$ from top-left to
bottom-right panel. }
\label{fig:nout8121_1_0_ms_baryonslice2}
\vspace*{\floatsep}
\mbox{
\subfloat{\includegraphics[width=0.245\linewidth]{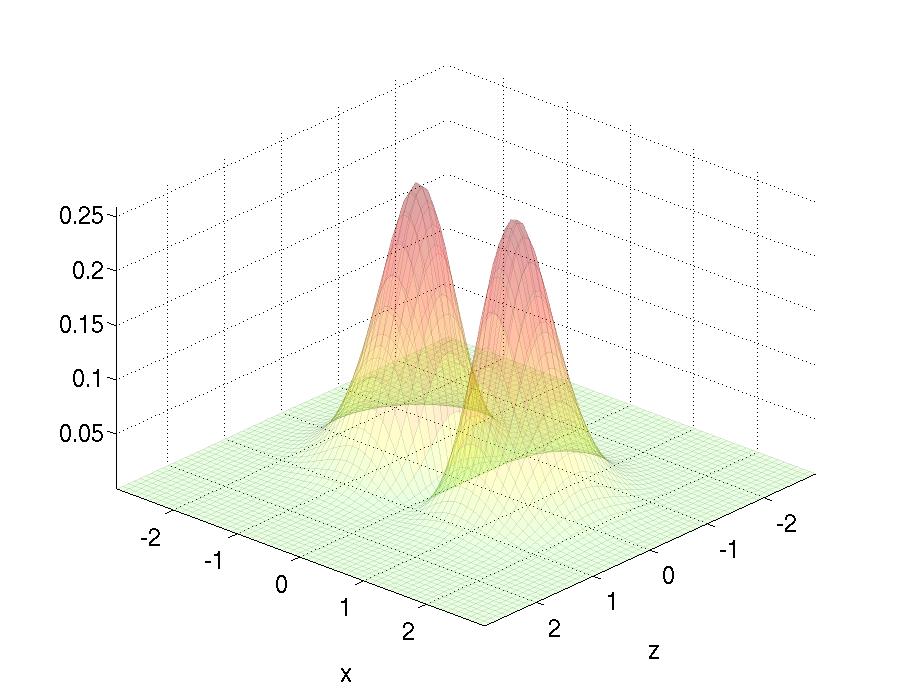}}
\subfloat{\includegraphics[width=0.245\linewidth]{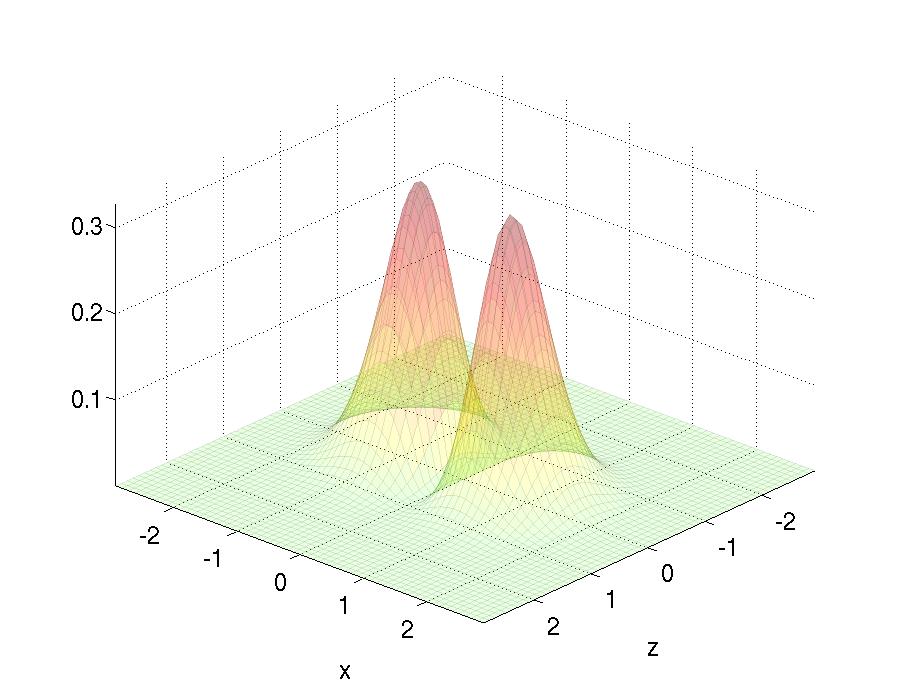}}
\subfloat{\includegraphics[width=0.245\linewidth]{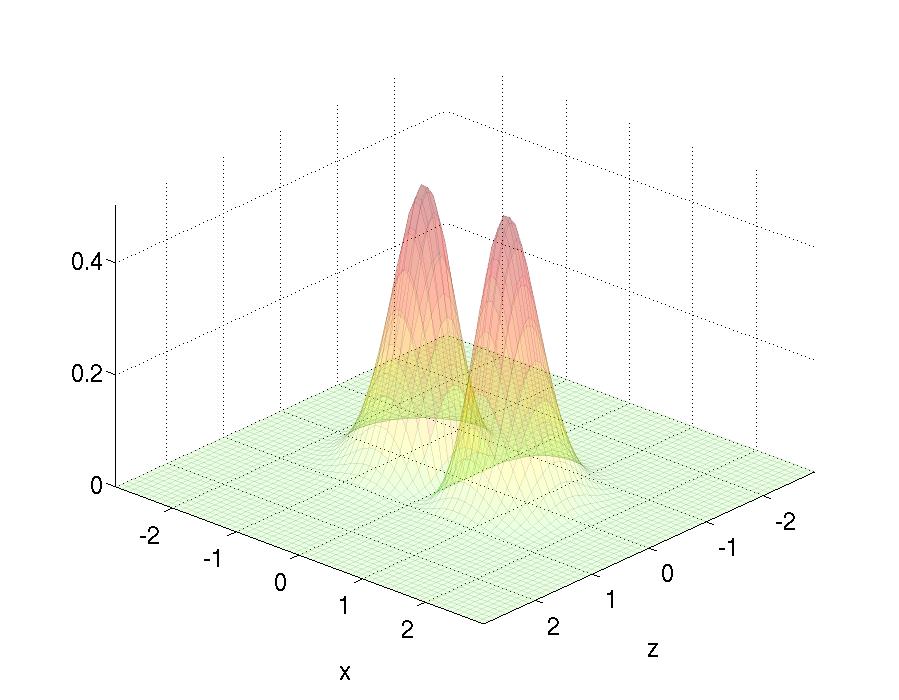}}
\subfloat{\includegraphics[width=0.245\linewidth]{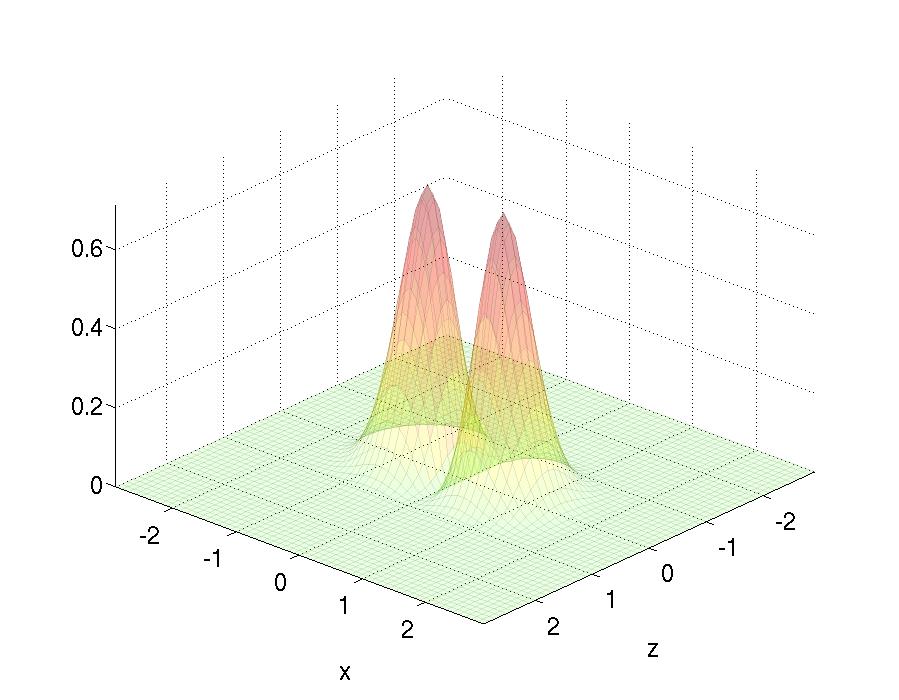}}}
\mbox{
\subfloat{\includegraphics[width=0.245\linewidth]{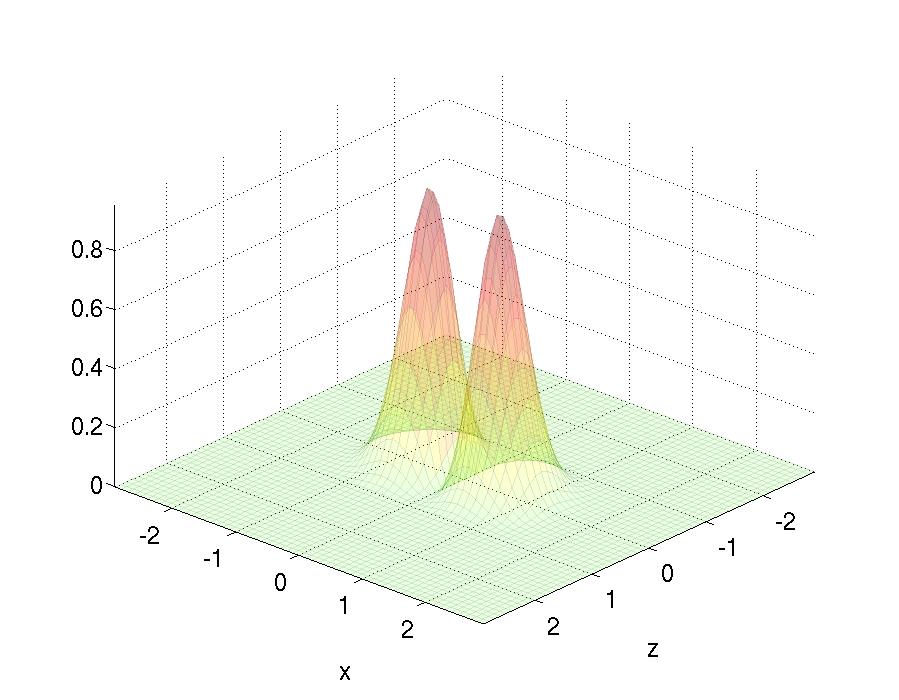}}
\subfloat{\includegraphics[width=0.245\linewidth]{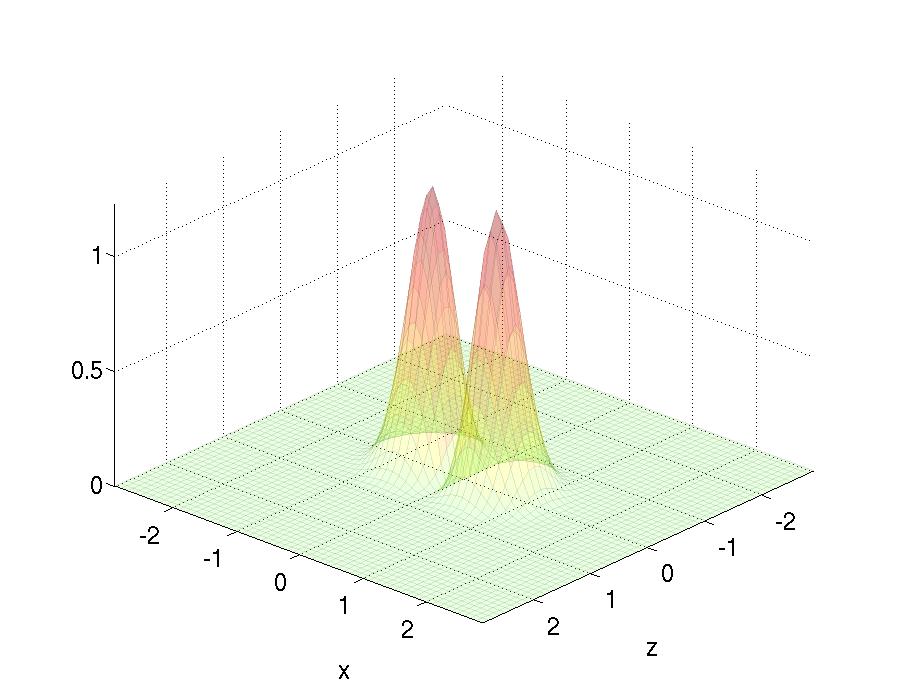}}
\subfloat{\includegraphics[width=0.245\linewidth]{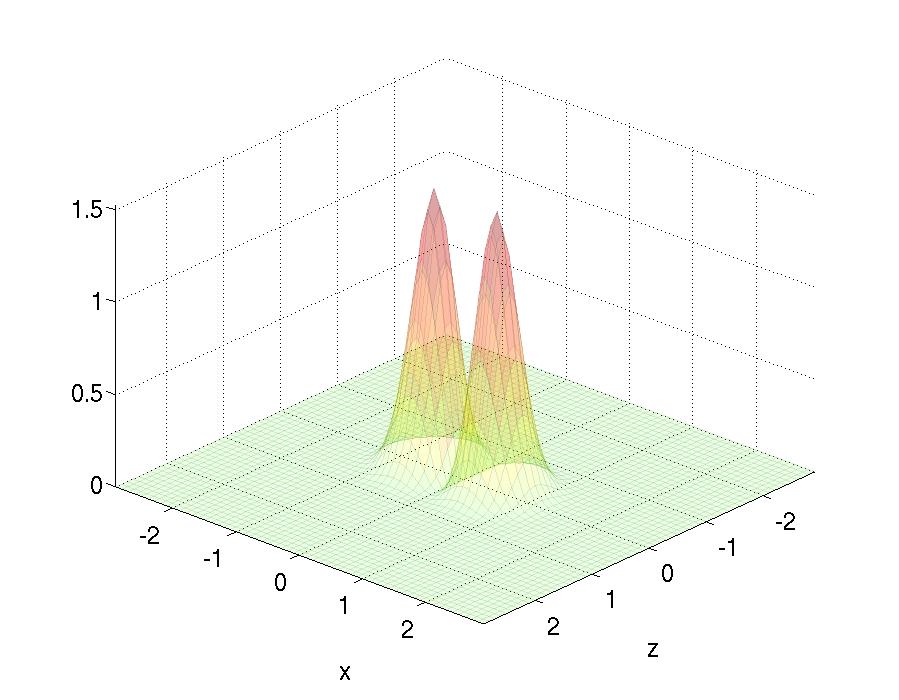}}
\subfloat{\includegraphics[width=0.245\linewidth]{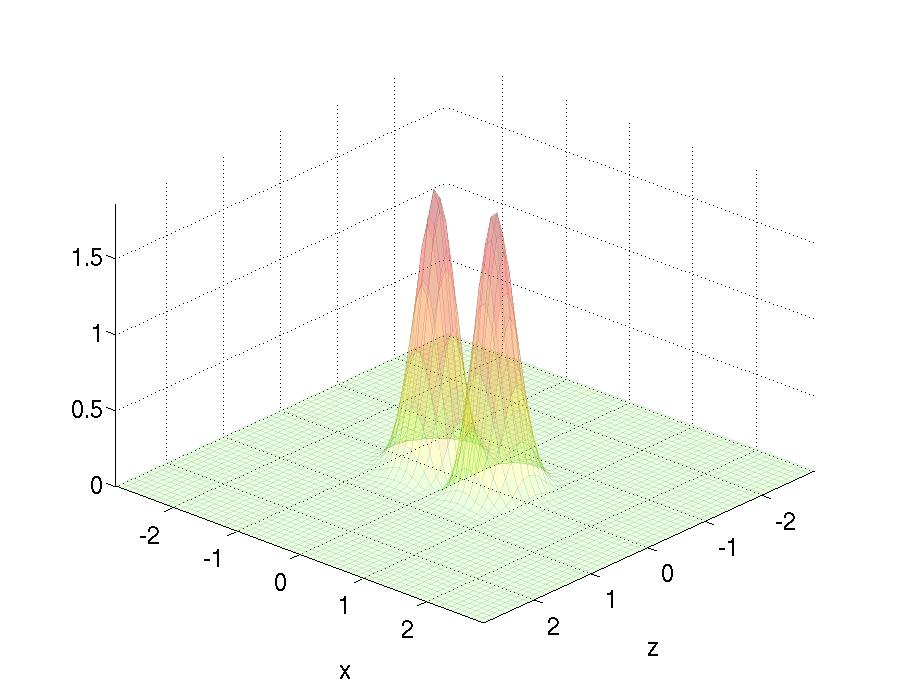}}}
\caption{A $B=2$ vortex ring in the 2+4 model. The figure is an
$xz$-slice of the baryon charge density at $y=0$ for the vortex
potential mass parameter $m=0,1,2,3,4,5,6,7$ from top-left to
bottom-right panel.}
\label{fig:nout8121_1_0_ms_baryonslice}
\end{center}
\end{figure}

\begin{figure}[!tp]
\begin{center}
\mbox{
\subfloat{\includegraphics[width=0.245\linewidth]{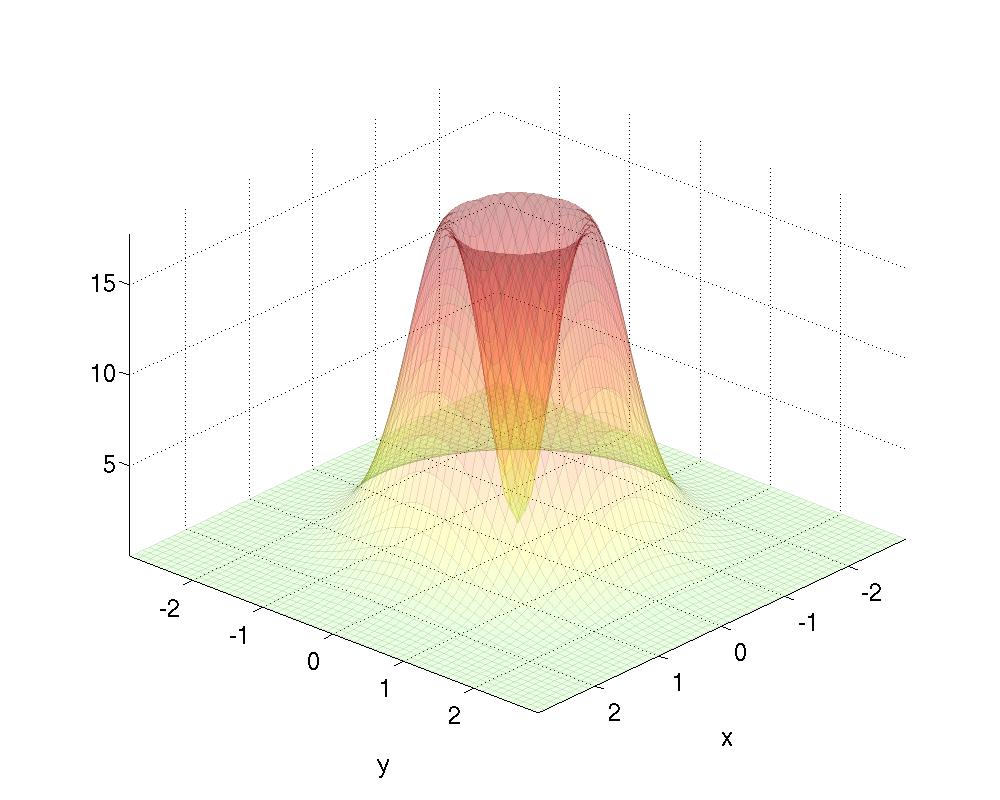}}
\subfloat{\includegraphics[width=0.245\linewidth]{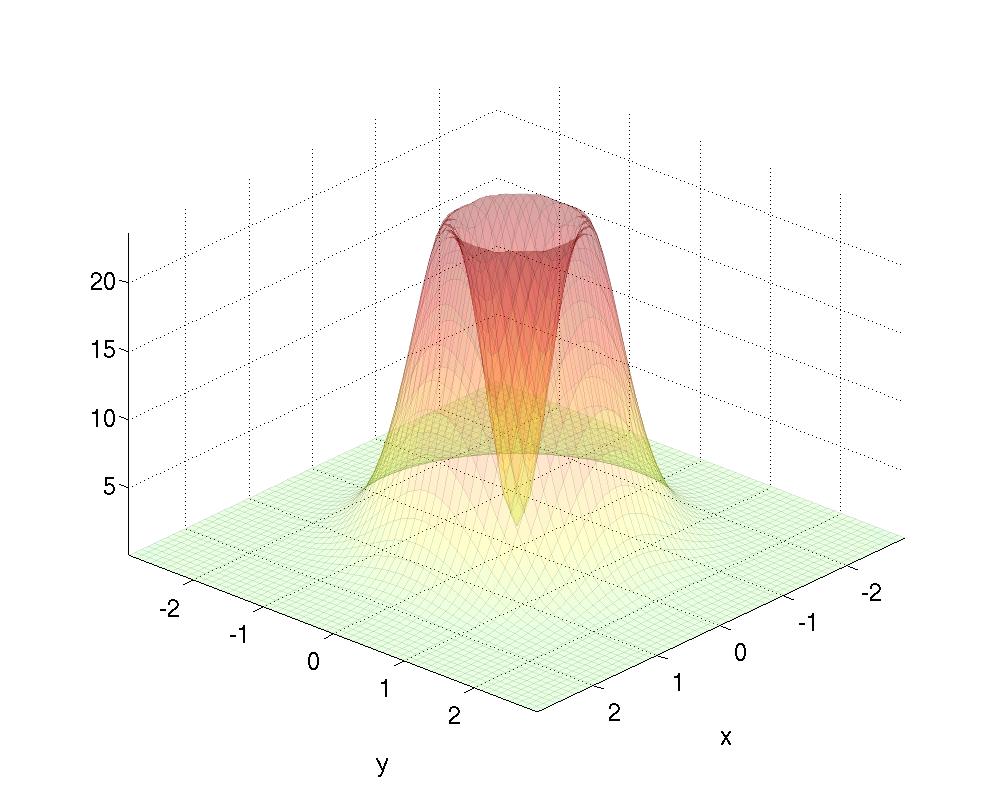}}
\subfloat{\includegraphics[width=0.245\linewidth]{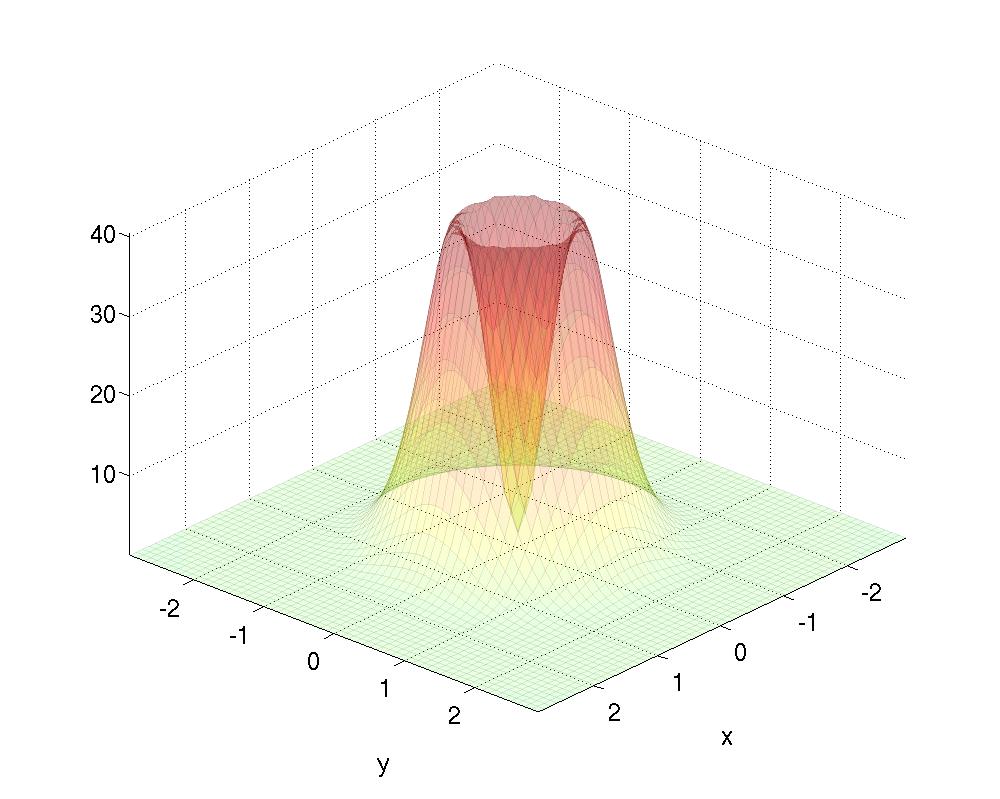}}
\subfloat{\includegraphics[width=0.245\linewidth]{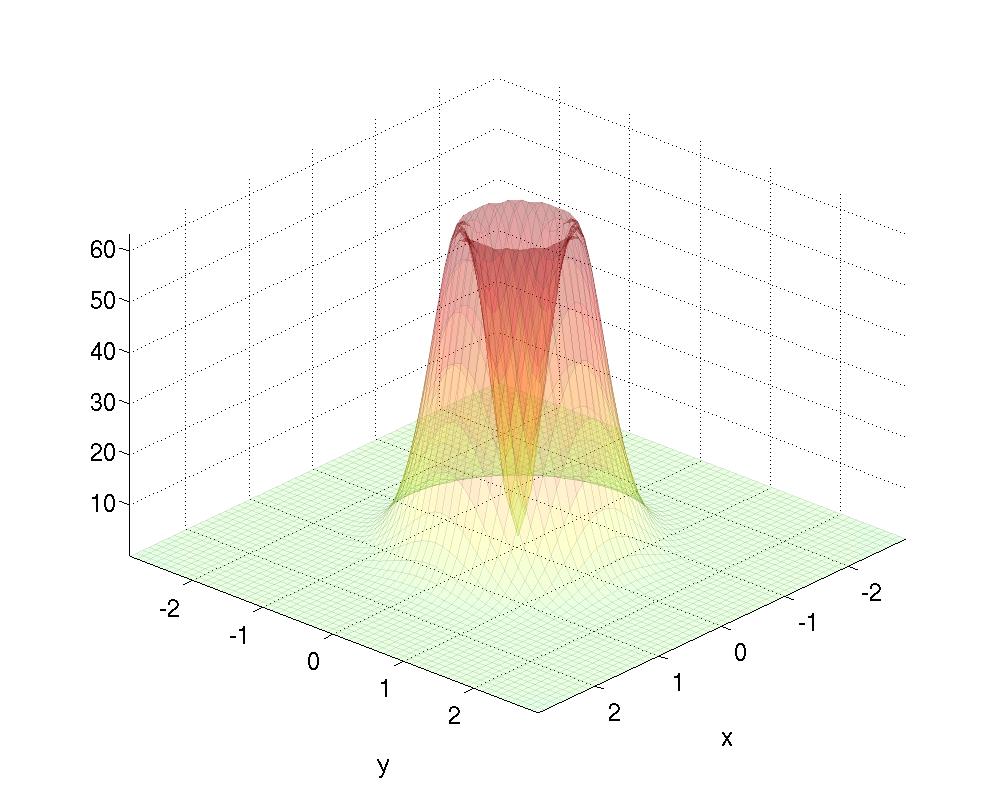}}}
\mbox{
\subfloat{\includegraphics[width=0.245\linewidth]{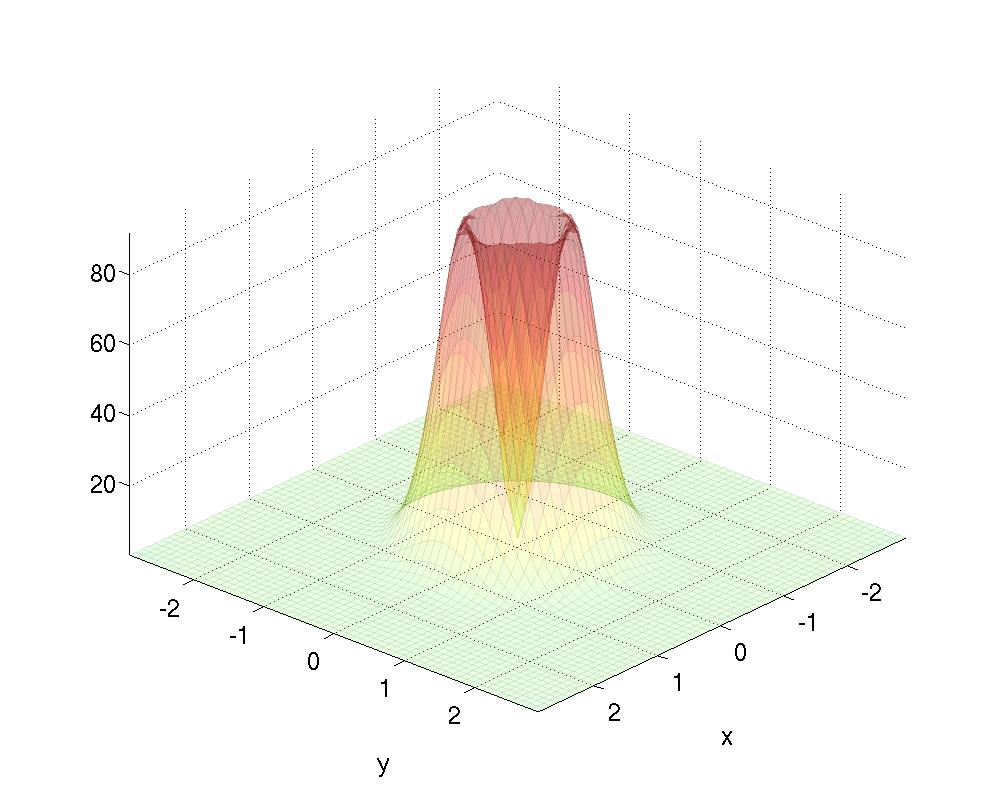}}
\subfloat{\includegraphics[width=0.245\linewidth]{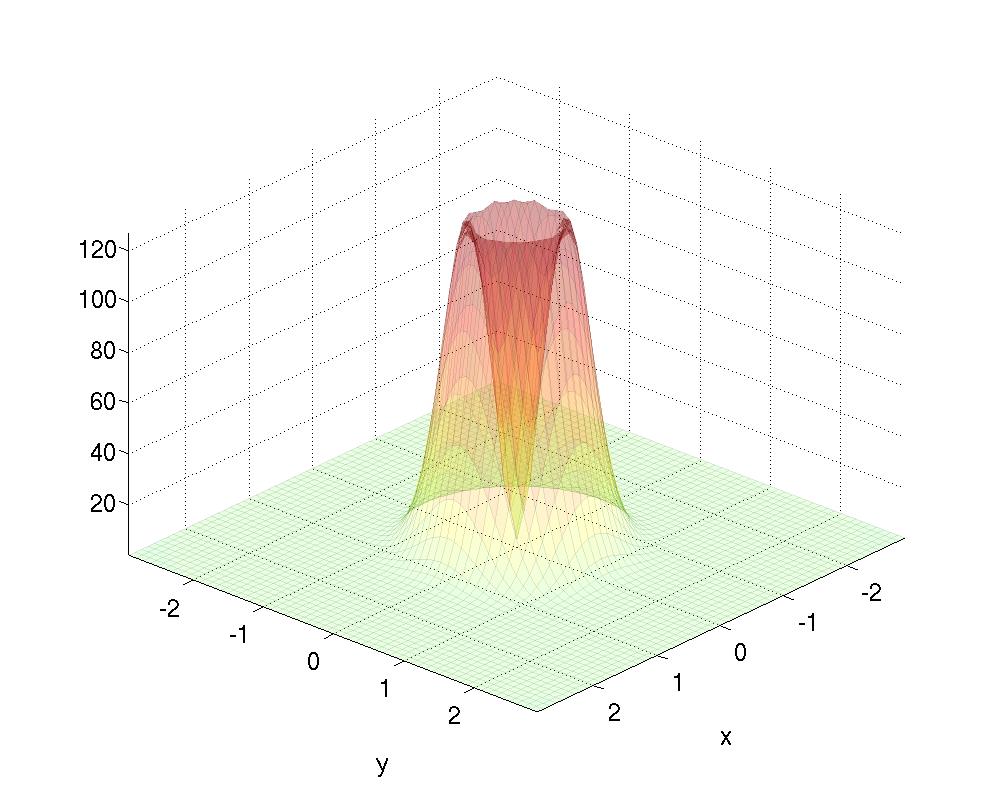}}
\subfloat{\includegraphics[width=0.245\linewidth]{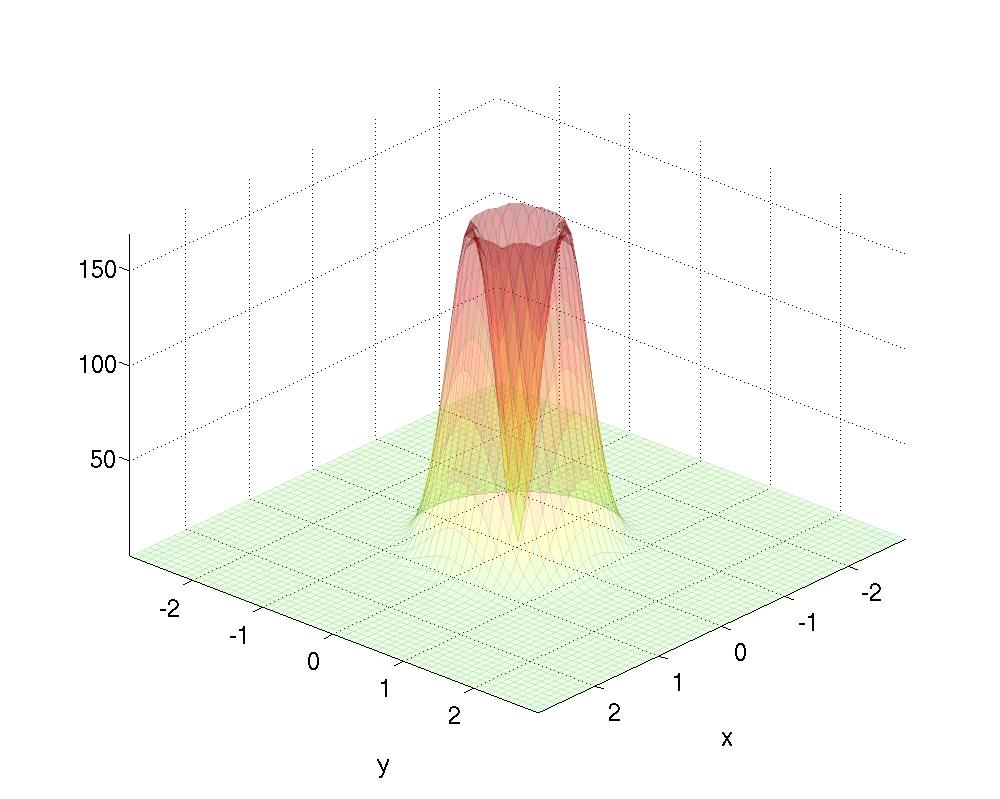}}
\subfloat{\includegraphics[width=0.245\linewidth]{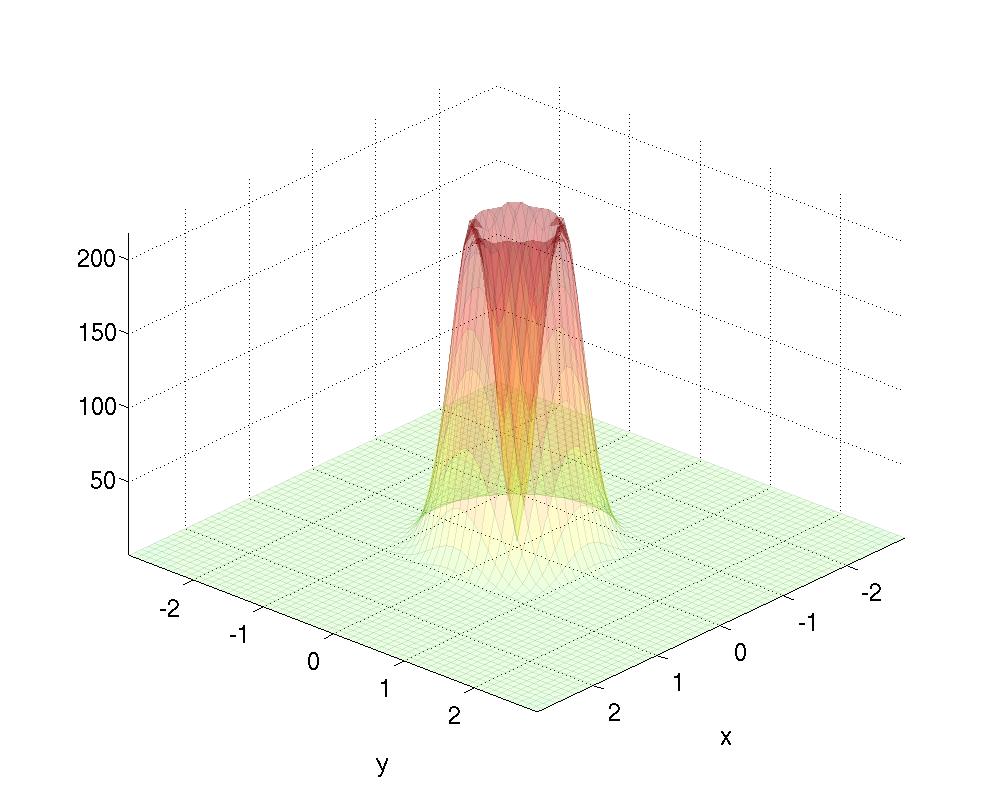}}}
\caption{A $B=2$ vortex ring in the 2+4 model. The figure is an
$xy$-slice of the energy density at $z=0$ for the vortex potential
mass parameter $m=0,1,2,3,4,5,6,7$ from top-left to bottom-right
panel. } 
\label{fig:nout8121_1_0_ms_energyslice2}
\vspace*{\floatsep}
\mbox{
\subfloat{\includegraphics[width=0.245\linewidth]{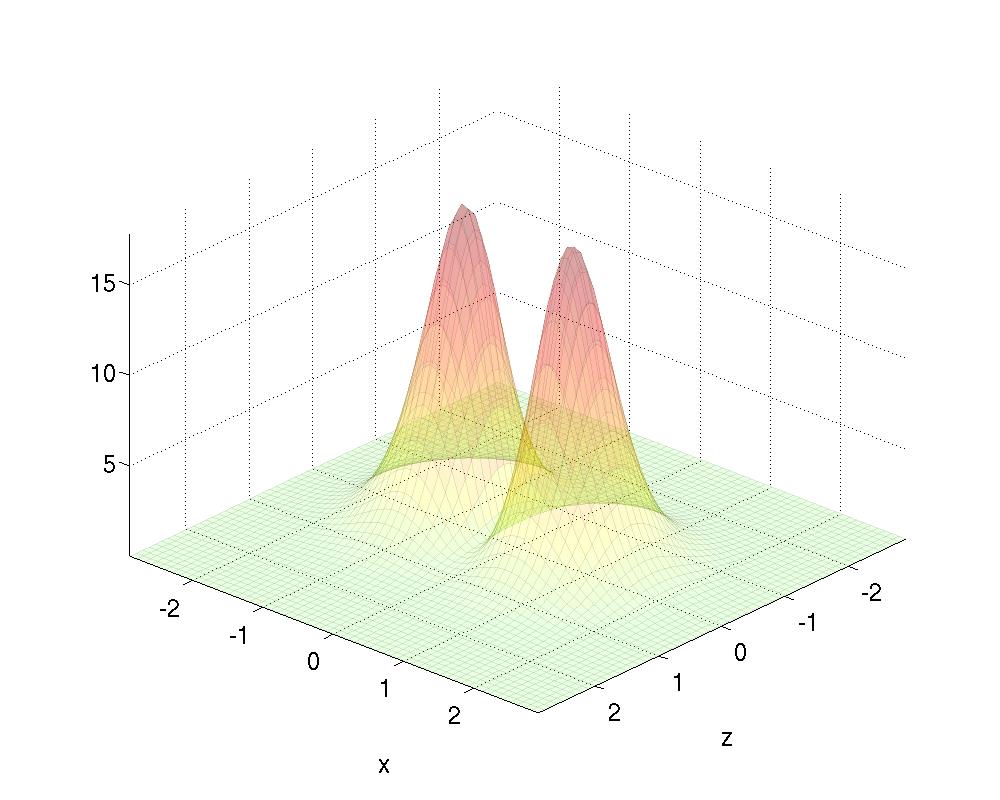}}
\subfloat{\includegraphics[width=0.245\linewidth]{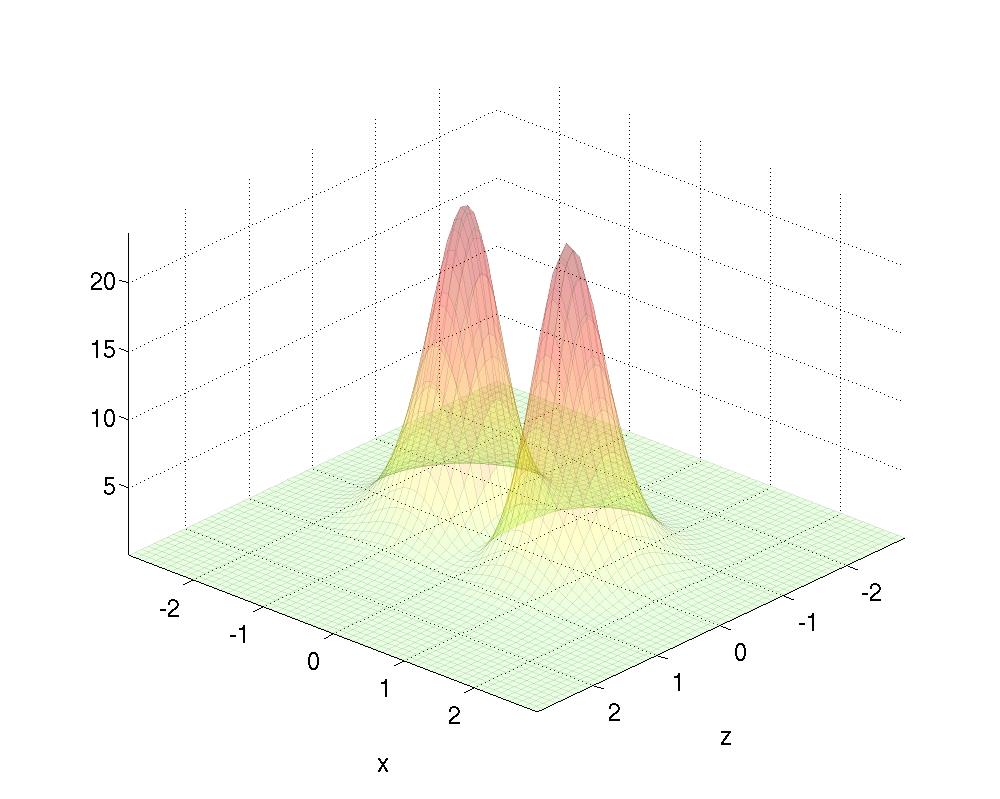}}
\subfloat{\includegraphics[width=0.245\linewidth]{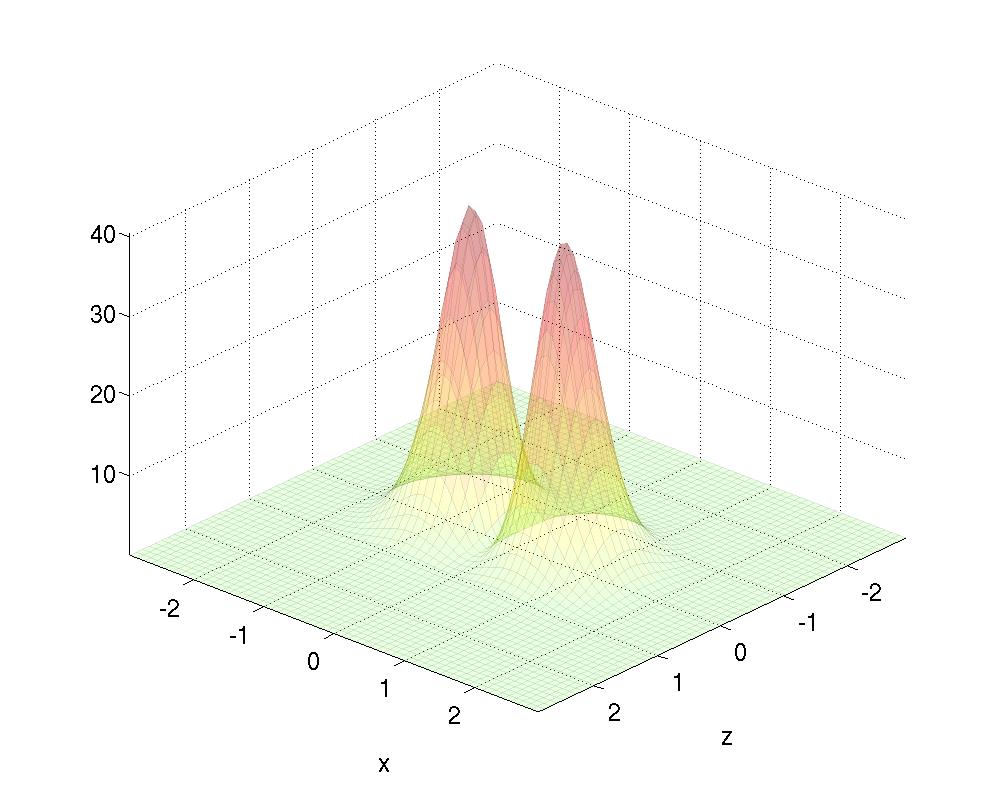}}
\subfloat{\includegraphics[width=0.245\linewidth]{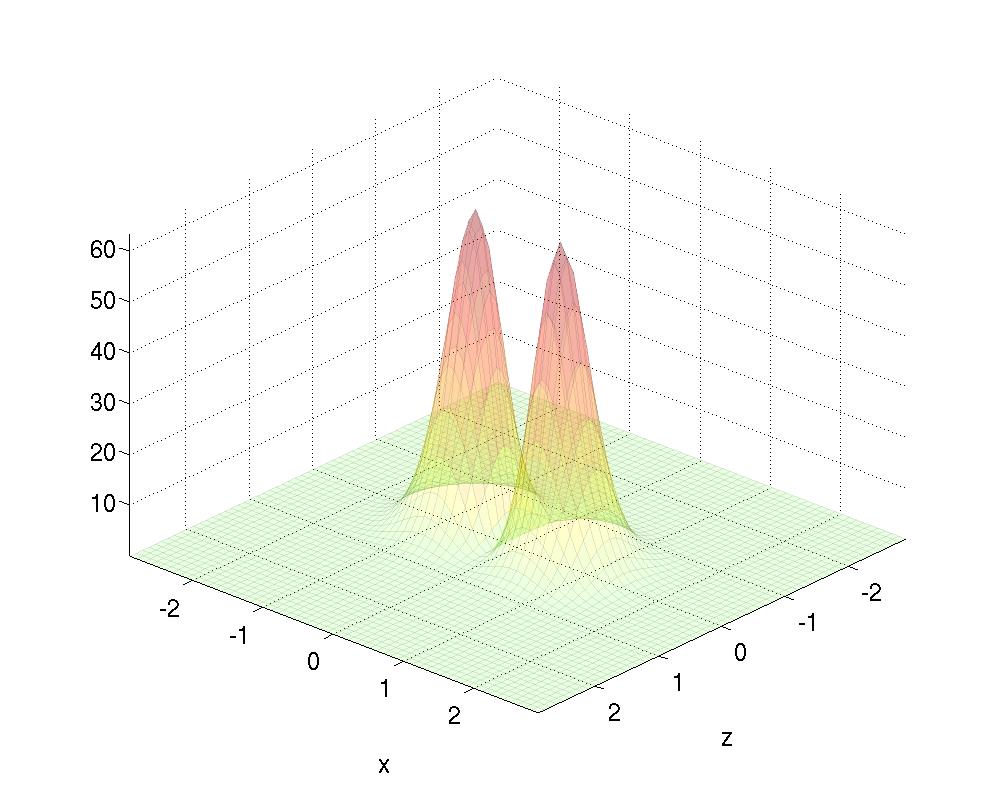}}}
\mbox{
\subfloat{\includegraphics[width=0.245\linewidth]{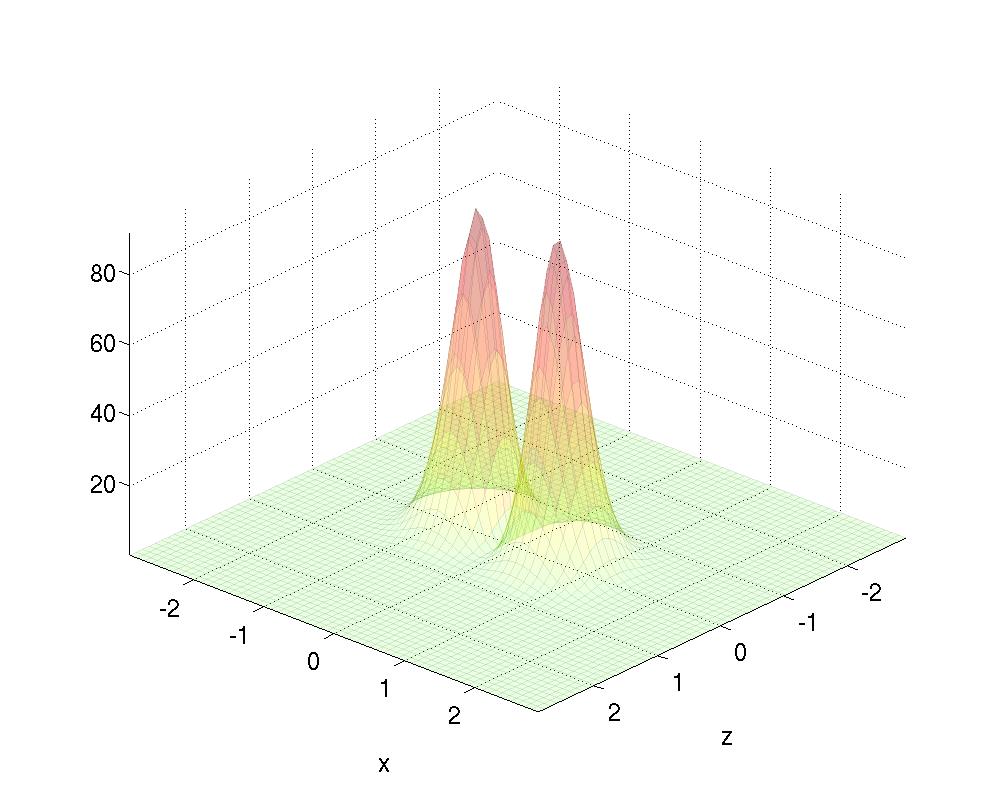}}
\subfloat{\includegraphics[width=0.245\linewidth]{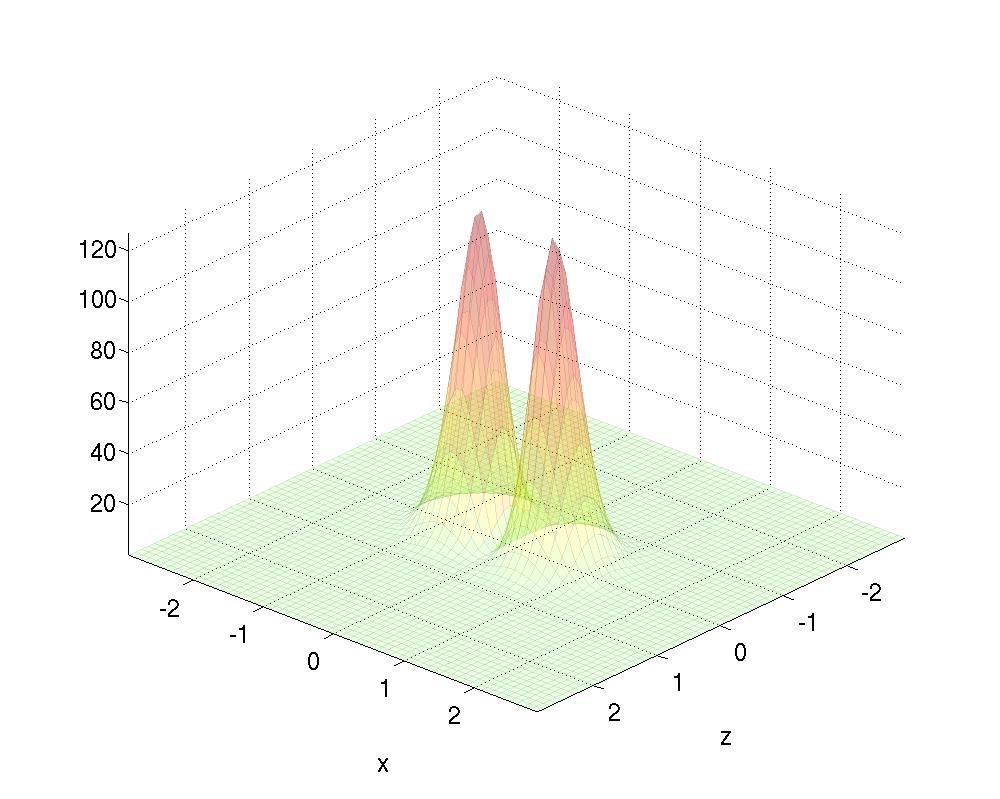}}
\subfloat{\includegraphics[width=0.245\linewidth]{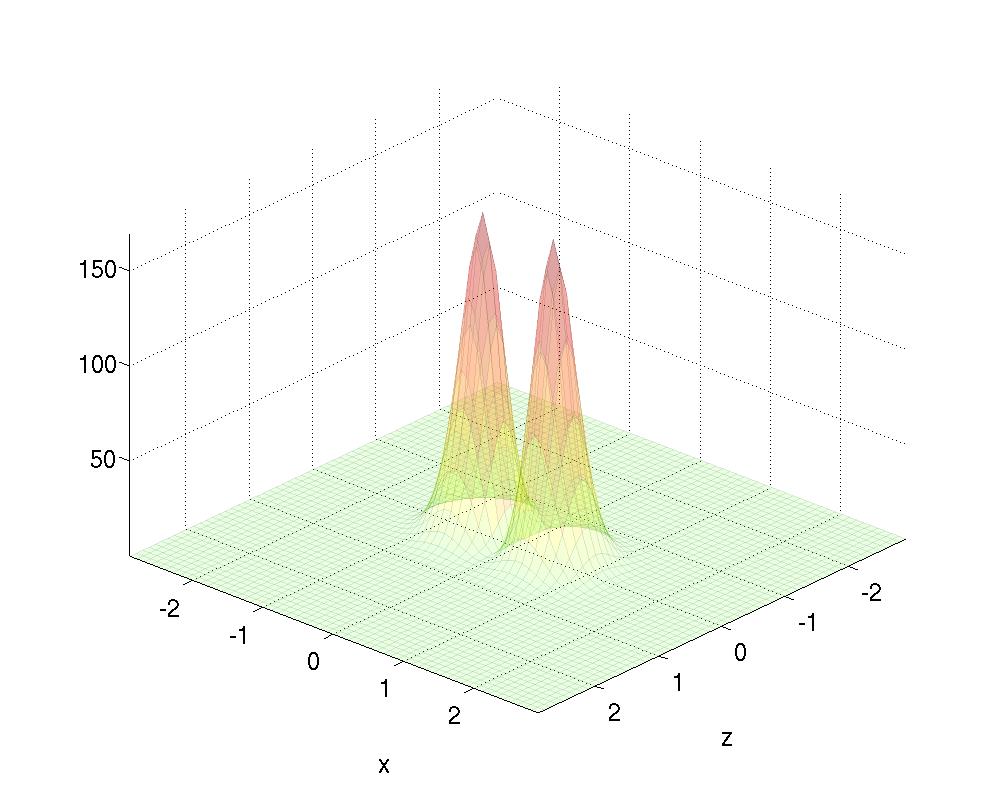}}
\subfloat{\includegraphics[width=0.245\linewidth]{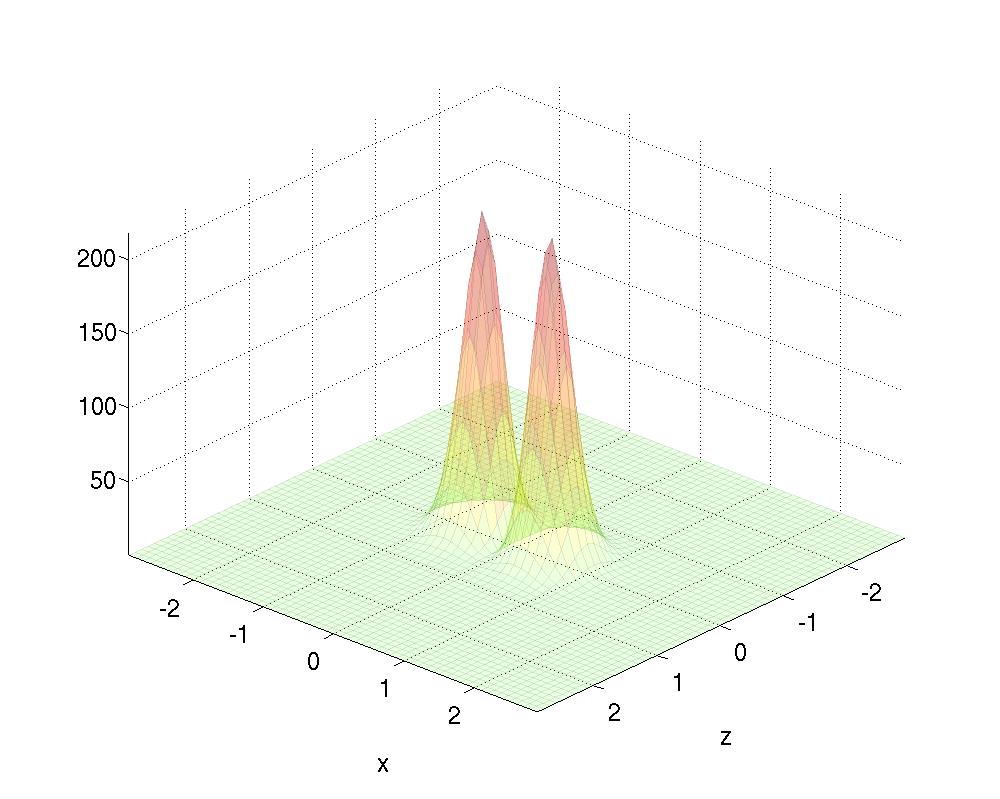}}}
\caption{A $B=2$ vortex ring in the 2+4 model. The figure is an
$xz$-slice of the energy density at $y=0$ for the vortex potential
mass parameter $m=0,1,2,3,4,5,6,7$ from top-left to bottom-right
panel.} 
\label{fig:nout8121_1_0_ms_energyslice}
\end{center}
\end{figure}

In Figs.~\ref{fig:nout8121_1_0_ms_baryonslice2}
and \ref{fig:nout8121_1_0_ms_baryonslice} we show $xy$-slices at $z=0$
and $xz$-slices at $y=0$, respectively, of the baryon charge densities
for the $B=2$ vortex ring in the 2+4 model for various values of
$m=0,1,\ldots,7$, while in Figs.~\ref{fig:nout8121_1_0_ms_energyslice2}
and \ref{fig:nout8121_1_0_ms_energyslice} we show the corresponding
$xy$-slices at $z=0$ and $xz$-slices at $y=0$, respectively, of the
energy densities.
We see from the figures that the baryon charge densities and the
energy densities are practically the same also in the $B=2$ case and
so from now on we can focus on the baryon charge densities in the 2+4
model.
We also note that turning on the vortex potential in the $B=2$ sector
has a quite small effect on the 2+4 model; it merely shrinks the
vortex ring and as a good approximation looks simply like a scale
transformation on the $B=2$ Skyrmion without the vortex potential. 

\begin{figure}[!tp]
\begin{center}
\mbox{
\subfloat{\includegraphics[width=0.245\linewidth]{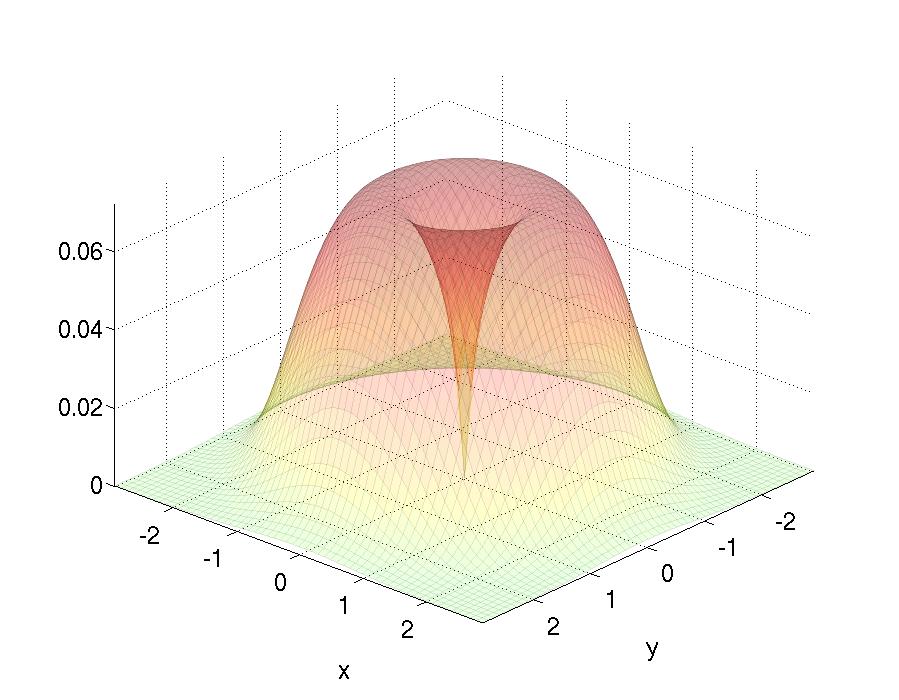}}
\subfloat{\includegraphics[width=0.245\linewidth]{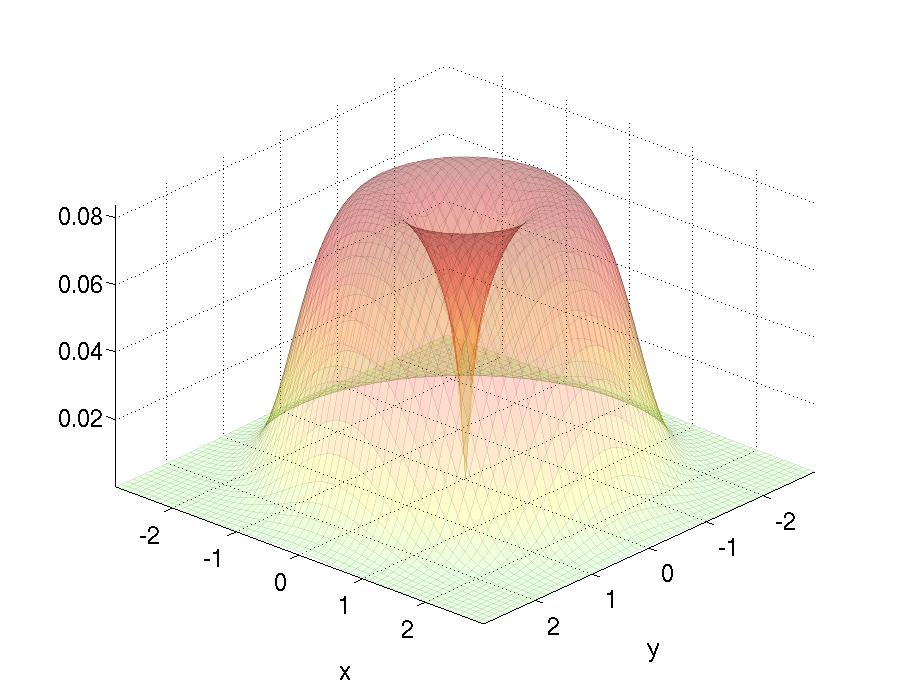}}
\subfloat{\includegraphics[width=0.245\linewidth]{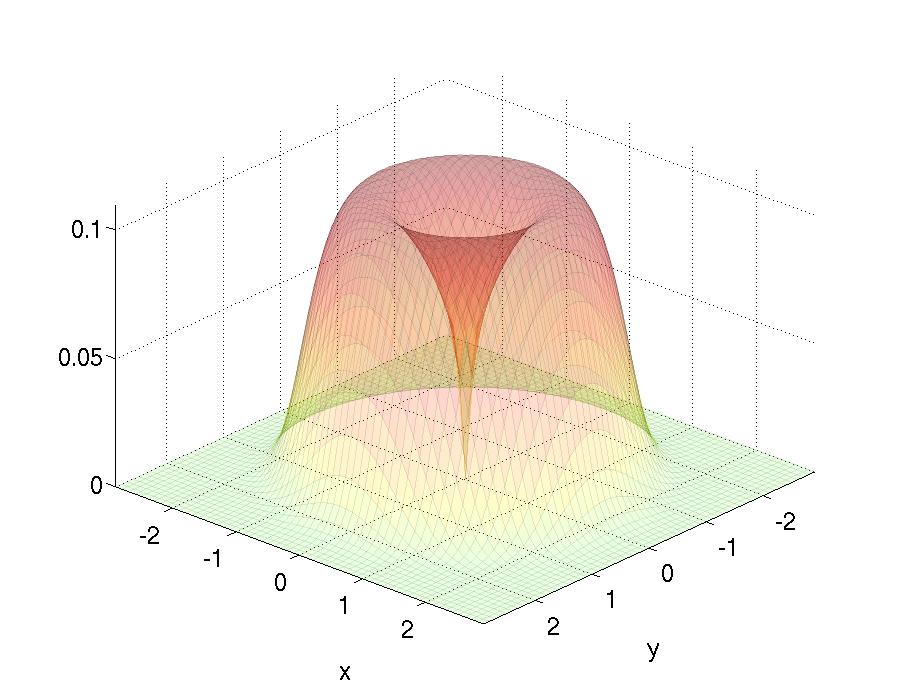}}
\subfloat{\includegraphics[width=0.245\linewidth]{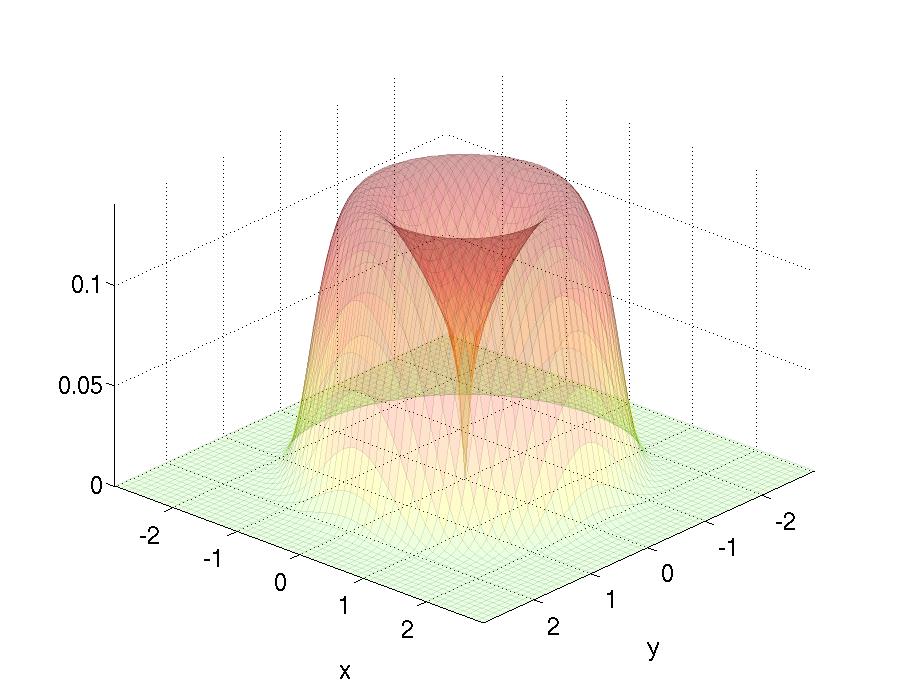}}}
\mbox{
\subfloat{\includegraphics[width=0.245\linewidth]{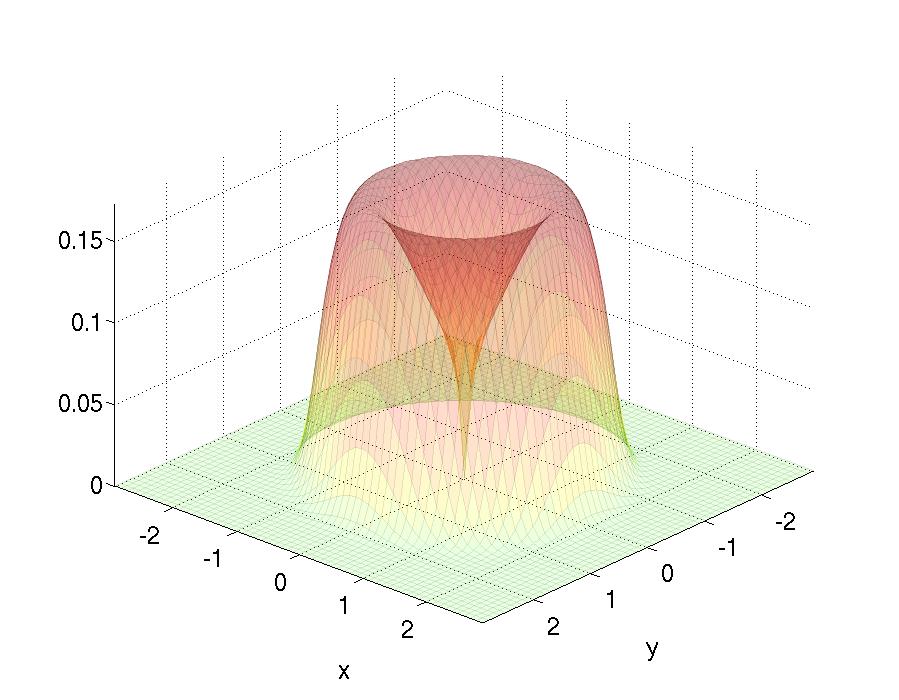}}
\subfloat{\includegraphics[width=0.245\linewidth]{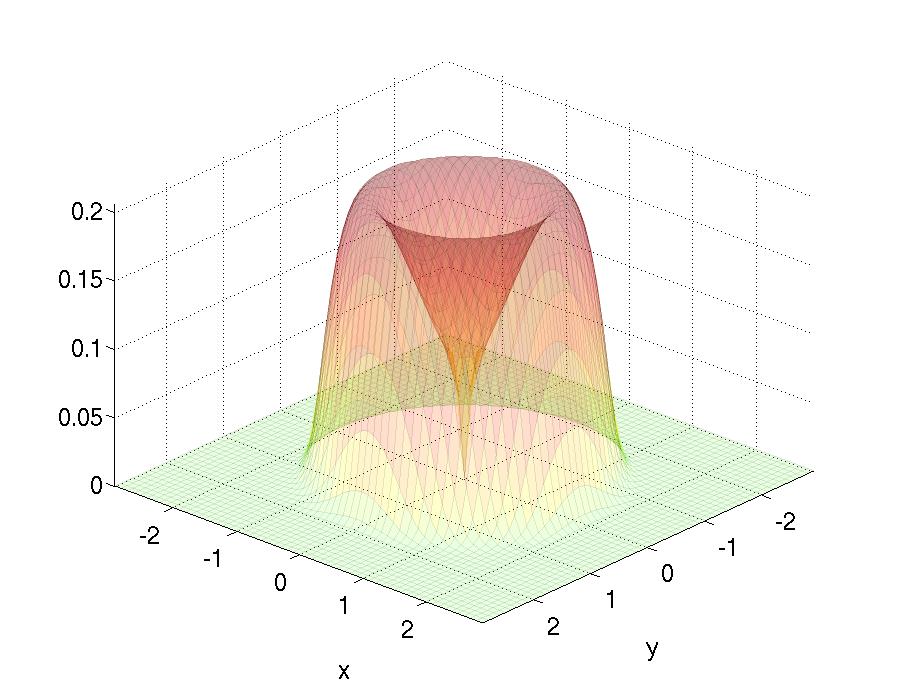}}
\subfloat{\includegraphics[width=0.245\linewidth]{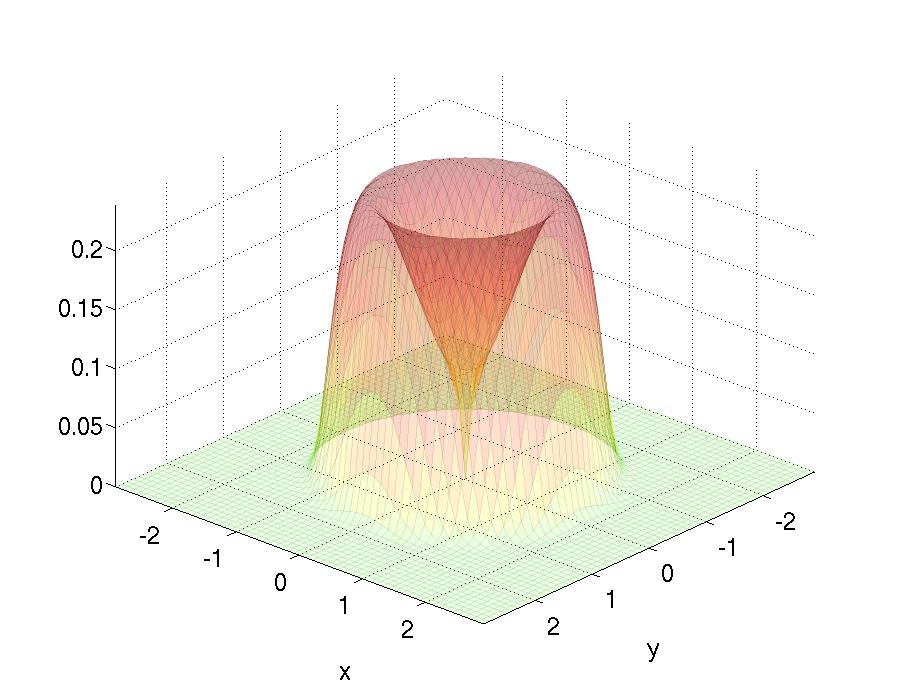}}
\subfloat{\includegraphics[width=0.245\linewidth]{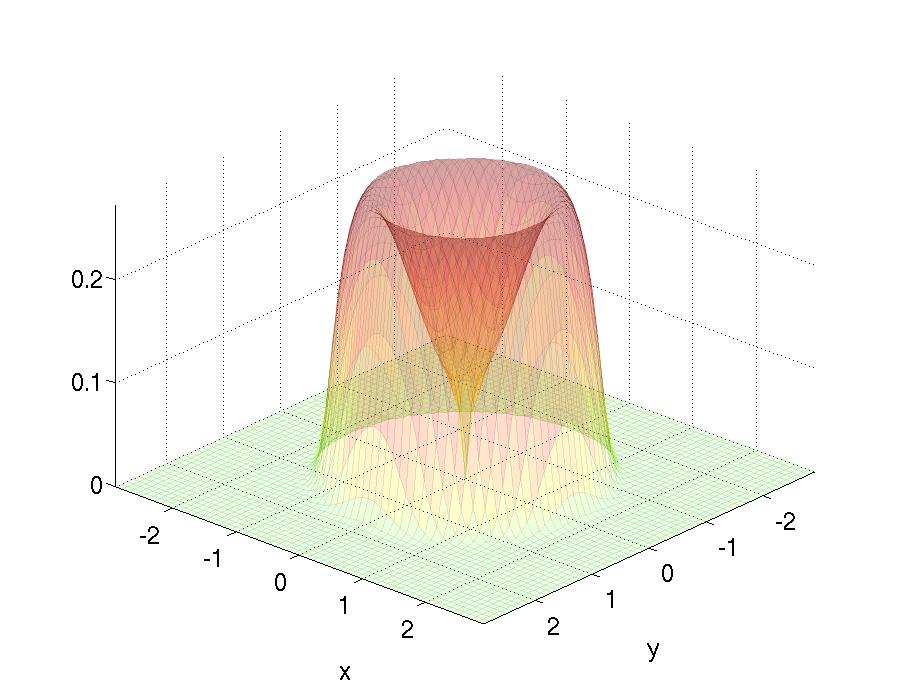}}}
\caption{A $B=2$ vortex ring in the 2+6 model. The figure is an
$xy$-slice of the baryon charge density at $z=0$ for the vortex
potential mass parameter $m=0,1,2,3,4,5,6,7$ from top-left to
bottom-right panel. }
\label{fig:nout8121_0_1_ms_baryonslice2}
\vspace*{\floatsep}
\mbox{
\subfloat{\includegraphics[width=0.245\linewidth]{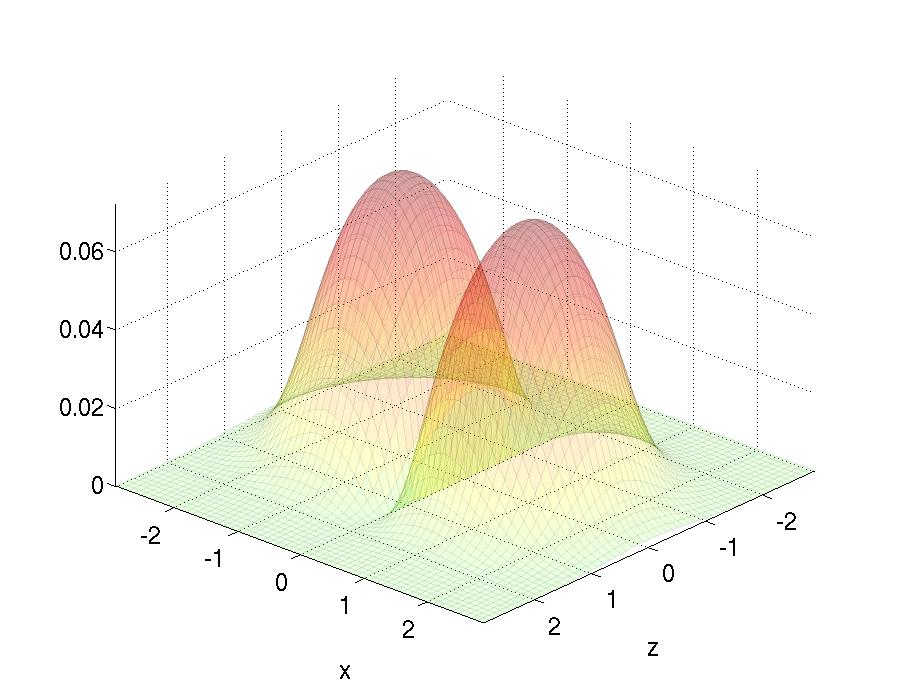}}
\subfloat{\includegraphics[width=0.245\linewidth]{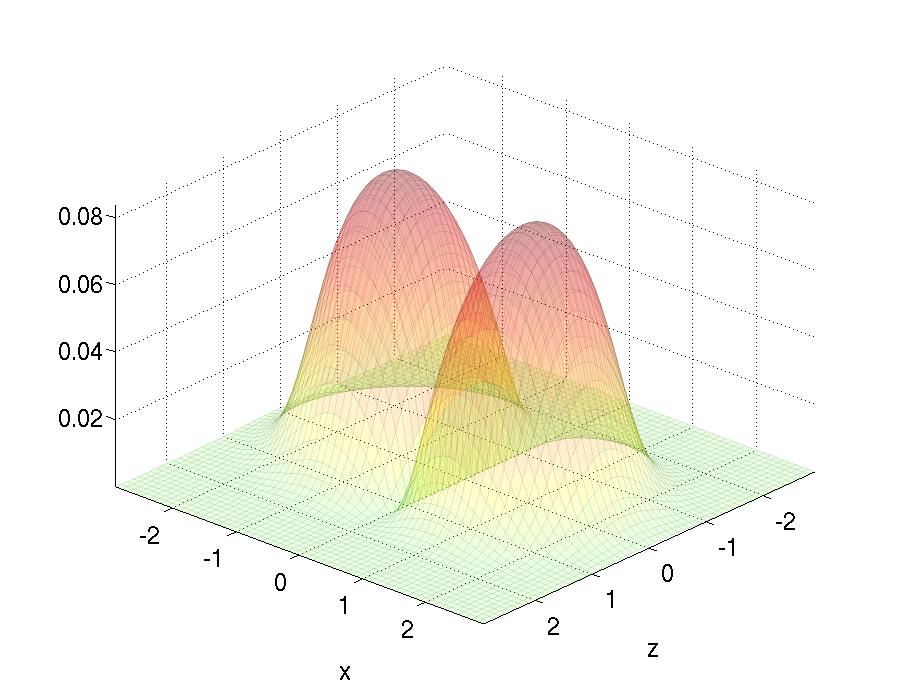}}
\subfloat{\includegraphics[width=0.245\linewidth]{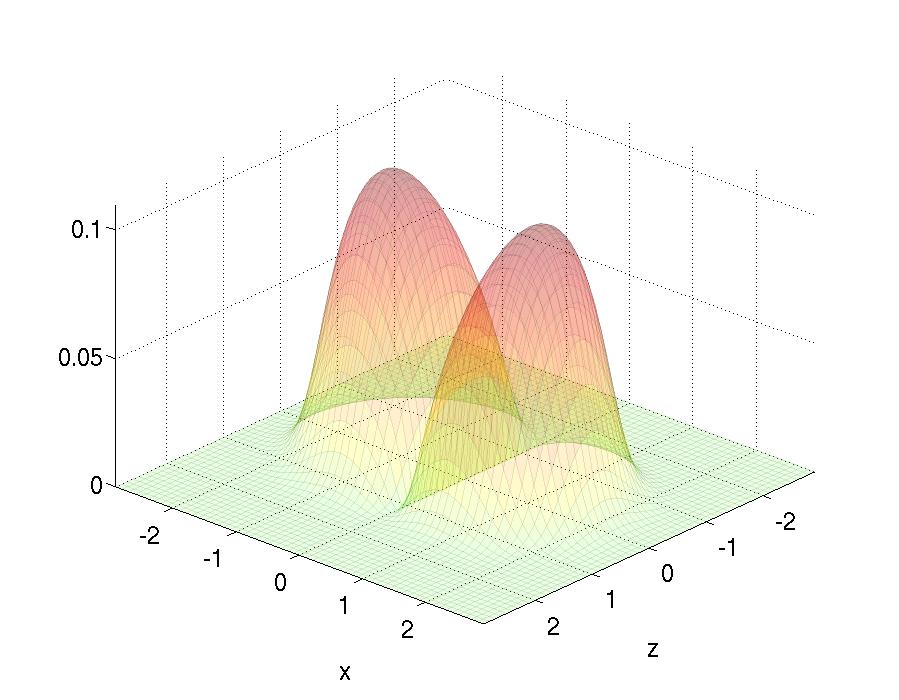}}
\subfloat{\includegraphics[width=0.245\linewidth]{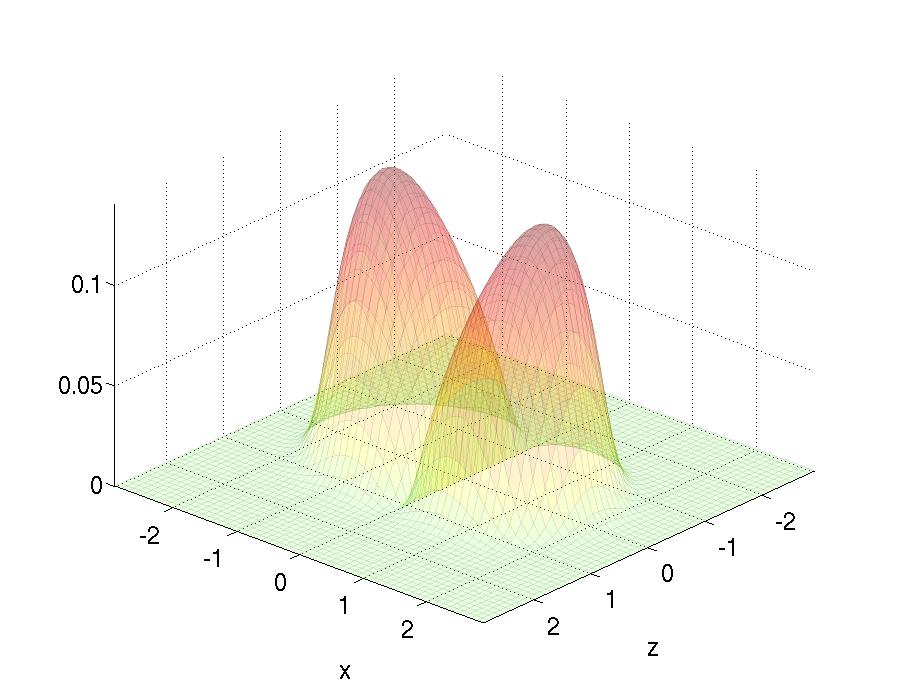}}}
\mbox{
\subfloat{\includegraphics[width=0.245\linewidth]{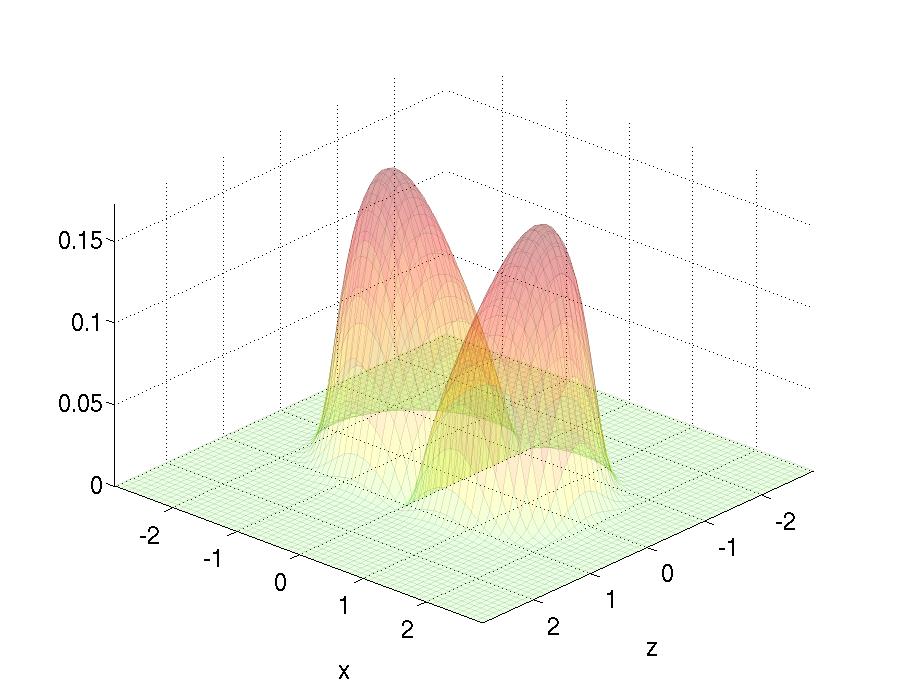}}
\subfloat{\includegraphics[width=0.245\linewidth]{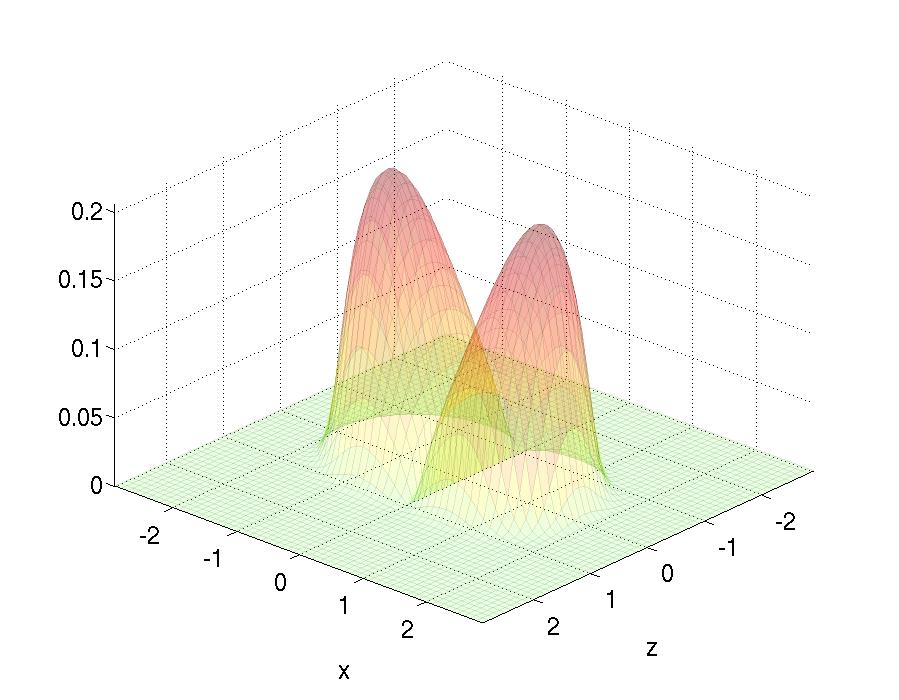}}
\subfloat{\includegraphics[width=0.245\linewidth]{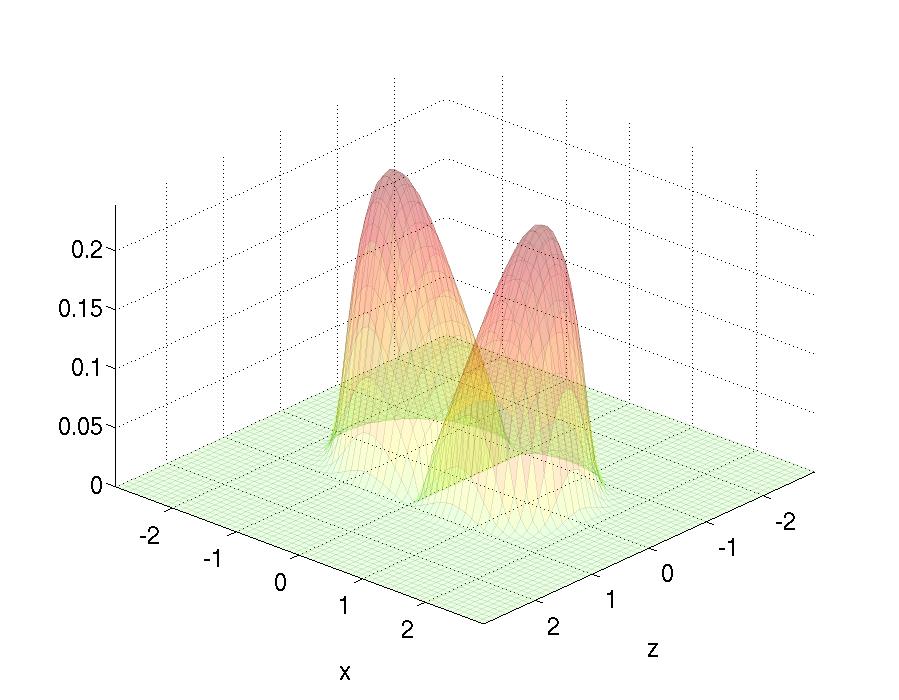}}
\subfloat{\includegraphics[width=0.245\linewidth]{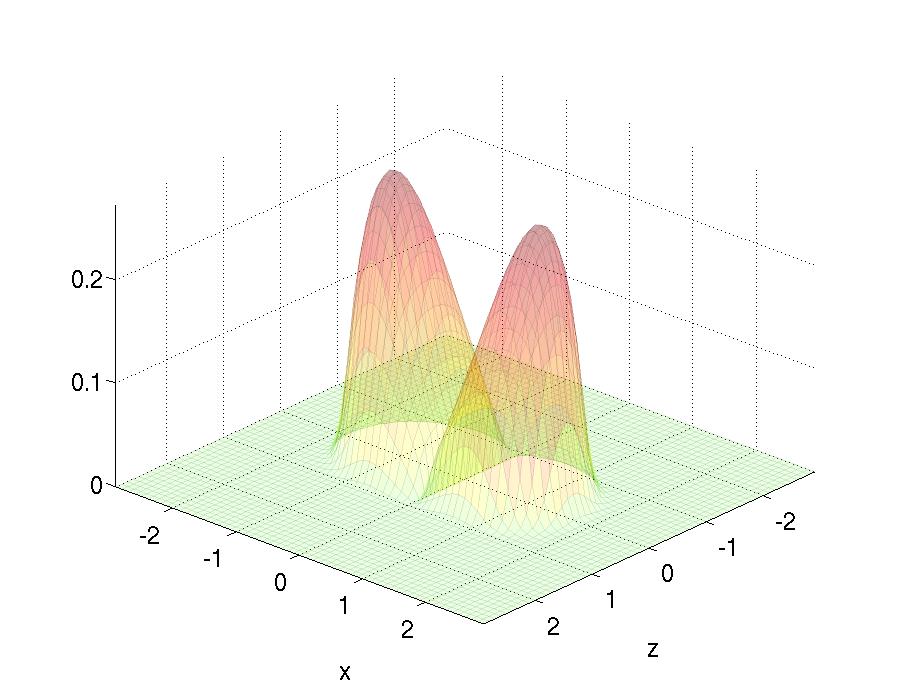}}}
\caption{A $B=2$ vortex ring in the 2+6 model. The figure is an
$xz$-slice of the baryon charge density at $y=0$ for the vortex
potential mass parameter $m=0,1,2,3,4,5,6,7$ from top-left to
bottom-right panel.}
\label{fig:nout8121_0_1_ms_baryonslice}
\end{center}
\end{figure}

\begin{figure}[!tp]
\begin{center}
\mbox{
\subfloat{\includegraphics[width=0.245\linewidth]{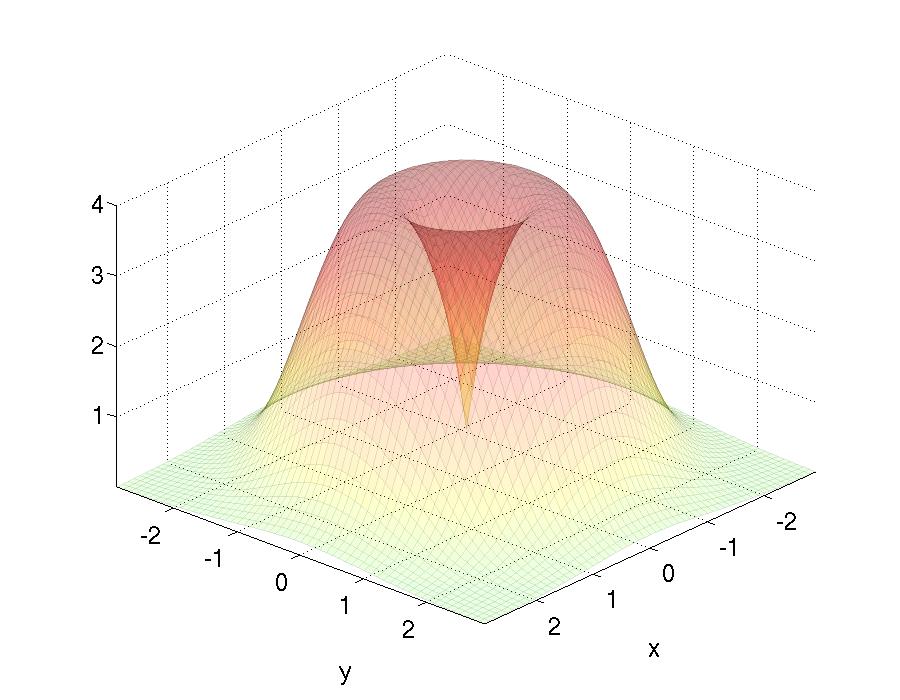}}
\subfloat{\includegraphics[width=0.245\linewidth]{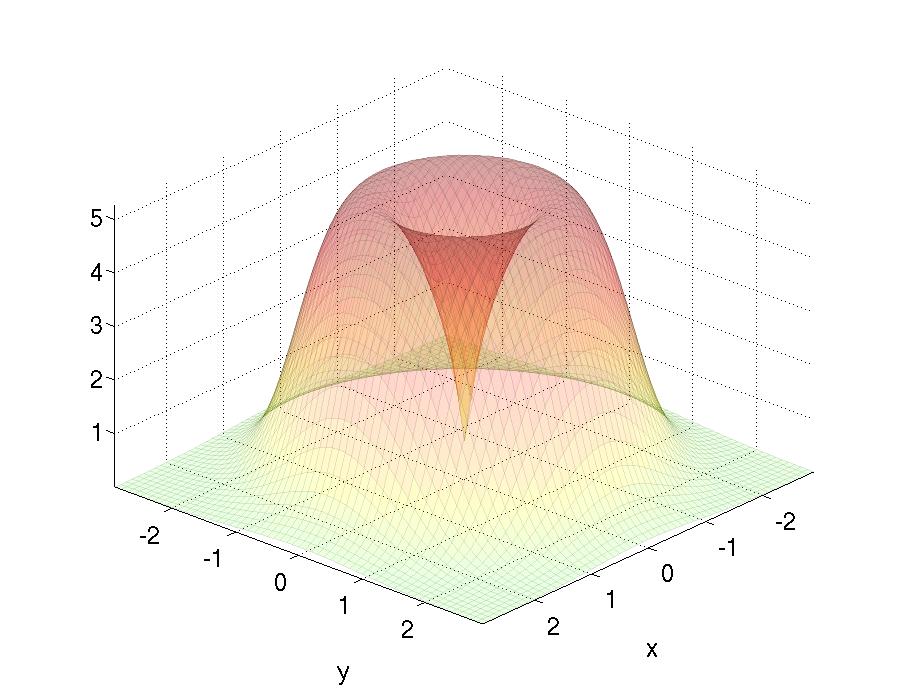}}
\subfloat{\includegraphics[width=0.245\linewidth]{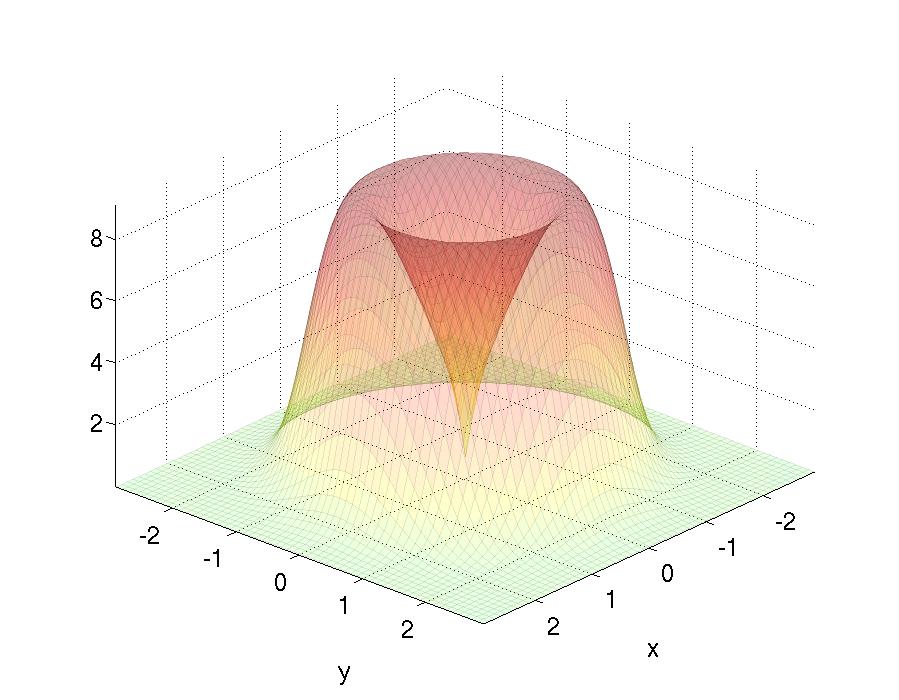}}
\subfloat{\includegraphics[width=0.245\linewidth]{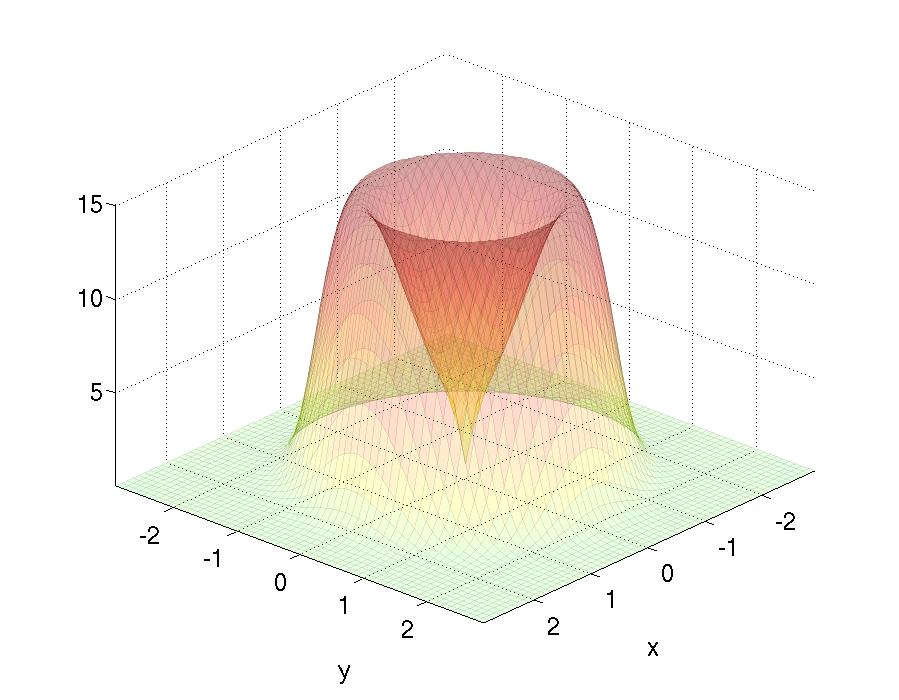}}}
\mbox{
\subfloat{\includegraphics[width=0.245\linewidth]{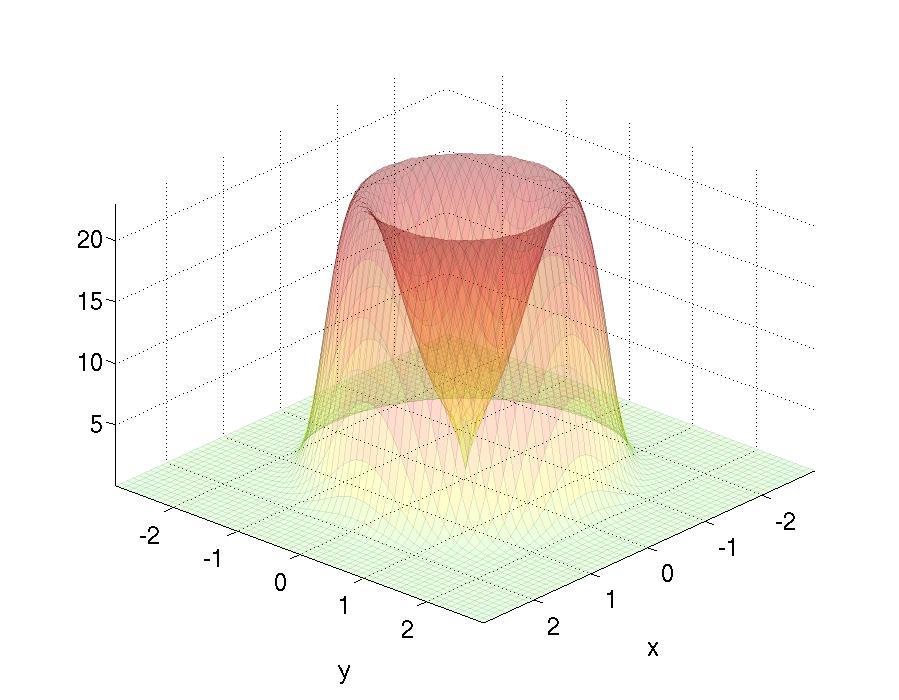}}
\subfloat{\includegraphics[width=0.245\linewidth]{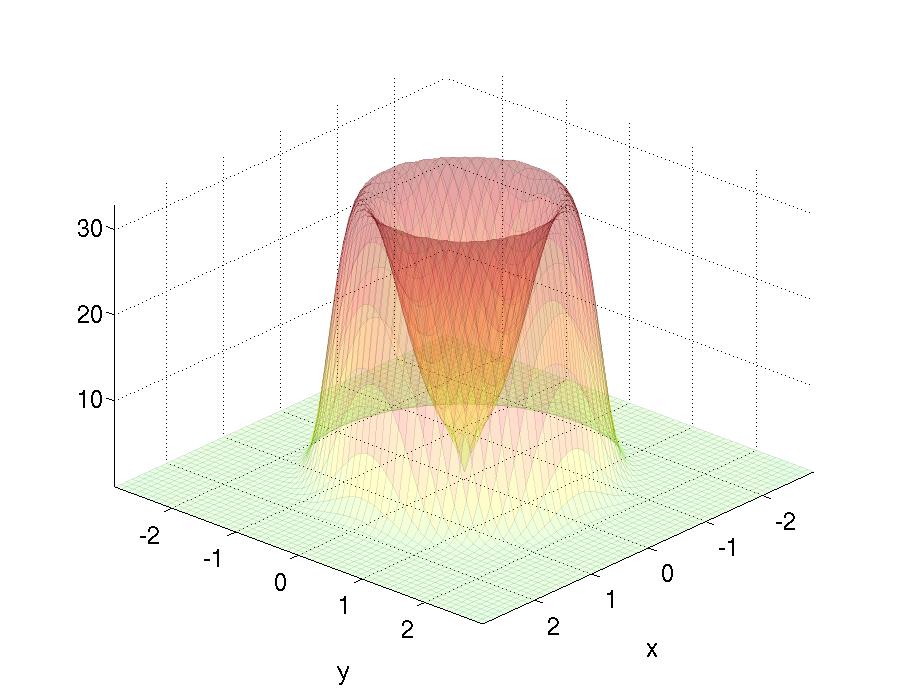}}
\subfloat{\includegraphics[width=0.245\linewidth]{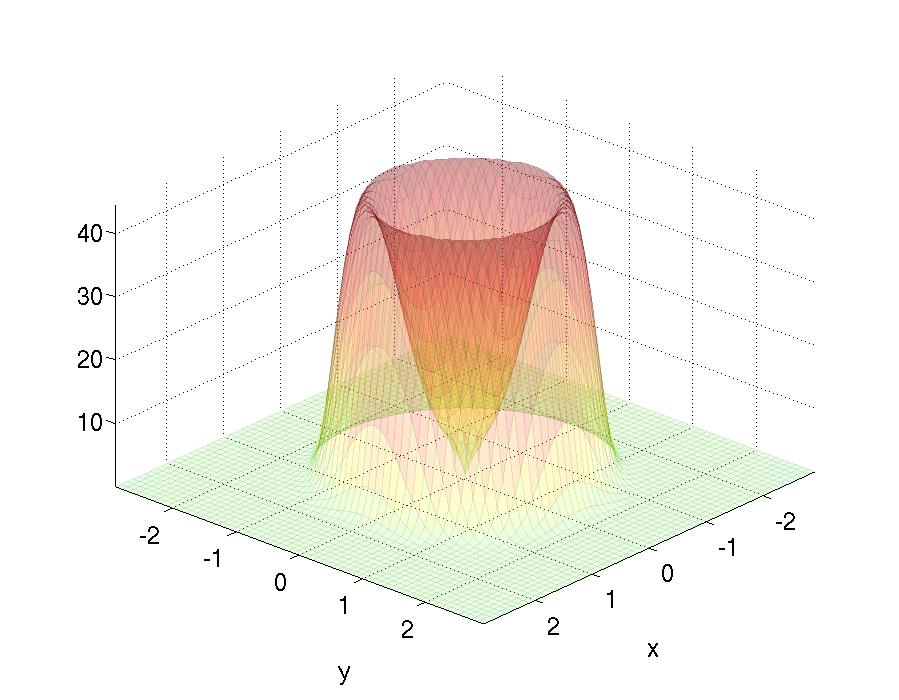}}
\subfloat{\includegraphics[width=0.245\linewidth]{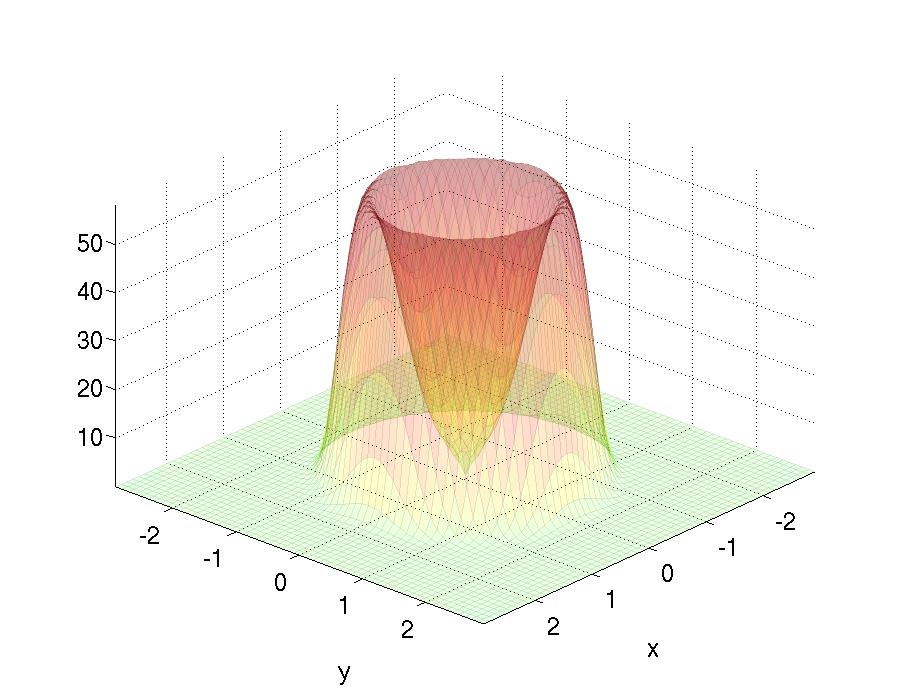}}}
\caption{A $B=2$ vortex ring in the 2+6 model. The figure is an
$xy$-slice of the energy density at $z=0$ for the vortex potential
mass parameter $m=0,1,2,3,4,5,6,7$ from top-left to bottom-right
panel. } 
\label{fig:nout8121_0_1_ms_energyslice2}
\vspace*{\floatsep}
\mbox{
\subfloat{\includegraphics[width=0.245\linewidth]{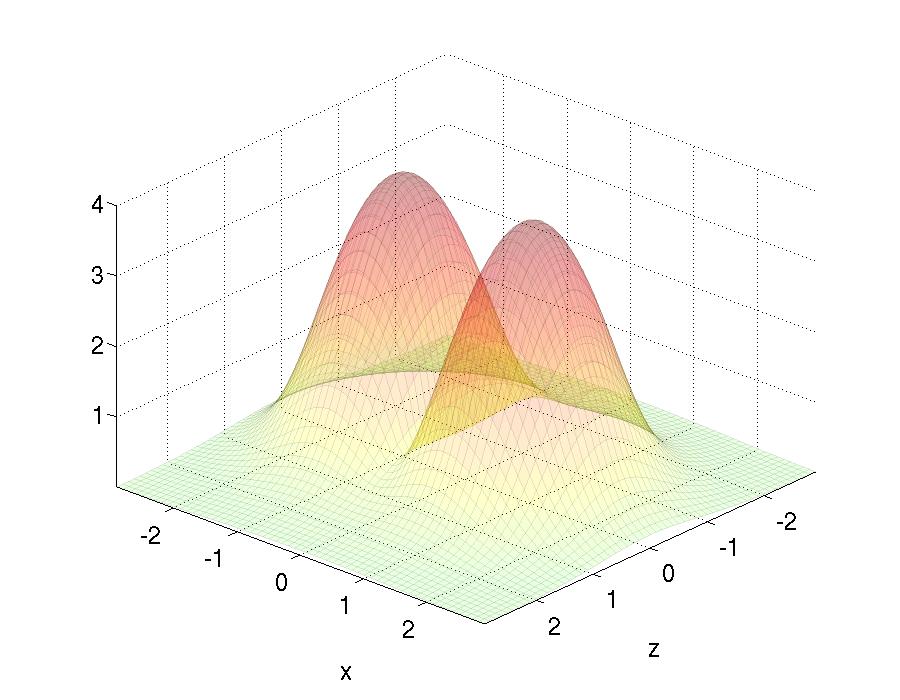}}
\subfloat{\includegraphics[width=0.245\linewidth]{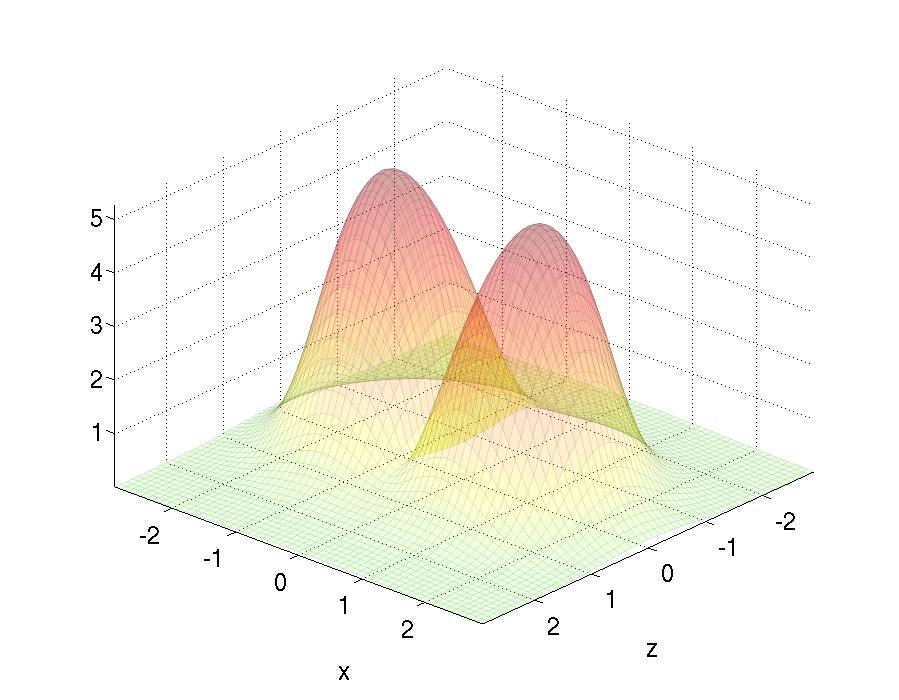}}
\subfloat{\includegraphics[width=0.245\linewidth]{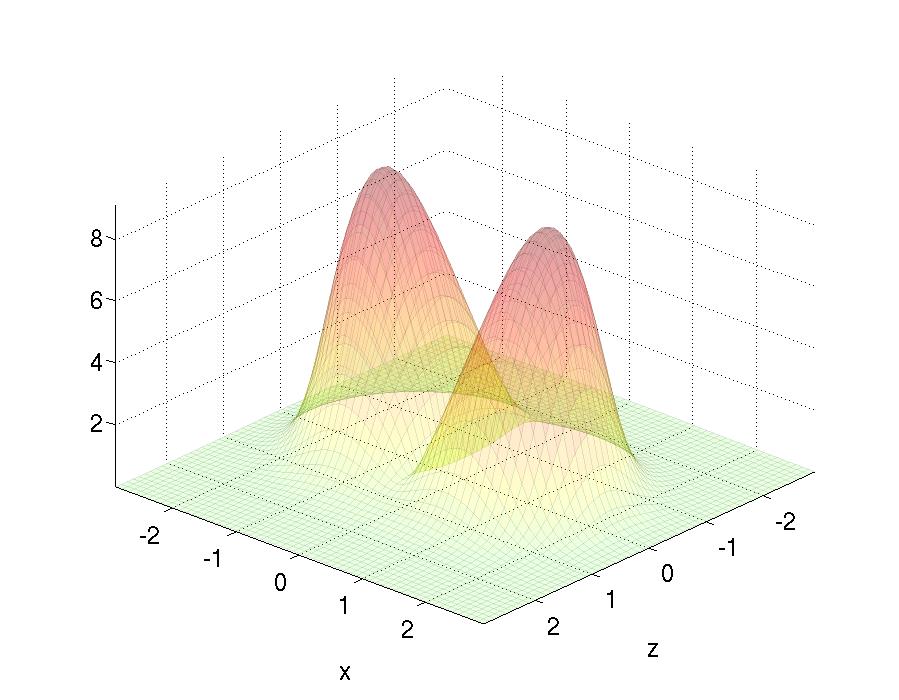}}
\subfloat{\includegraphics[width=0.245\linewidth]{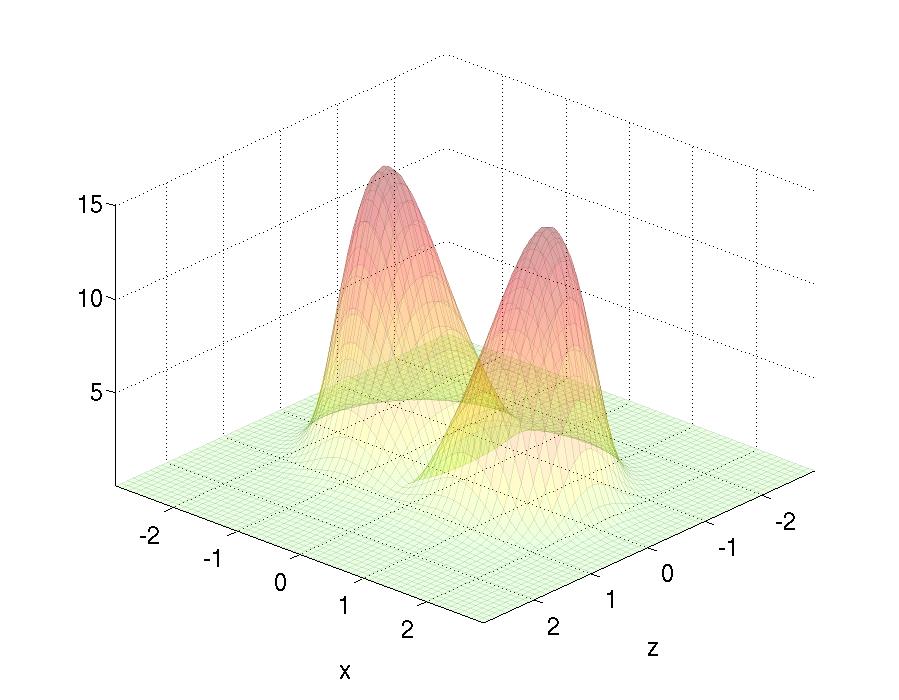}}}
\mbox{
\subfloat{\includegraphics[width=0.245\linewidth]{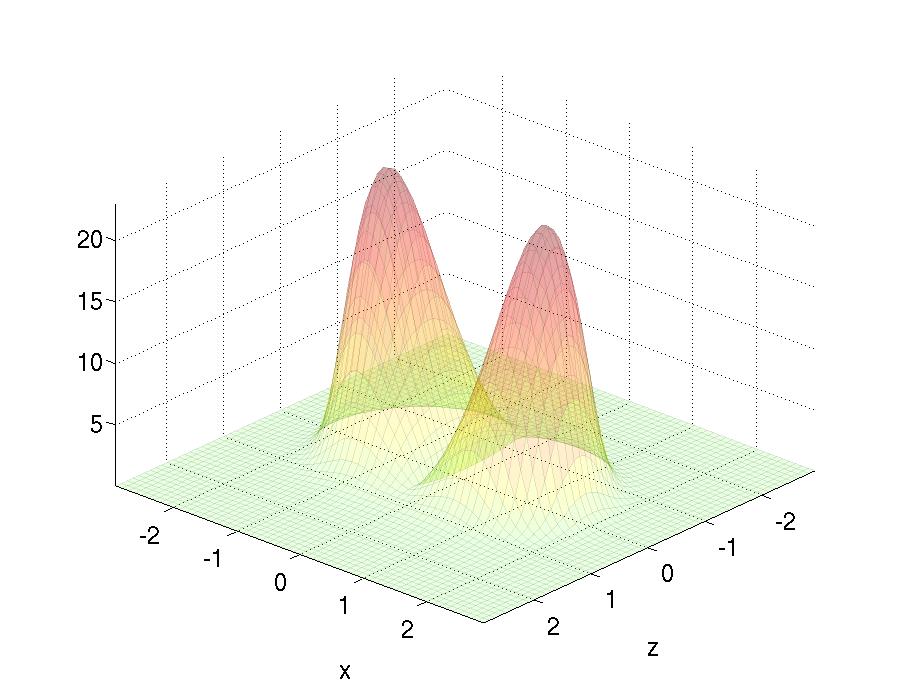}}
\subfloat{\includegraphics[width=0.245\linewidth]{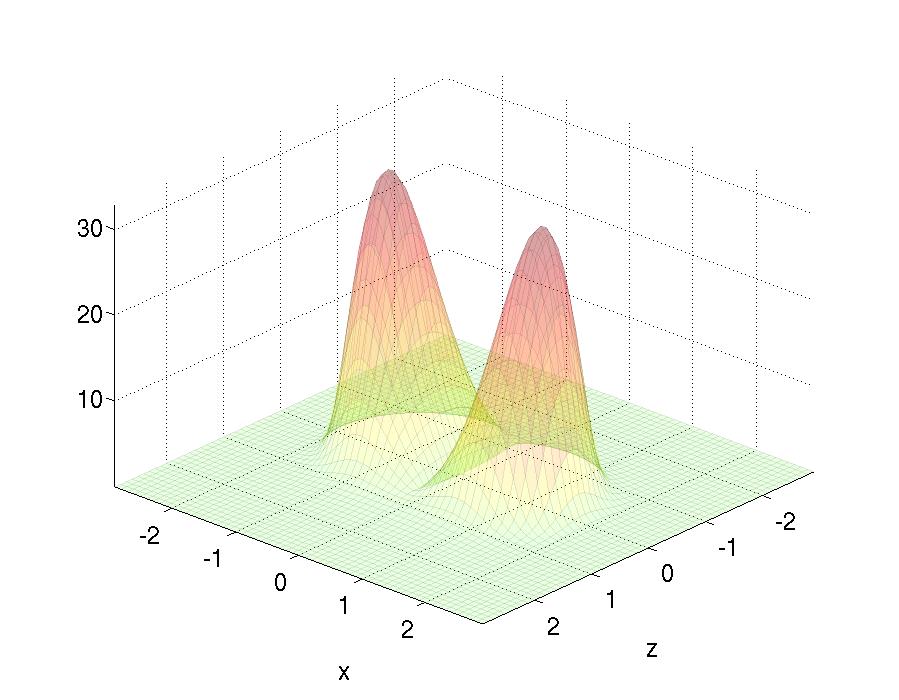}}
\subfloat{\includegraphics[width=0.245\linewidth]{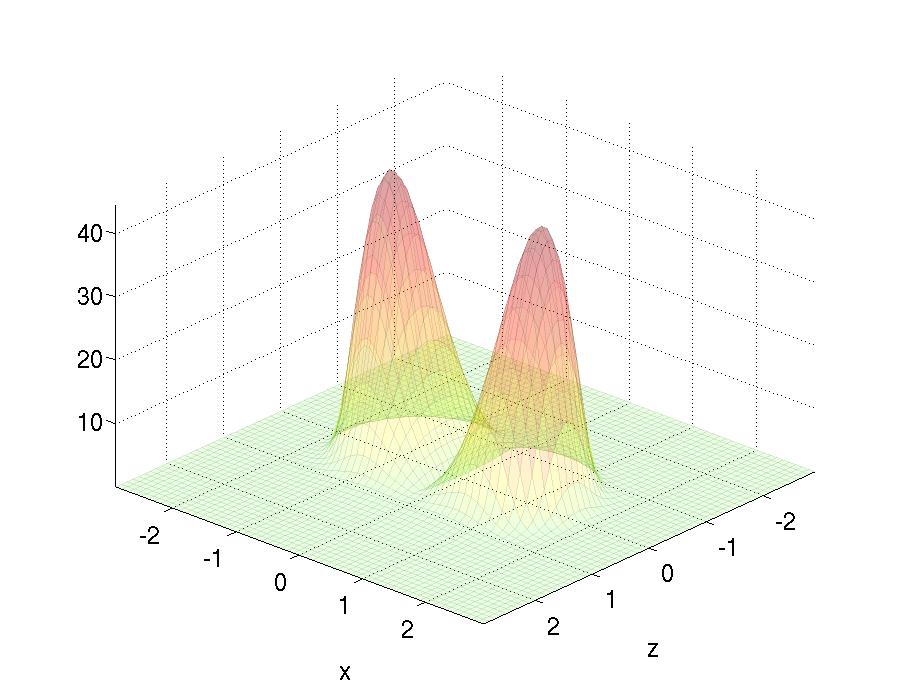}}
\subfloat{\includegraphics[width=0.245\linewidth]{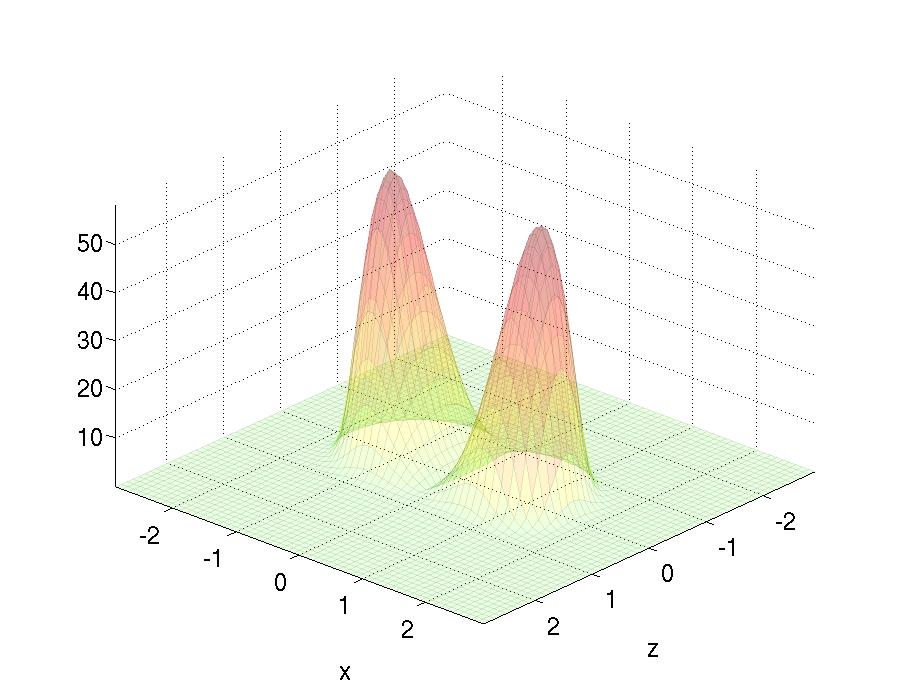}}}
\caption{A $B=2$ vortex ring in the 2+6 model. The figure is an
$xz$-slice of the energy density at $y=0$ for the vortex potential
mass parameter $m=0,1,2,3,4,5,6,7$ from top-left to bottom-right
panel.} 
\label{fig:nout8121_0_1_ms_energyslice}
\end{center}
\end{figure}

Next we turn to the 2+6 model and
Figs.~\ref{fig:nout8121_0_1_ms_baryonslice2} 
and \ref{fig:nout8121_0_1_ms_baryonslice} show $xy$-slices at $z=0$
and $xz$-slices at $y=0$, respectively, of the baryon charge densities
for the $B=2$ vortex ring for various values of $m=0,1,\ldots,7$,
while in Figs.~\ref{fig:nout8121_0_1_ms_energyslice2} 
and \ref{fig:nout8121_0_1_ms_energyslice} show the corresponding
slices of the energy densities.

Let us first warn the reader that the surface plots shown in the
figures are cropped with respect to the calculations (which are done
on far larger grids than shown here); so one should not worry about
boundary effects from the lattice being a finite one. 

As in the case of the 2+4 model, increasing the mass of the vortex
potential $m$ has the effect of shrinking the vortex ring, however, in
the 2+6 the hole in the torus actually grows, which is a big
difference between the 2+6 and 2+4 models.
Let us also note that the baryon charge densities have a quite
distinct difference from their respective energy densities.
The holes in the baryon charge densities are convex, whereas the holes
in the energy densities turn from convex to concave as $m$ is
increased.
This is also a difference between the 2+6 and the 2+4 model. 

One last consideration for the $B=2$ sector is whether the doubly
wound vortex ($k=2$), twisted once is stable or not.
In Ref.~\cite{Gudnason:2014jga} this case was studied in the BEC type
vortex potential (as opposed to the simpler vortex potential
considered here) and the findings were negative; that is, the $k=2$
vortex twisted once would decay into two separate $k=1$ vortices.
The initial condition for the $k=2$ vortex ring twisted once is given
by \cite{Gudnason:2014jga}
\beq
\phi_{\rm initial}^{\rm T} = \left(
\sqrt{1 - \sin^2 f\sin^2\theta}\,e^{i2\arctan(\tan f\cos\theta)},
\sin f\sin\theta\right).
\label{eq:B2k2initialcond}
\eeq
All our numerical attempts at finding a $k=2$ vortex ring twisted once
were unstable, so we conclude that the simpler vortex potential share
the same instability as that of BEC type, see
App.~\ref{app:stringsplitting}.

\section{Confined Skyrmions}\label{sec:quadpot}

In this section we will consider the full
potential \eqref{eq:Vtotal}; i.e.~the vortex potential plus the kink
potential. 
In particular, this potential allows for two kinks to be absorbed and 
confined on the vortex world sheet and in turn yielding a full unit of
baryon charge from the bulk point of view.

In this section, we will begin with the straight vortex and then turn
to embedding halfkinks on the vortex worldsheet. Then we curl up the
vortex strings and the result will be an even number of kinks on the
vortex ring; more precisely $2B$ kinks on a single vortex ring (with
$B$ twists). 

\subsection{Straight vortex}

It is straightforward to construct the straight vortex -- without a
kink on its worldsheet -- but with the kink potential turned on. It
suffices to see that with the Ansatz \eqref{eq:vortex_ansatz}, the
mass parameter $m$ in Sec.~\ref{sec:straightvortex} should simply be
replaced by 
\beq
m_{\rm vortex}^2 \equiv m^2 - m_2^2 > 0.
\label{eq:mvortex}
\eeq
All the discussion of Sec.~\ref{sec:straightvortex} thus goes through. 

The next step is to add the kink to the vortex world sheet, which we
will turn to in the next subsection. First, however, we will compare
the 3-dimensional solutions to 1-dimensional ones found in
Sec.~\ref{sec:straightvortex}.

\subsubsection{Comparison of the PDE and ODE solutions for the vortex}

Since the kink breaks translational symmetry along the vortex string,
we need at least to consider a 2-dimensional partial differential
equation (PDE). We choose however to work in Cartesian coordinates as
in the last section, and calculate the vortex-kink system again on 
a cubic lattice using the relaxation method. Here we use a $121^3$
lattice with the smallest stepsize
$\Delta x=\Delta y\simeq 0.025+0.017c_4$ and the largest stepsize is
$\Delta z\simeq 1/(30m_2)$. 

\begin{figure}[!thp]
\begin{center}
\includegraphics[width=0.49\linewidth]{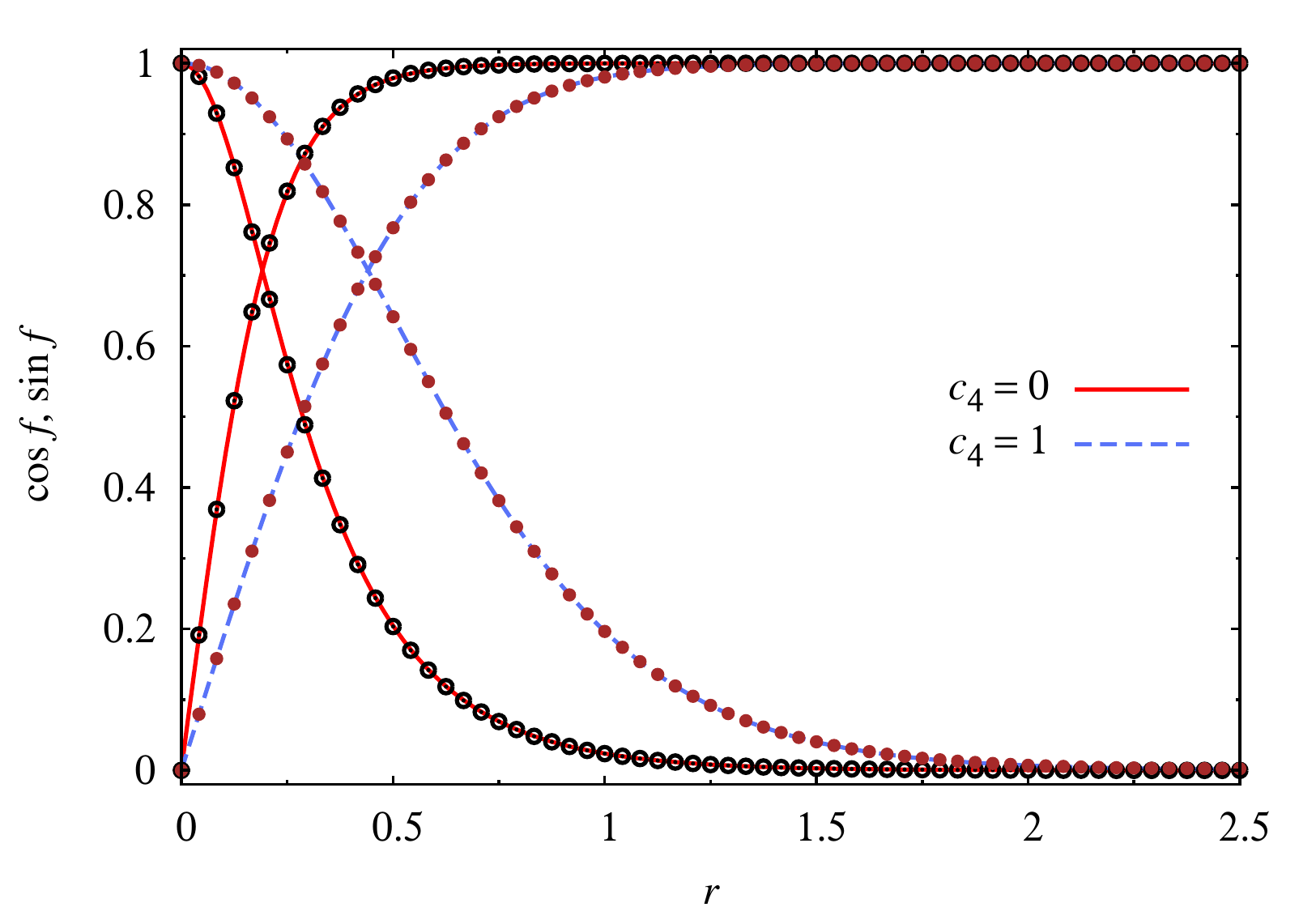}
\caption{Comparison of vortex condensates between the ODE (points) and
PDE (lines) calculations. Here the mass parameters are $m=4$ and
$m_2=1$ giving the effective vortex mass
$m_{\rm vortex}=\sqrt{15}\simeq 3.873$. }
\label{fig:compare_vortex}
\end{center}
\end{figure}

Far from the kink, the vortex should be unaltered with respect to that
found in Sec.~\ref{sec:straightvortex}, except for the fact that the
vortex mass $m$ is replaced with $m_{\rm vortex}$ in
Eq.~\eqref{eq:mvortex}. 
We therefore -- as a cross check on our calculations -- compare the
solutions found in Sec.~\ref{sec:straightvortex} using the ordinary
differential equation (ODE) and the Runge-Kutta method to the full 3D
PDE solutions used in this section as bases for the kinks.
The two vortex profiles are shown in Fig.~\ref{fig:compare_vortex},
where the lines are ODE solutions and the points are PDE data.
As we can see the lattice calculation gives exactly the same
solution.

\subsection{Halfkinks on a vortex as half-Skyrmions}\label{sec:singlehalfkinks}

In this section we twist the U(1) modulus of the vortex string, which
gives rise to a kink on the vortex worldline and to a Skyrmion in
the bulk volume (all space).

\subsubsection{A single halfkink on a vortex as a half-Skyrmion}

As mentioned already in the introduction, the fundamental kink in the
(quadratic) kink potential is a halfkink, meaning that in terms of a
U(1) modulus it winds only $\pi$ (as opposed to a full winding of
$2\pi$). 
The fact that the modulus winds only $\pi$ makes the calculation very
easy since the halfkink is fixed by the boundary conditions at
$z\to\pm\infty$ and the solution is simply the interpolation that
minimizes the energy of the configuration connecting these two vacua. 

The halfkink actually exists even in the case without higher
derivative terms, i.e.~in the case $c_4=c_6=0$; we will call this case
the 2 model
\beq
2\;{\rm model}: \quad c_4=0, \quad c_6=0.
\eeq
The string itself, asymptotically far away from the halfkink is the
same as that in the 2+6 model. This is because the BPS-Skyrme term is 
sixth order in derivatives and due to the anti-symmetrization, it
vanishes on the (straight) string. This statement is equivalent to the
fact that there is no baryon charge on the string itself before the
kink is embedded to its worldsheet.
In the 2+4 model, on the other hand, the string feels the fourth-order 
derivative term and gets widened by it; see Figs.~\ref{fig:skvtx1}
and \ref{fig:compare_vortex}. 

We are now ready to embed the halfkink to the string worldsheet.
Let us consider the Ansatz \eqref{eq:vortex_ansatz}, but with
$\chi=\chi(z)$; in particular let us consider as a guess
\beq
\chi_{\rm initial}(z) = \frac{\pi}{2}\left(\tanh[m_2 z]-1\right),
\label{eq:kinkAnsatz}
\eeq
which corresponds to the boundary conditions $\chi(-\infty)=-\pi$ and
$\chi(\infty)=0$. 

The baryon charge with the vortex Ansatz \eqref{eq:vortex_ansatz} and
$\chi=\chi(z)$ is nontrivial only for a nonzero kink number ($k>0$)
and it reads 
\begin{align}
B = -\frac{1}{2\pi} \int drdz\; \sin(2f) f_r \chi_z 
= \frac{1}{4\pi} \left[\cos 2f\right]_{r=0}^{r=\infty}
\left[\chi\right]_{z=-\infty}^{z=\infty} 
= \frac{\chi(-\infty) - \chi(\infty)}{2\pi},
\end{align}
where we have plugged in the boundary conditions for the
vortex \eqref{eq:BCvortex} in the last equality.
Using now the boundary conditions for the kink corresponding to the
Ansatz \eqref{eq:kinkAnsatz}, we get $B=-1/2$; i.e.~an anti-halfkink.
A parity transformation turns the anti-halfkink into a halfkink, so we
will not care about the charge being negative.

\begin{figure}[!tp]
\begin{center}
\mbox{
\subfloat[]{\includegraphics[width=0.24\linewidth]{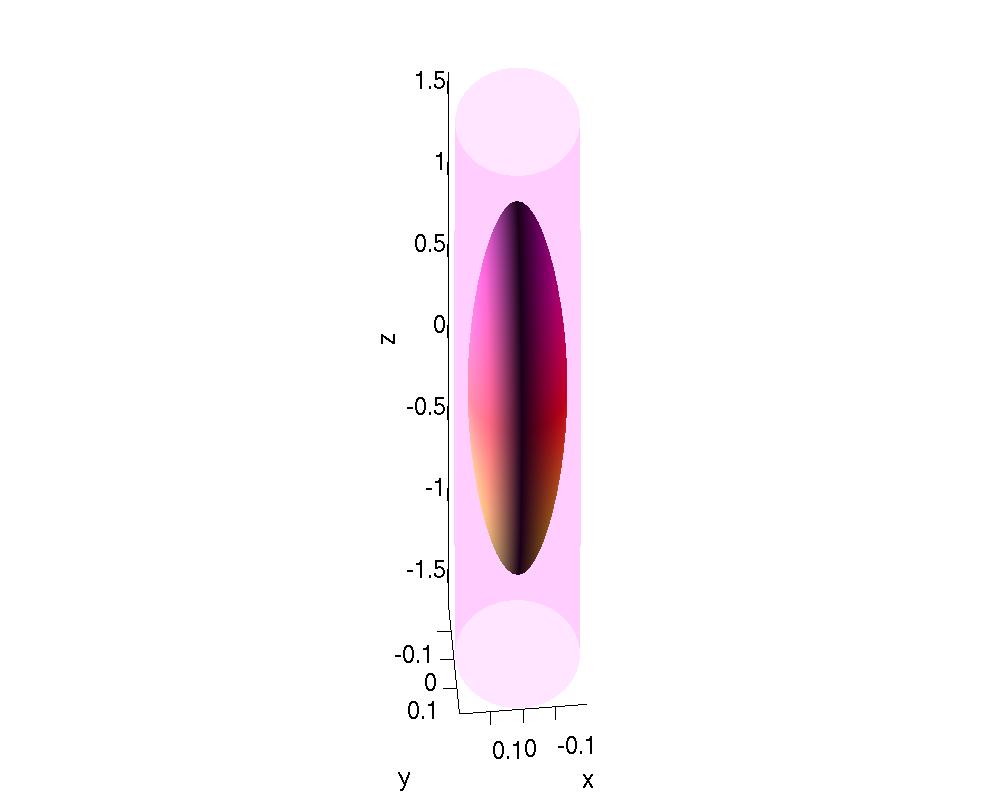}}
\subfloat[]{\includegraphics[width=0.24\linewidth]{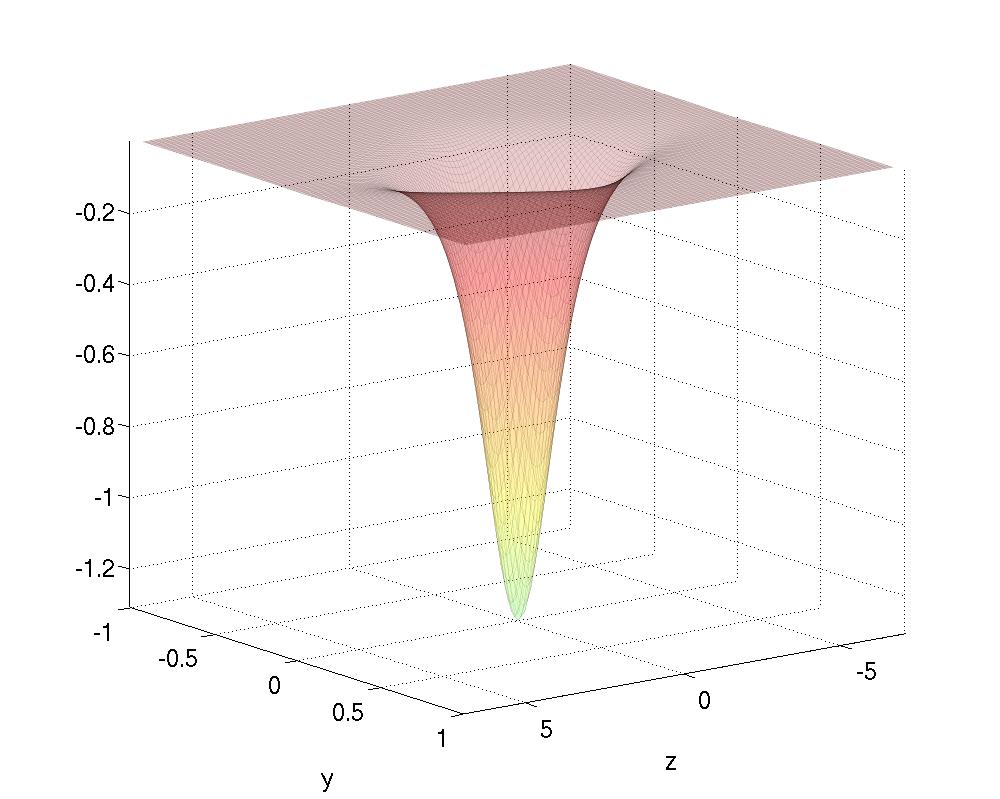}}
\subfloat[]{\includegraphics[width=0.24\linewidth]{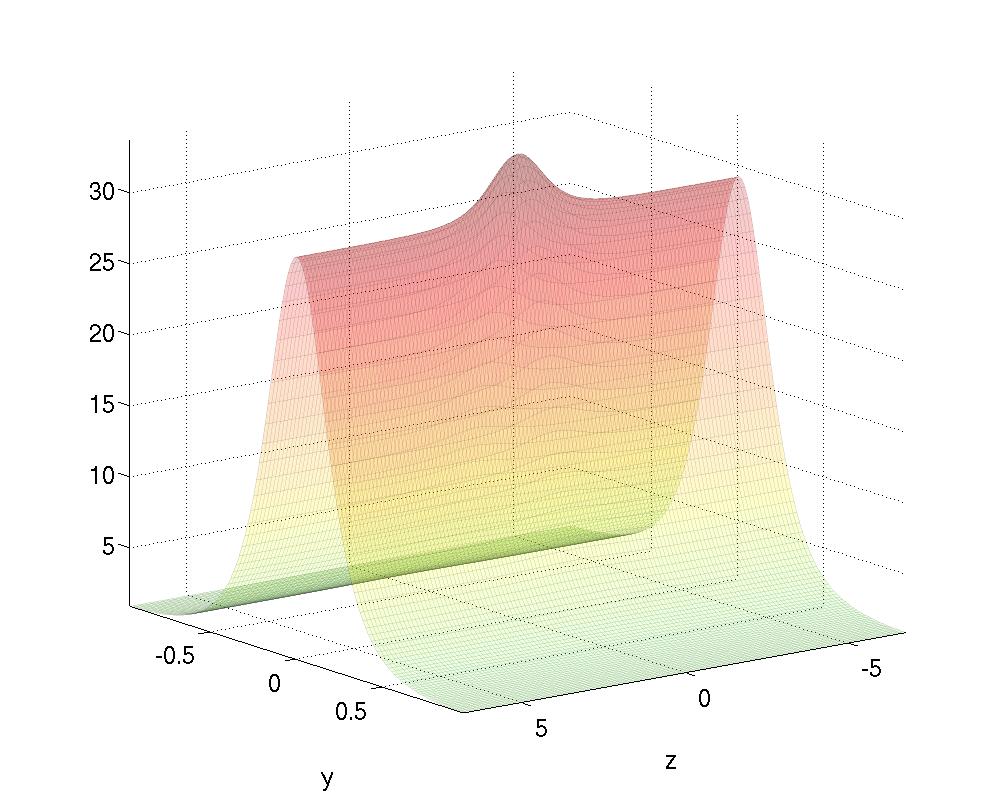}}
\subfloat[]{\includegraphics[width=0.24\linewidth]{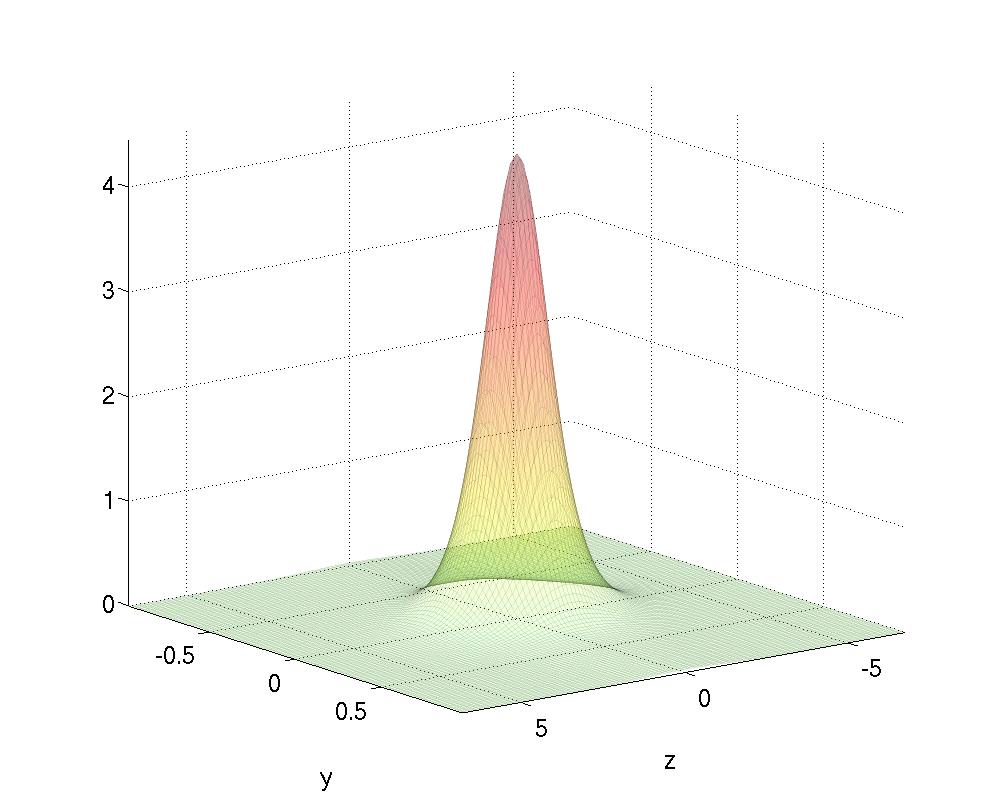}}}
\caption{A halfkink in the 2 model; (a) isosurfaces of energy density
(pink) and baryon charge density (colored), (b) $yz$-slice of baryon
charge density, (c) $yz$-slice of total energy density and (d)
$yz$-slice of kink energy density, all at $x=0$.
The baryon charge is calculated to be $B^{\rm numerical}=-0.4993$.
For coloring of the baryon charge isosurface, see the text.
In this figure $m=4$ and $m_2=1$. }
\label{fig:hk121_0_0_m2_1}
\end{center}
\end{figure}

\begin{figure}[!tp]
\begin{center}
\mbox{
\subfloat[]{\includegraphics[width=0.24\linewidth]{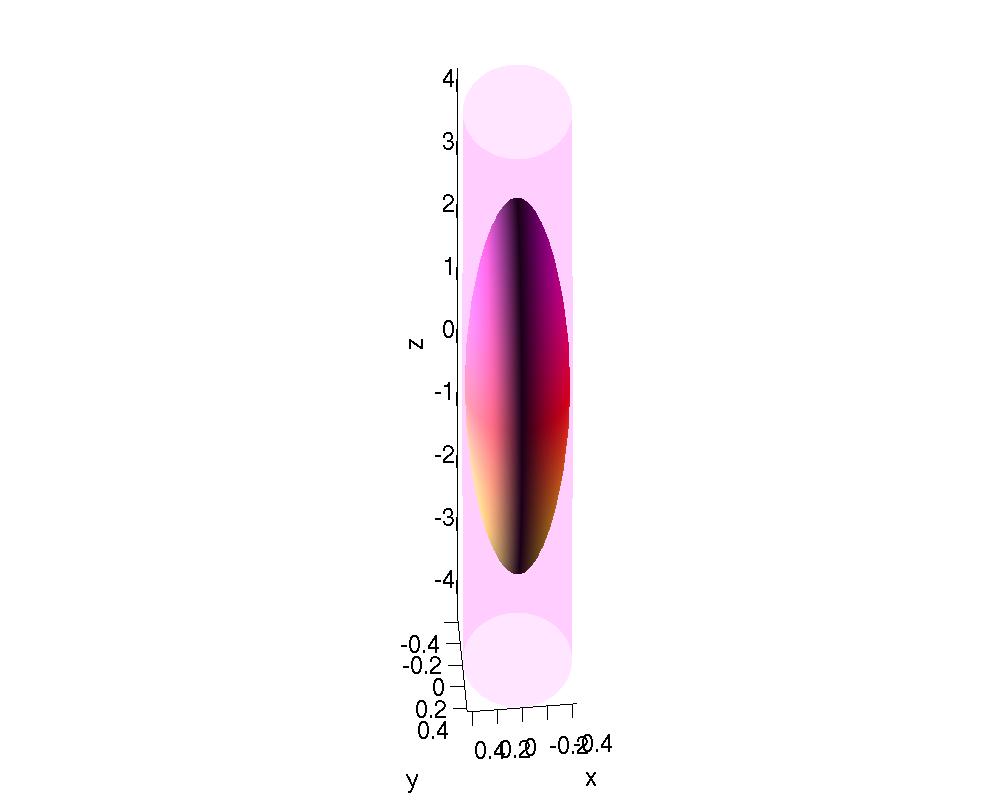}}
\subfloat[]{\includegraphics[width=0.24\linewidth]{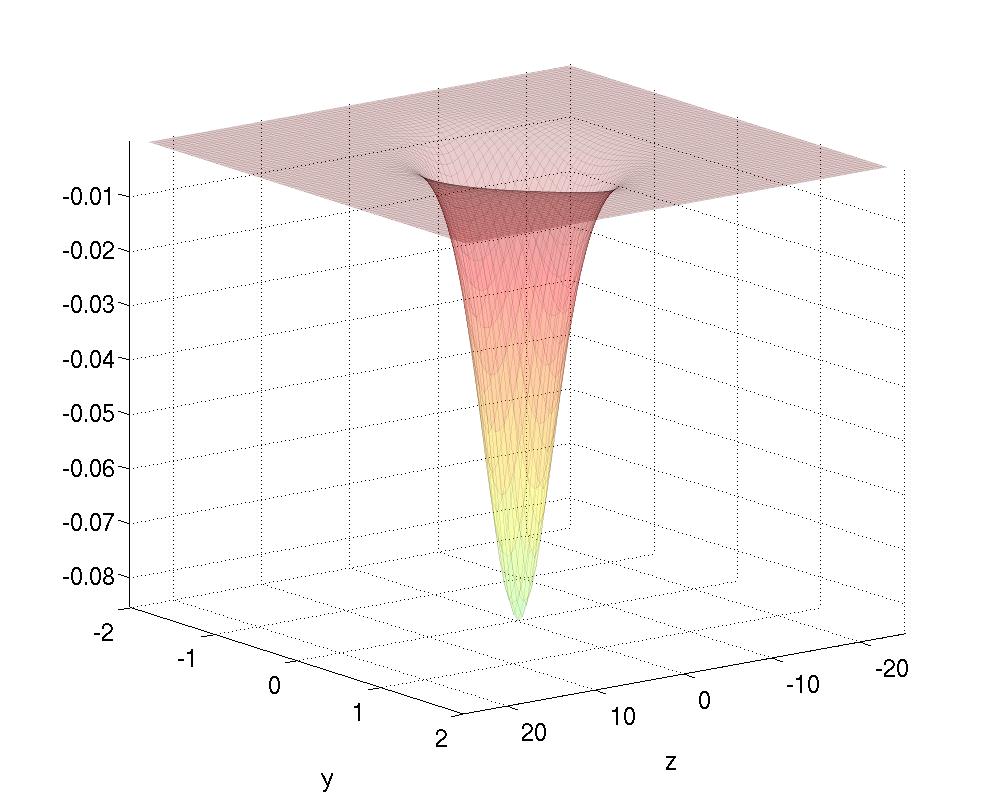}}
\subfloat[]{\includegraphics[width=0.24\linewidth]{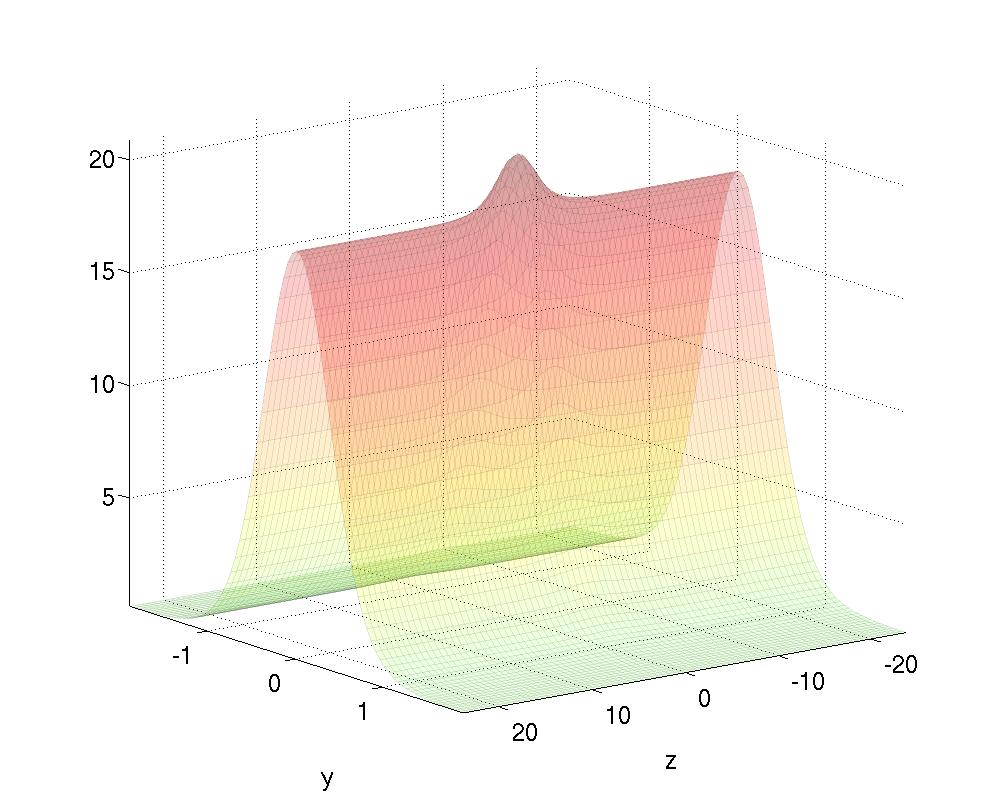}}
\subfloat[]{\includegraphics[width=0.24\linewidth]{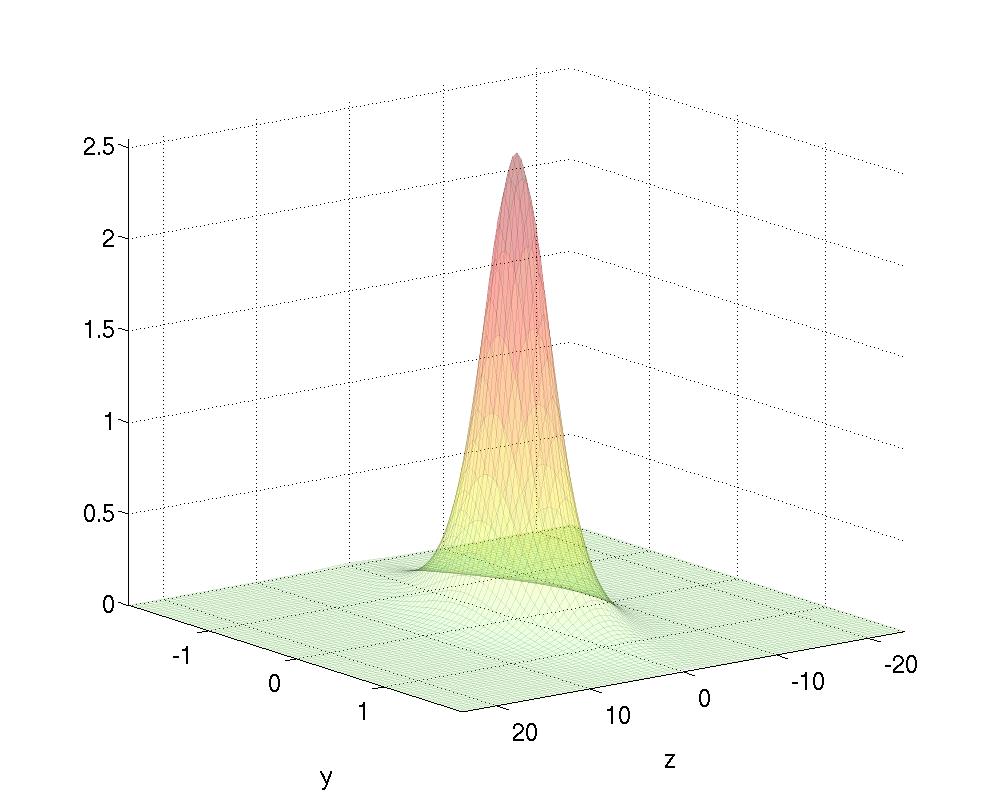}}}
\caption{A halfkink in the 2+4 model; (a) isosurfaces of energy density
(pink) and baryon charge density (colored), (b) $yz$-slice of baryon
charge density, (c) $yz$-slice of total energy density and (d)
$yz$-slice of kink energy density, all at $x=0$.
The baryon charge is calculated to be $B^{\rm numerical}=-0.4998$.
For coloring of the baryon charge isosurface, see the text.
In this figure $m=4$ and $m_2=1$. }
\label{fig:hk81_1_0_m2_1}
\end{center}
\end{figure}

\begin{figure}[!tp]
\begin{center}
\mbox{
\subfloat[]{\includegraphics[width=0.24\linewidth]{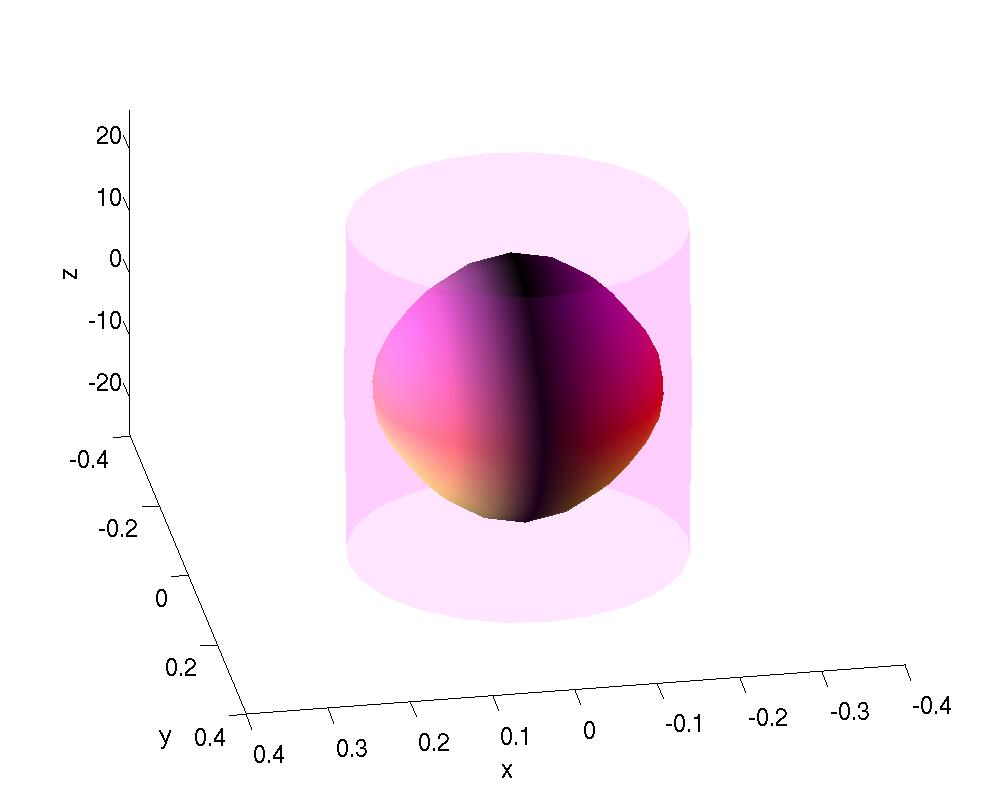}}
\subfloat[]{\includegraphics[width=0.24\linewidth]{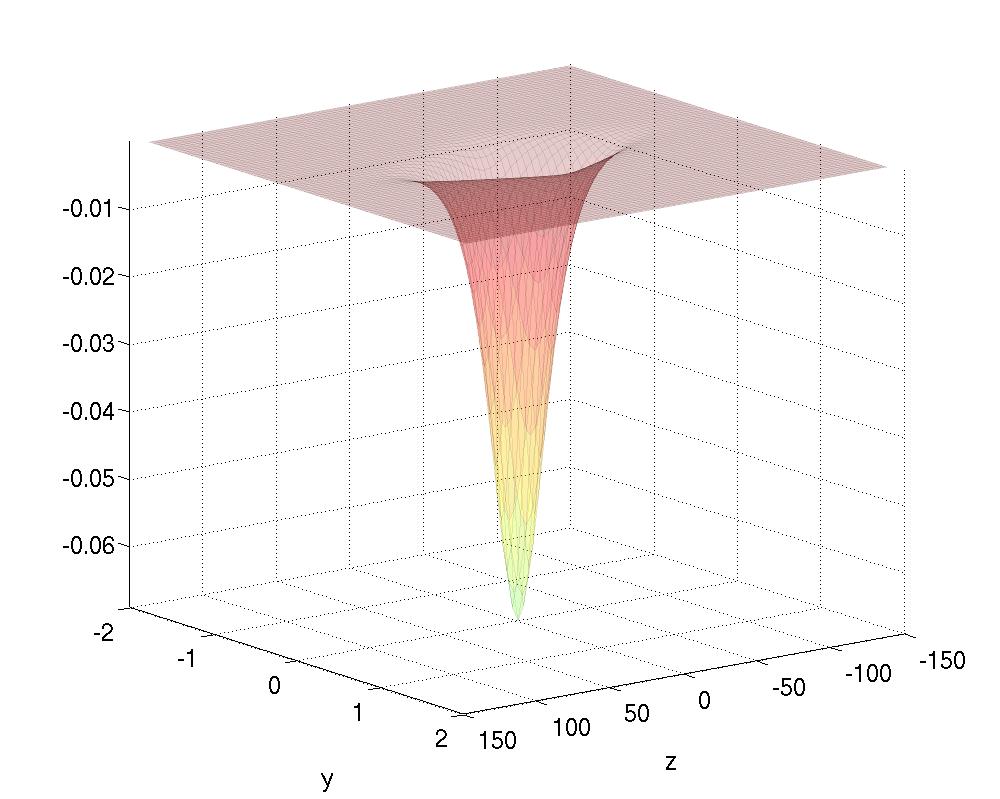}}
\subfloat[]{\includegraphics[width=0.24\linewidth]{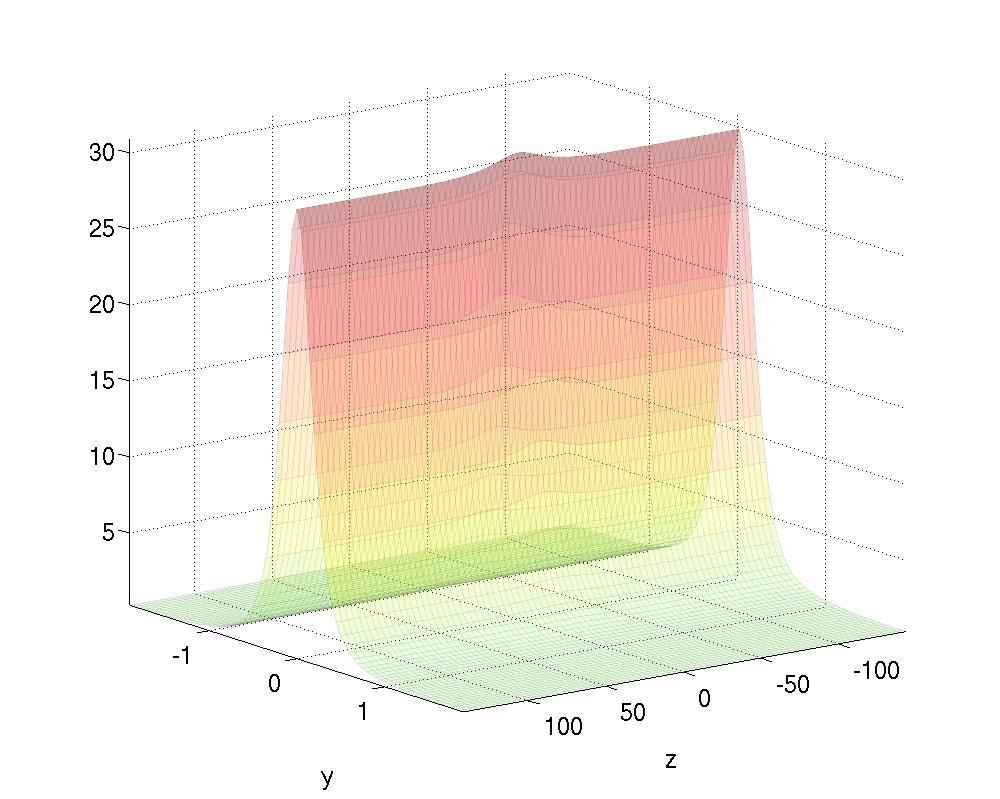}}
\subfloat[]{\includegraphics[width=0.24\linewidth]{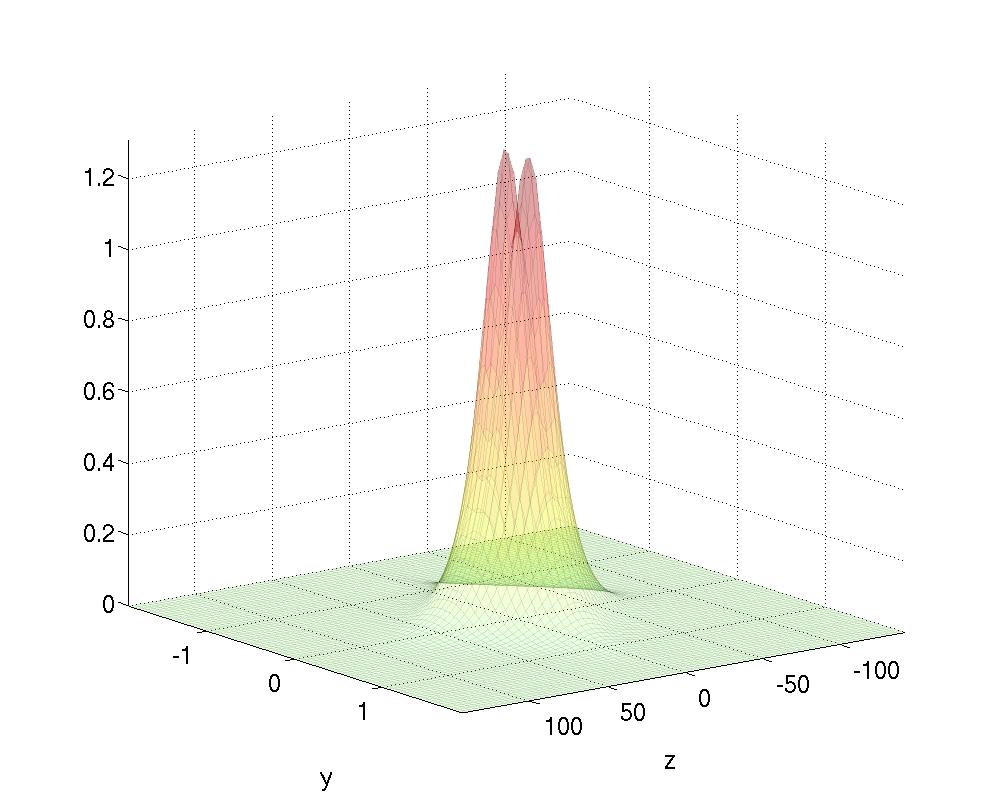}}}
\caption{A halfkink in the 2+6 model; (a) isosurfaces of energy density
(pink) and baryon charge density (colored), (b) $yz$-slice of baryon
charge density, (c) $yz$-slice of total energy density and (d)
$yz$-slice of kink energy density, all at $x=0$.
The baryon charge is calculated to be $B^{\rm numerical}=-0.4997$.
Note that the isosurfaces in (a) are not shown with the proper scale
ratio; it is widened to render the figure readable. 
For coloring of the baryon charge isosurface, see the text.
In this figure $m=4$ and $m_2=1$. }
\label{fig:hk81_0_1_m2_1}
\end{center}
\end{figure}

In Figs.~\ref{fig:hk121_0_0_m2_1}, \ref{fig:hk81_1_0_m2_1}
and \ref{fig:hk81_0_1_m2_1} are shown the halfkink solutions in terms
of the isosurfaces, baryon charge densities and energy densities, for
the 2 model, 2+4 model and 2+6 model, respectively.
The isosurfaces of energy densities and baryon charge densities are
all shown at half-maximum values of their respective quantities.
The coloring scheme used for the baryon charge isosurfaces is
made by constructing a normalized 3-vector
$\mathbf{n}=(\Im\phi_2,\Re\phi_2,\Im\phi_1)/|(\Im\phi_2,\Re\phi_2,\Im\phi_1)|$
of length 1.
The first two components are then mapped to the color hue circle where
$n_1+in_2=e^{i\theta}$ and $\theta=0,\pi/3,2\pi/3$
corresponds to red, green, blue.
$|n_3|$ corresponds to the lightness with $n_3=0$ being black and
$|n_3|=1$ white.
The (d) panel of the figures displays the kink energy, which we
calculated by subtracting the asymptotic string energy from the total
energy, leaving roughly the energy density of the kink on top of the
string worldsheet, see Eq.~\eqref{eq:Ekink}. 

First we note that the length of the halfkinks differ between the
different models and the shortest is that of the 2 model, while the
longest halfkink is that of the 2+6 model.
Next we can see that the string width is larger in the 2+4 model than
in the other two models, as expected.
Finally, we observed from the kink energy densities that in the 2
and 2+4 models, the shape is peak-like, whereas in the 2+6 model the
shape of the energy density is like a twin peak, where the two peaks
sit on each side of the string. Since this figure is a $yz$-slice,
the kink energy density is hence slightly torus-like.
The dip in energy in the center of the kink is however only slightly
less than the maximum value. 

\begin{table}[!th]
\begin{center}
\caption{Lengths of halfkinks in various models. }
\label{tab:halfkinklengths}
\begin{tabular}{l||r@{.}lr@{.}lr@{.}lr@{.}l}
model & \multicolumn{2}{c}{$z_B^{\rm numerical}$}
& \multicolumn{2}{c}{$z_{E^{\rm kink}}^{\rm numerical}$}
& \multicolumn{2}{c}{$z_{E^{\rm kink}}^{\rm eff}$}
& \multicolumn{2}{c}{$1.570z_{E^{\rm kink}}^{\rm eff}$}\\
\hline\hline
2   & 1&475 &  1&424   & $\pi/\sqrt{12}\simeq$ 0&9069 & 1&424\\
2+4 & 3&365 &  3&080   & 2&128                      & 3&358\\
2+6 & 22&48 &  20&14 & 15&04                      & 23&62
\end{tabular}
\end{center}
\end{table}

In this subsection, we have numerically calculated the solutions of
the halfkinks in three different models. The lengths of these
halfkinks are given in Tab.~\ref{tab:halfkinklengths}, where we have
defined the kink length as $z_B$ in Eq.~\eqref{eq:xiBsq}.
This definition uses the baryon charge density as a measure, which is
argued in the recent paper \cite{Adam:2015zhc} to be a natural
definition. 
We will alternatively perform the same calculation of the halfkink
lengths using the kink energy density as a measure
\beq
z_{E^{\rm kink}}^2 \equiv \frac{\int d^3x \; z^2 \mathcal{E}^{\rm
kink}}{\int d^3x \; \mathcal{E}^{\rm kink}},
\eeq
where the kink energy density is defined as the total energy density
with that of the vortex subtracted off
\beq
\mathcal{E}^{\rm kink}\equiv \mathcal{E} - \mathcal{E}^{\rm vortex}.
\label{eq:Ekink}
\eeq
Note that the integral of the kink energy density is convergent.

In the next section, we will, as an example, calculate the lengths of
the halfkinks using an effective field theory approach.

\subsubsection{Effective theory on the straight vortex}

Now we will consider the effective theory on vortex string, following
Ref.~\cite{Gudnason:2014gla}. The idea is simple, namely integrating 
out the vortex to obtain the leading-order effective theory of the 
U(1) modulus
\begin{align}
-\mathcal{L}^{\rm eff,kink} = \left[\frac{a_{2,0,0}}{m_{\rm vortex}^2}
    + c_4(a_{2,2,0} + a_{2,0,2}) 
    + 2c_6 a_{2,2,2} m_{\rm vortex}^2\right] (\p_\alpha\chi)^2 
  + \frac{a_{2,0,0} m_2^2}{m_{\rm vortex}^2} \sin^2\chi,
  \label{eq:Leffkink}
\end{align}
where $m_{\rm vortex}$ is the mass scale of the vortex defined in
Eq.~\eqref{eq:mvortex}, $\alpha=t,z$ and the dimensionless
coefficients in the effective Lagrangian density read 
\begin{align}
a_{{\sf k},\ell,{\sf m}}
\equiv \pi m_{\rm vortex}^{2-\ell-{\sf m}} \int dr \; r^{1-\ell} \cos^{\sf k} f
  \sin^\ell f (f_r)^{\sf m}.
\label{eq:vortex_coefficients}
\end{align}
Since we cannot solve the vortex profile analytically, the above
integrals, $a$, have to be evaluated numerically. As they are unitless
numbers, they do not depend on the vortex mass scale when $c_4$ is
turned off ($c_4=0$). However, when $c_4>0$, they depend on the
combination $c_4m_{\rm vortex}^2$.
The numerically evaluated integrals are shown in
Tab.~\ref{tab:aintegrals}. 
Note that all the integrals in Tab.~\ref{tab:aintegrals} are
convergent. As we observed in the last section, basically the only
divergent integral (for positive values of the indices) is
$a_{0,2,0}$ whose integrant goes only like $1/r$. Once a (positive)
factor of $\cos f$ or $f_r$ is turned on, there is an exponential
falloff of the integrant and hence the integral converges. 

\begin{table}[!htp]
\begin{center}
\caption{Numerically evaluated integrals, $a$, defined in
Eq.~\eqref{eq:vortex_coefficients}. }
\label{tab:aintegrals}
\begin{tabular}{c||r@{.}lr@{.}lr@{.}lr@{.}l}
$c_4 m_{\rm vortex}^2$ & \multicolumn{2}{c}{0} & \multicolumn{2}{c}{3}
& \multicolumn{2}{c}{8}
& \multicolumn{2}{c}{15}\\ 
\hline\hline
$a_{2,0,0}$ & 1&5708  & 3&6391  & 5&3157  & 6&9352\\
$a_{2,2,0}$ & 1&2178  & 1&1397  & 1&1232  & 1&1150\\
$a_{2,0,2}$ & 0&92328 & 0&97555 & 0&98732 & 0&99329\\
$a_{2,2,2}$ & 0&95707 & 0&36546 & 0&24404 & 0&18479
\end{tabular}
\end{center}
\end{table}

Using the effective theory approach, we can calculate the halfkink
analytically from the effective Lagrangian \eqref{eq:Leffkink},
\beq
\chi = -2\arctan e^{-m_{\rm eff} z},
\label{eq:halfkink_sol}
\eeq
where the effective mass is given by
\beq
m_{\rm eff}^2 =
\frac{a_{2,0,0}m_2^2}
{a_{2,0,0} + c_4 m_{\rm vortex}^2 (a_{2,2,0} + a_{2,0,2}) + 2c_6
m_{\rm vortex}^4 a_{2,2,2}}.
\label{eq:meff}
\eeq
Note that for $c_4=c_6=0$, the effective kink mass simplifies to
$m_{\rm eff} = m_2$.
The solution satisfies the boundary conditions
\beq
\chi(-\infty) = -\pi, \qquad
\chi(\infty) = 0.
\eeq
As all the dimensionless coefficients, $a$, are positive definite, the 
effect of turning on $c_4$ or $c_6$ is a decrease in the effective
kink mass and hence a prolongation of the kink length. 
Only the kink mass term has the effect of contraction of the kink and
hence a shortening of the kink length. This is exactly what one would
expect in a $(1+1)$-dimensional effective theory.

Now with the exact kink solution \eqref{eq:halfkink_sol} and the
effective mass \eqref{eq:meff} at hand, we can calculate the kink
length in the effective field theory framework.
The energy density is
\beq
\mathcal{E}^{\rm kink} = \frac{2m_2^2 a_{2,0,0}}{m_{\rm vortex}^2}
\sech^2(m_{\rm eff} z),
\eeq
from which we can directly calculate
\beq
z_E^2 = \frac{\int dz\; z^2 \sech^2(m_{\rm eff} z)}
{\int dz\; \sech^2(m_{\rm eff} z)}
= \frac{\pi^2}{12 m_{\rm eff}^2},
\eeq
giving the length $L\sim \pi/(2\sqrt{3}m_{\rm eff})$.
Using this result, we calculate the theoretical result for the kink
lengths of the halfkinks presented in Sec.~\ref{sec:singlehalfkinks};
the result is shown as the second last column in
Tab.~\ref{tab:halfkinklengths}. 
The last column in the table is the same calculation, but with a
rescaled prefactor matched to fit the numerical integration.
We can see that the lengths calculated within the effective field
theory framework matches the full PDE calculation qualitatively, but
quantitatively only within about the 60\% level. 

We should warn the reader -- as explained in
Ref.~\cite{Gudnason:2014gla} -- that this leading order effective
field theory calculation neglects the backreaction of the kink to
the vortex background.

\subsection{Vortex rings with halfkinks as Skyrmions}

In this section we will consider the vortex rings of
Sec.~\ref{sec:vortexrings}, but in the presence of the quadratic kink
potential.

\subsubsection{Singly twisted vortex rings}

We start with the single vortex ring, which is the vortex string
twisted once, yielding baryon number $B=1$. 
Since single the vortex ring is naturally twisted by $2\pi$ (in order
to close it), the quadratic kink potential will induce two halfkinks
on the vortex ring.
Since the halfkinks are repulsive to one another, they will naturally
reside on opposite sides of the ring.
This incarnation of the Skyrmion is the final possibility and has not
been considered before in the literature. 

Now this configuration is quite nontrivial, because the single vortex
ring does not generally exist; only in the case of the 2+6 model and
for large enough vortex potential mass $m$. 
We will therefore only consider the 2+6 model with $m=7$ and then
turn on the quadratic kink potential $m_2>0$. 

\begin{figure}[!tp]
\begin{center}
\mbox{
\subfloat[]{\includegraphics[width=0.24\linewidth]{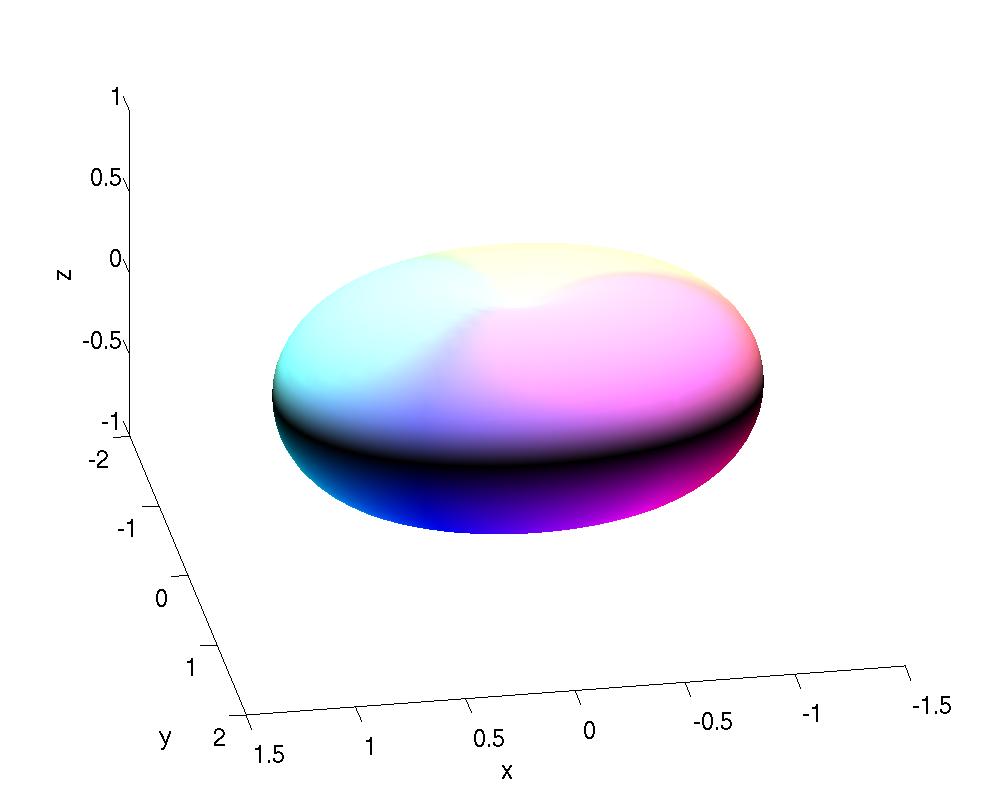}}
\subfloat[]{\includegraphics[width=0.24\linewidth]{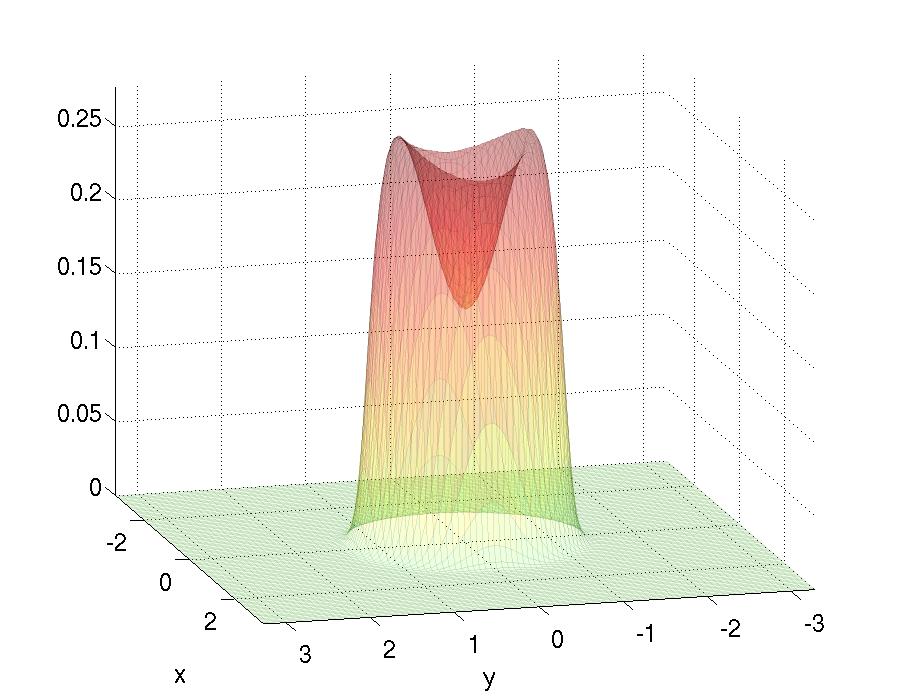}}
\subfloat[]{\includegraphics[width=0.24\linewidth]{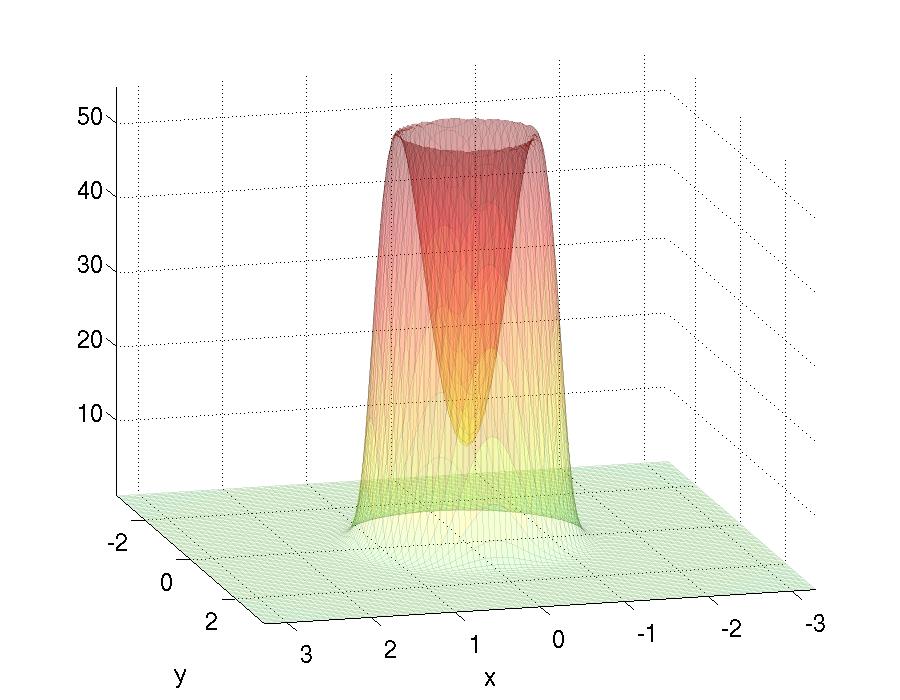}}
\subfloat[]{\includegraphics[width=0.24\linewidth]{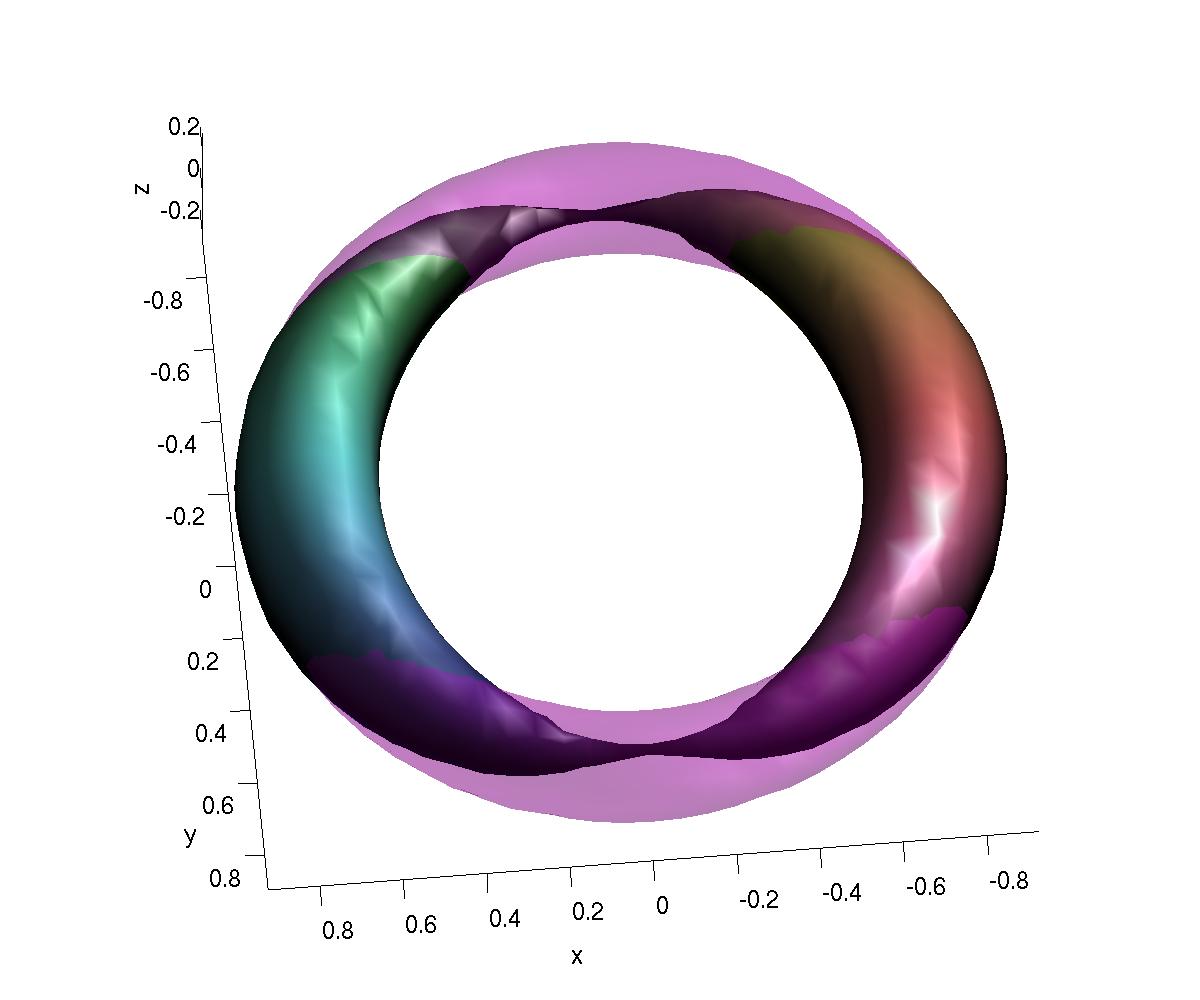}}}
\caption{A vortex ring with two halfkinks on its worldsheet in the 2+6
model; (a) baryon charge isosurface at half-maximum value of the
baryon charge density (the energy density is slightly smaller and
quite similar in shape and hence not shown here), (b) $xy$-slice of
baryon charge density at $z=0$, (c) $xy$-slice of energy density at
$z=0$, (d) isosurfaces of baryon charge density (colored) and energy
density (pink) both at 92\% of the maximum value of their respective
densities.
The baryon charge is calculated to be $B^{\rm numerical}=0.99986$.
For coloring of the baryon charge isosurfaces, see the text.
In this figure $m=7$ and $m_2=3$. }
\label{fig:rhk121_0_1_m7_m2_3}
\end{center}
\end{figure}

In Fig.~\ref{fig:rhk121_0_1_m7_m2_3} is shown the vortex ring in the
2+6 model for $m=7$ and with the quadratic kink potential turned on:
$m_2=3$.
At the half-maximum level isosurfaces, the ring looks basically
unperturbed by the quadratic kink potential.
However, at a closer look, one can see from the $xy$-slice of the
baryon charge density at $z=0$ (Fig.~\ref{fig:rhk121_0_1_m7_m2_3}b)
that the maximum of the baryon charge density now does not form a
ring, but has two maxima; these represent the centers of the two
halfkinks.
It is quite interesting to compare this figure to the $xy$-slice of
the energy density at $z=0$ (Fig.~\ref{fig:rhk121_0_1_m7_m2_3}c),
which remains unperturbed by the presence of the quadratic kink
potential; the maximum energy density retains the form of a ring. 
Finally, in Fig.~\ref{fig:rhk121_0_1_m7_m2_3}d is shown the
isosurfaces at 92\% of the maximum values of the baryon charge density
(colored) and energy density (pink), respectively.
Also this figure shows that the energy density retains the form of a
torus, whereas the baryon charge density oscillates from a larger
value -- which we interpret as the center of a halfkink -- to a
smaller value.\footnote{The shape of the configuration -- like beads
on a ring -- looks on the surface quite similar to a different model,
where we have constructed half-Skyrmions \cite{Gudnason:2015nxa}. The
similarity between the two configurations is, however, only visual. }
The halfkinks looking like beads on a ring are half-Skyrmions.

One may consider the effective field theory approach also in this case
in order to describe the halfkinks.
However, the fact that the energy density remains unperturbed is a
clear indication that there is a strong backreaction from the kink
onto the host vortex ring (Skyrmion).

\subsubsection{Doubly twisted vortex rings}

We now turn to the single vortex ring, twisted twice, yielding baryon
number $B=2$.
As mentioned in Sec.~\ref{sec:B2rings}, this is the normal Skyrmion of
charge $B=2$; here however we turn on both the vortex potential as
well as the kink potential \eqref{eq:Vtotal}.
Since the $B=2$ vortex ring is twisted twice along the $\phi$
direction and the kink potential induces halfkinks, we expect the
resulting vortex ring to host four halfkinks and again evenly
distributed around the ring. 
In some sense this vortex ring is simpler than the $B=1$ vortex ring,
because it already takes the shape of torus without the addition of
the vortex potential \eqref{eq:Vvortex}.
However, in order to ensure the correct symmetry breaking -- as
discussed in Sec.~\ref{sec:model} -- we turn on the vortex potential
with a larger coefficient than that of the kink potential: i.e.~we
take $m>m_2$.
For the sake of comparison with the singly twisted vortex ring kinks,
we choose to use the same values of the potential parameters, namely
$m=7$ and $m_2=3$; but other parameters will also yield vortex rings
with four kinks for this $B=2$ case. 

\begin{figure}[!tp]
\begin{center}
\mbox{
\subfloat[]{\includegraphics[width=0.24\linewidth]{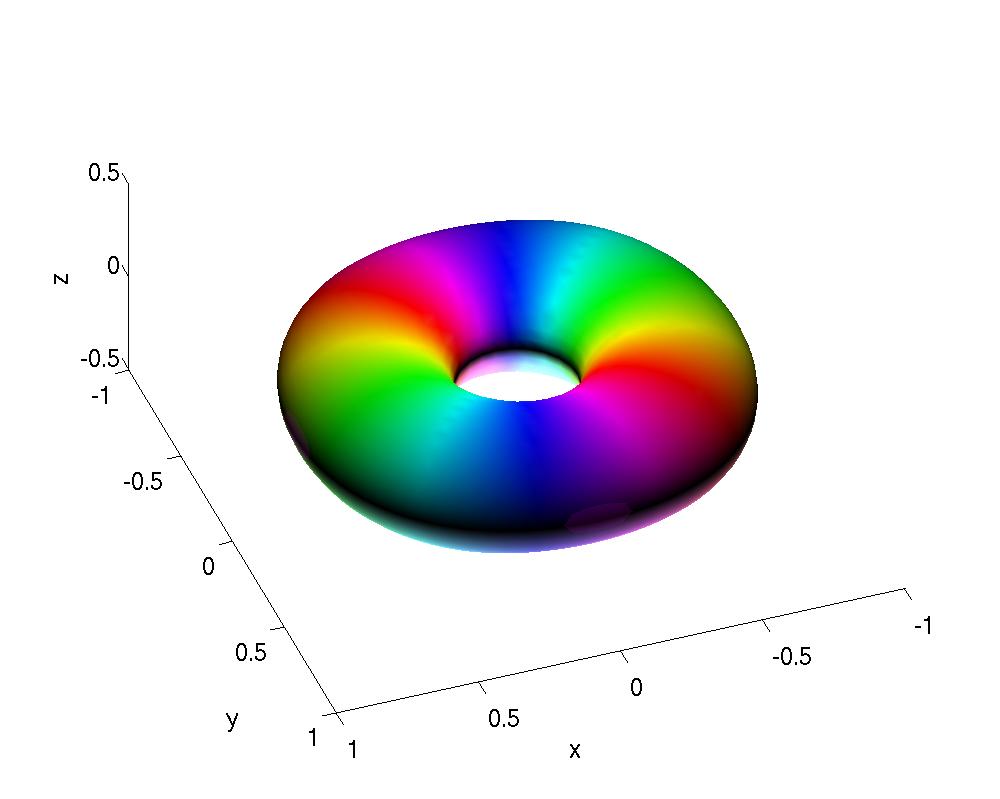}}
\subfloat[]{\includegraphics[width=0.24\linewidth]{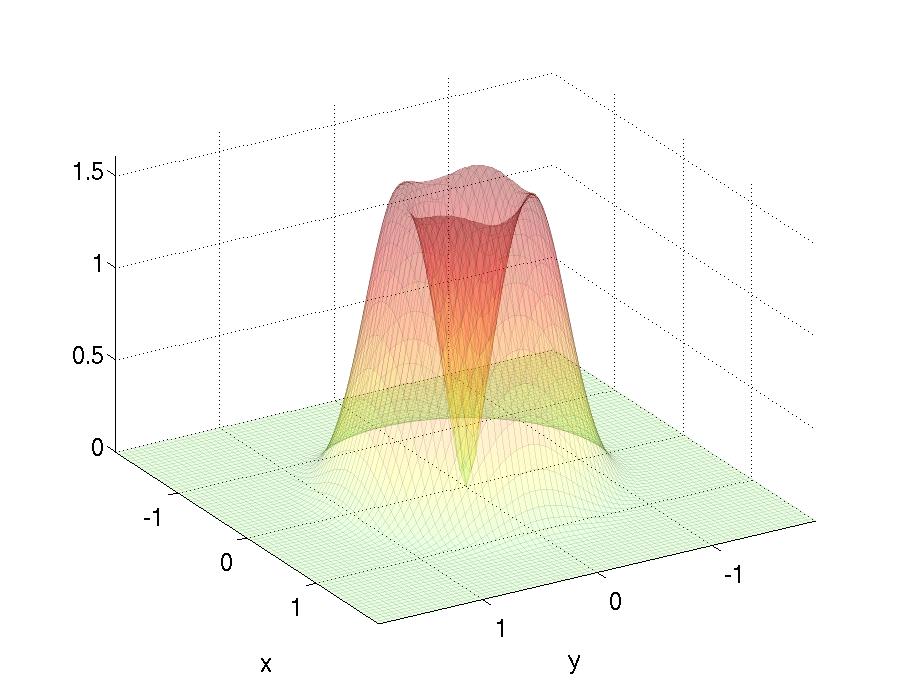}}
\subfloat[]{\includegraphics[width=0.24\linewidth]{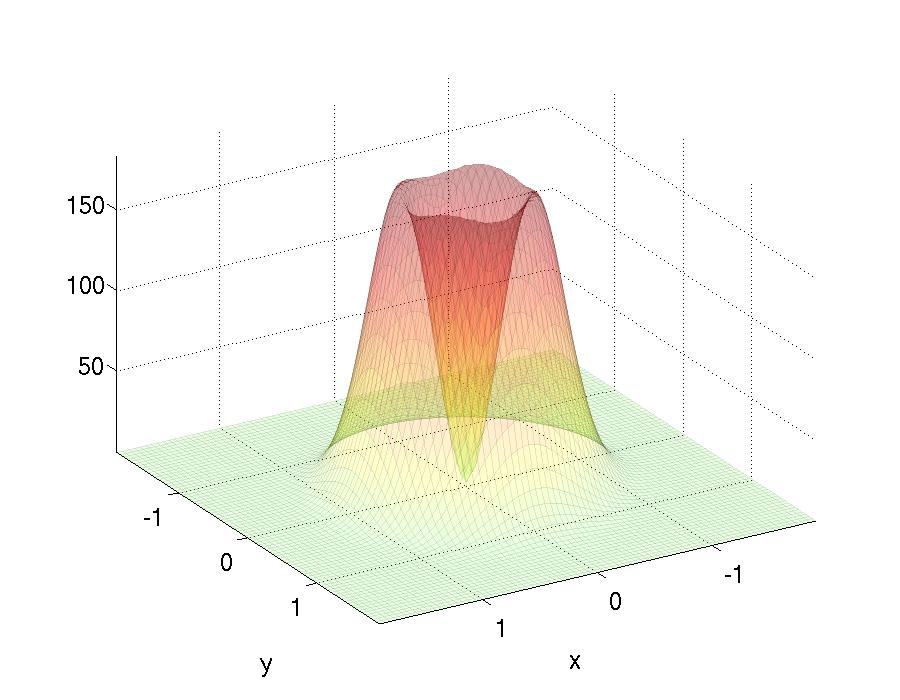}}
\subfloat[]{\includegraphics[width=0.24\linewidth]{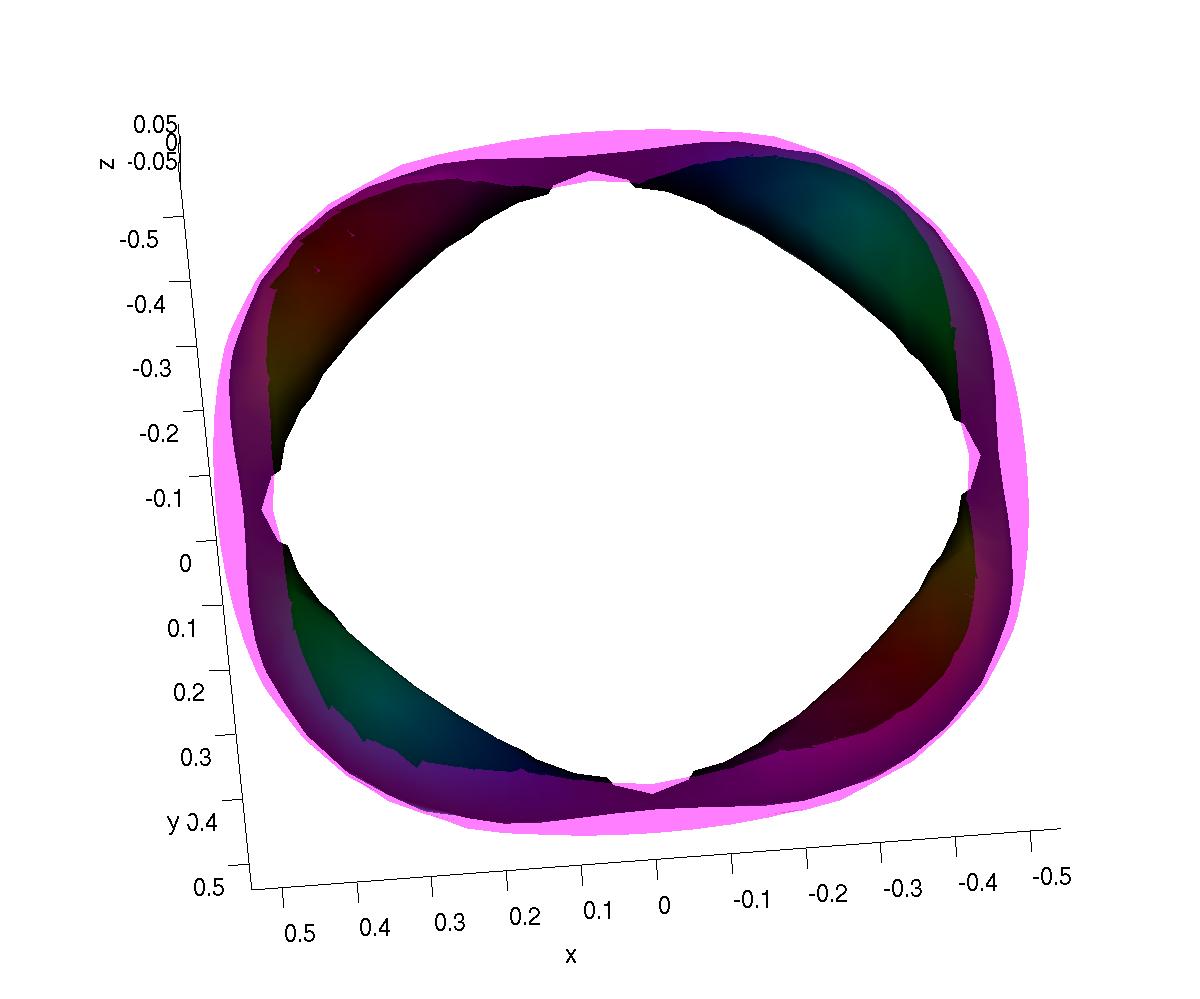}}}
\caption{A doubly twisted vortex ring with four halfkinks on its
worldsheet in the 2+4 model; (a) baryon charge isosurface at
half-maximum value of the baryon charge density (the energy density is
slightly smaller and quite similar in shape and hence not shown here),
(b) $xy$-slice of baryon charge density at $z=0$, (c) $xy$-slice of
energy density at $z=0$, (d) isosurfaces of baryon charge density
(colored) and energy density (pink) both at 95.5\% of the maximum
value of their respective densities.
The baryon charge is calculated to be $B^{\rm numerical}=1.9997$.
For coloring of the baryon charge isosurfaces, see the text.
In this figure $m=7$ and $m_2=3$. }
\label{fig:8121_1_0_m7_m2_3}
\end{center}
\end{figure}

\begin{figure}[!tp]
\begin{center}
\mbox{
\subfloat[]{\includegraphics[width=0.24\linewidth]{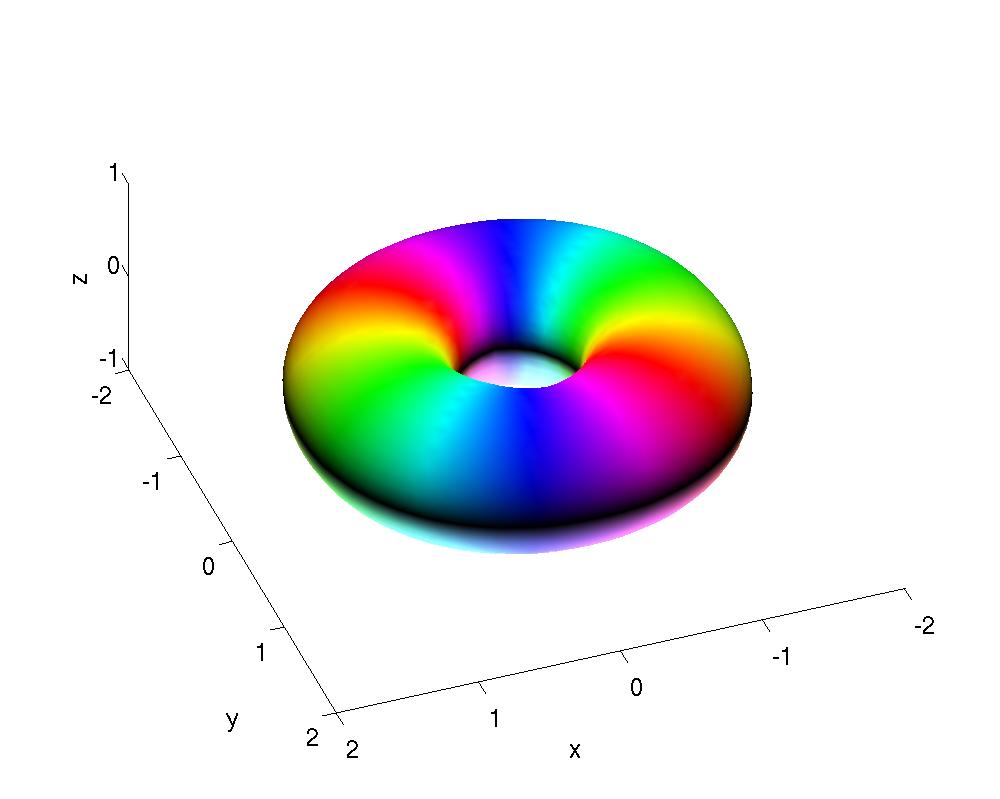}}
\subfloat[]{\includegraphics[width=0.24\linewidth]{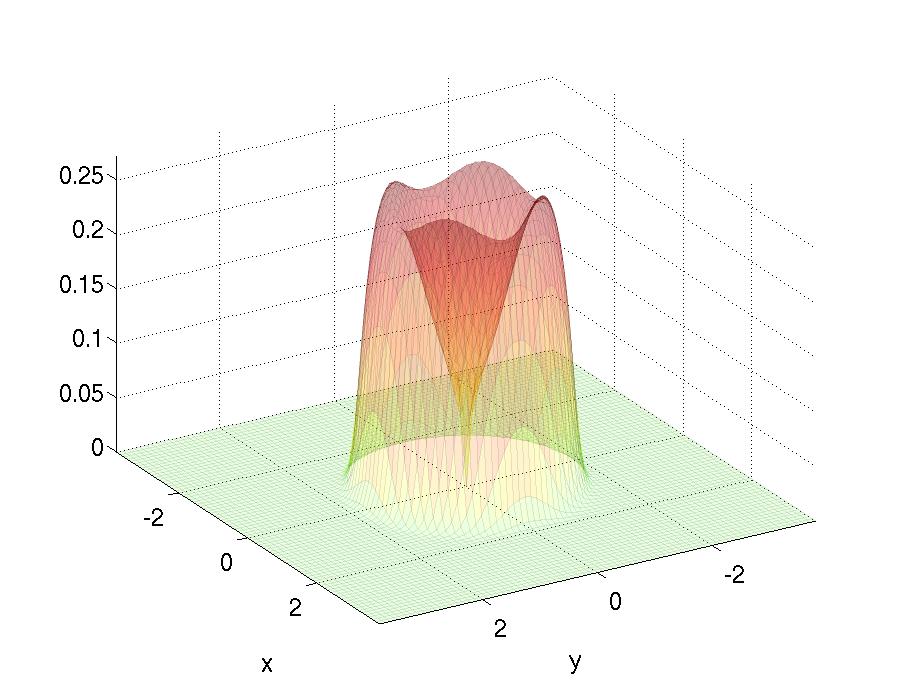}}
\subfloat[]{\includegraphics[width=0.24\linewidth]{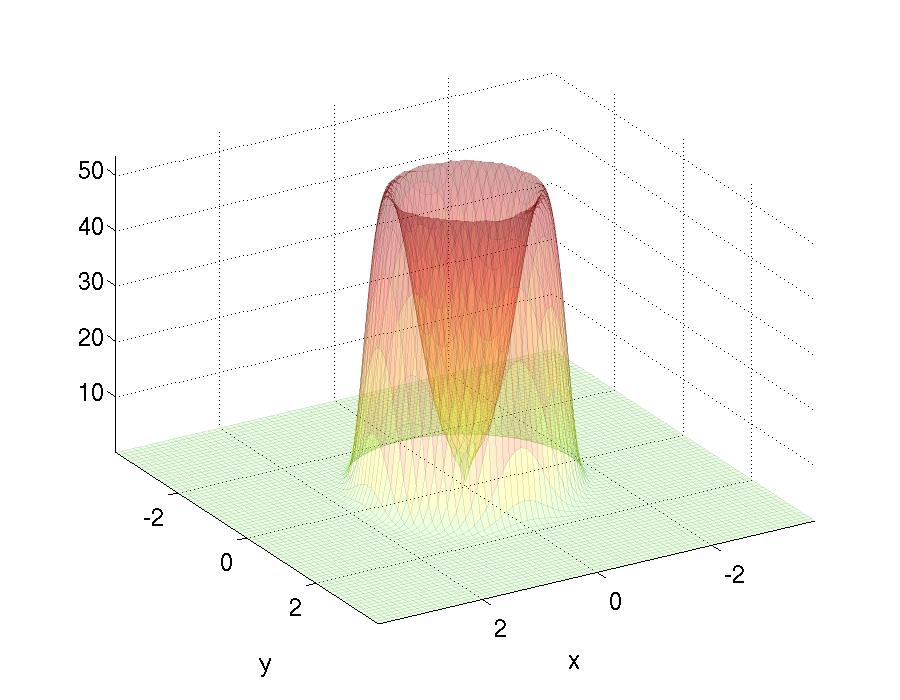}}
\subfloat[]{\includegraphics[width=0.24\linewidth]{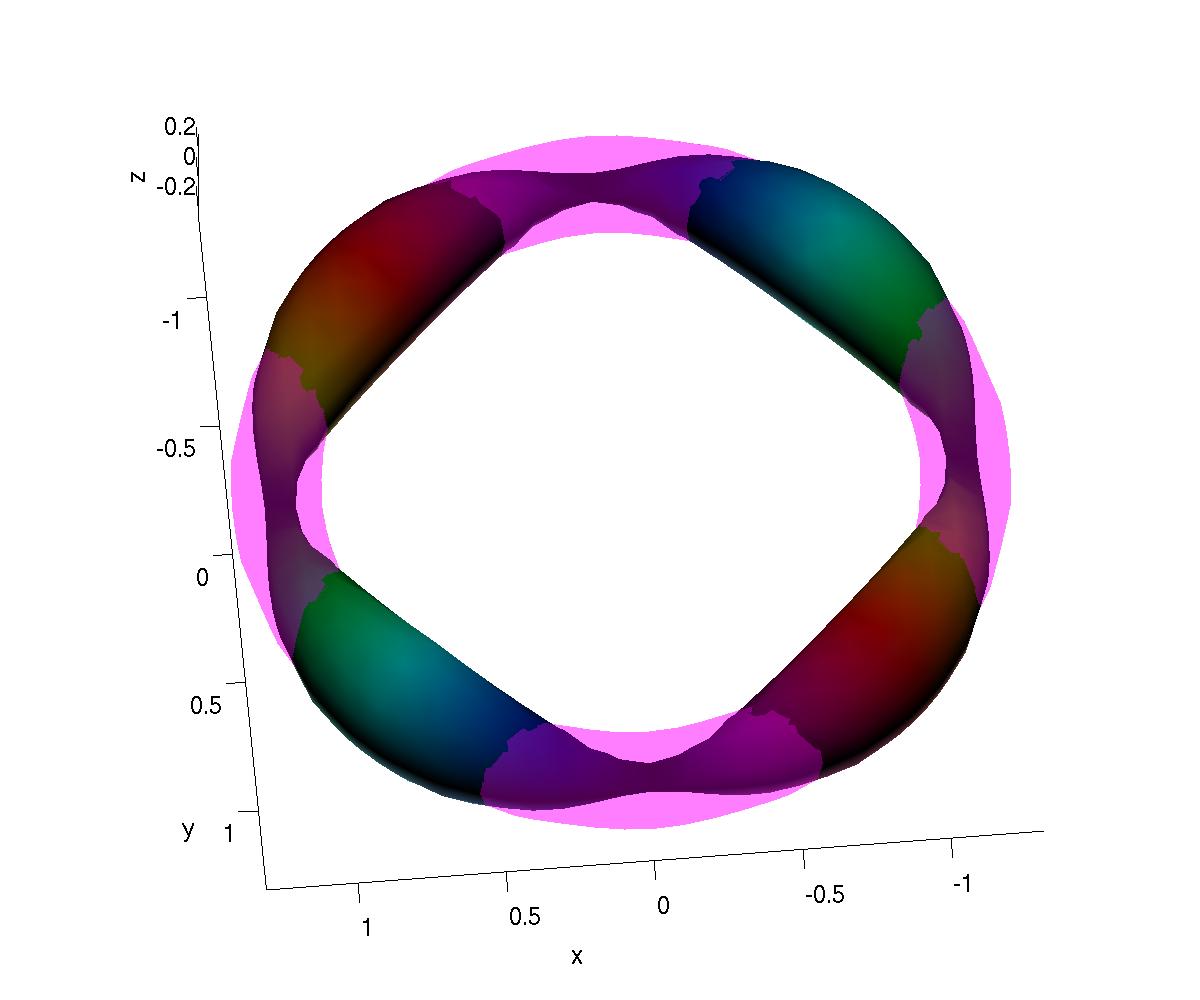}}}
\caption{A doubly twisted vortex ring with four halfkinks on its
worldsheet in the 2+6 model; (a) baryon charge isosurface at
half-maximum value of the baryon charge density (the energy density is
slightly smaller and quite similar in shape and hence not shown here),
(b) $xy$-slice of baryon charge density at $z=0$, (c) $xy$-slice of
energy density at $z=0$, (d) isosurfaces of baryon charge density
(colored) and energy density (pink) both at 91.5\% of the maximum value
of their respective densities.
The baryon charge is calculated to be $B^{\rm numerical}=1.9991$.
For coloring of the baryon charge isosurfaces, see the text.
In this figure $m=7$ and $m_2=3$. }
\label{fig:8121_0_1_m7_m2_3}
\end{center}
\end{figure}

In Fig.~\ref{fig:8121_1_0_m7_m2_3} and \ref{fig:8121_0_1_m7_m2_3} we
show doubly twisted vortex rings with four halfkinks on their
worldsheets in the 2+4 and 2+6 models, respectively.
As in the case of the singly twisted vortex rings, the kinks are
not quite visible at the isosurfaces at the half-maximum baryon charge
density level.
From the $xy$-slices (panel (b) of the figures) we can see that the
kink oscillations are at the 5-10\% level of the baryon charge
density.
The $xy$-slices of the energy density at $z=0$ show an interesting
difference between the 2+4 and the 2+6 model, namely the kinks are --
as in the singly twisted vortex ring case -- not visible from the
energy densities in the 2+6 model, whereas in the 2+4 model they are. 
Finally, we fix the isosurface level-set to match the oscillation
amplitude of the baryon charge density due to the kinks in panels
(d). We can confirm that in the 2+4 model the energy density
oscillates -- and hence shows the presence of the kinks -- whereas in
the 2+6 model, it does not.
We also see that for the same values of the potential parameters, $m$
and $m_2$, the oscillation in the baryon charge density in the 2+6
model is almost twice as large in amplitude as it is in the 2+4
model. 

In the next subsubsection we will try to compare these oscillations of
the baryon charge densities due to the kinks in the various cases in
the framework of the effective theory on the vortex ring.

\subsubsection{Effective theory on the vortex ring}

We will again calculate the effective theory for the ``modulus''
living on the vortex and for the first time present the effective
theory for the field on the torus. This case was mentioned in
Ref.~\cite{Gudnason:2014gla} but not derived.
We consider the same type of Ansatz as given in
Eq.~\eqref{eq:vortexring_ansatz}, but substituting the azimuthal
coordinate with a field $\Phi(\phi)$:
\beq
\phi_{\rm initial}^{\rm T} =
\left(\cos f - i\sin f\cos \theta,e^{i\Phi}\sin f\sin\theta\right),
\eeq
and $f$ is now promoted to be a function of the polar angle as well as
the spherical radius: $f=f(r,\theta)$. 
After the dust has settled, we can write
\begin{align}
-\mathcal{L}^{\rm eff,kink} &=
\bigg[
\frac{a_{2,0,0,0}}{m_{\rm vortex}}
+ c_4 m_{\rm vortex} \left(a_{4,0,0,0} + a_{2,2,0,0} + a_{2,0,2,0}\right)
+2c_6\, a_{4,2,0,0}m_{\rm vortex}^3 
\bigg](\p_\alpha\Phi)^2 \non
&\phantom{=\ }
+ \frac{m_2^2 \, a_{0,0,0,2}}{m_{\rm vortex}^3} \sin^2\Phi,
\label{eq:vortexring_efftheory}
\end{align}
where we have defined the coefficients
\beq
a_{{\sf k},\ell,{\sf m},{\sf n}} \equiv \frac{1}{2}
m_{\rm vortex}^{3-{\sf k}-\ell-{\sf m}} \int dr\,d\theta
\frac{\sin^{1+{\sf n}}\theta}{r^{{\sf k}+{\sf m}-2}}
\sin^{{\sf k}+{\sf n}}(f)f_r^{\ell}
f_\theta^{\sf m}.
\label{eq:vortexring_coefficients}
\eeq
Although the exact solutions to the equation of motion derived from
the effective theory \eqref{eq:vortexring_efftheory}
\beq
\Phi_{\phi\phi} - m_{\rm eff}^2 \sin 2\Phi = 0,
\label{eq:vortexring_efftheory_eom}
\eeq
are known in
terms of the Jacobi amplitude related to elliptic integrals,
adjusting the parameters in order for the solution to match the
boundary conditions
\beq
\Phi(0) = 0, \qquad
\Phi(2\pi) = 2\pi B,
\eeq
is somewhat intricate. 
We therefore choose to consider only numerical solutions to
Eq.~\eqref{eq:vortexring_efftheory_eom}. 
The effective mass (squared) in
Eq.~\eqref{eq:vortexring_efftheory_eom} is given by 
\beq
m_{\rm eff}^2 \equiv \left(\frac{m_2^2}{2m_{\rm vortex}^2}\right)
\frac{a_{0,0,0,2}}
{a_{2,0,0,0}
+ c_4 m_{\rm vortex}^2\left(a_{4,0,0,0} + a_{2,2,0,0} +
a_{2,0,2,0}\right)
+ 2c_6m_{\rm vortex}^4 a_{4,2,0,0}}.
\eeq

Instead of measuring the kink length, in this case we will consider
the baryon charge density in the effective field theory framework; in
particular, we see from the last section that the kink potential
induces an oscillation in the baryon charge density as we go around
the azimuthal angle.
In terms of the effective theory, we get
\beq
\mathcal{B} = -\frac{1}{\pi^2} a_{2,1,0,0} \Phi_\phi,
\eeq
from which the total baryon charge is calculated as
$B=\int d\phi\;\mathcal{B}$. 
We can easily confirm the qualitative behavior by inspecting the
effective theory \eqref{eq:vortexring_efftheory} and using the fact
that the baryon charge density is proportional to the azimuthal
derivative of $\Phi$; namely that there will be $2B$ maxima of the
baryon charge density on the vortex ring.

As a more quantitative comparison, let us calculate the relative
difference between the maximal and minimal baryon charge density on
the torus.
For this we need to evaluate the effective field theory
coefficients \eqref{eq:vortexring_coefficients}.
We use the numerical solution in Sec.~\ref{sec:vortexrings} as basis
for the evaluation of said coefficients and the results are given in
Tab.~\ref{tab:ringcoef}.

\begin{table}[!htp]
\begin{center}
\caption{Numerically evaluated coefficients, $a$, for the effective
theory. }
\label{tab:ringcoef}
\begin{tabular}{llllll}
\multicolumn{6}{c}{singly twisted vortex ring in the 2+6 model}\\
\hline\hline
$a_{2,0,0,0}$ & $a_{4,0,0,0}$ & $a_{2,2,0,0}$ & $a_{2,0,2,0}$ &
$a_{4,2,0,0}$ & $a_{0,0,0,2}$ \\
12.47 & -- & -- & -- & 0.03178 & 193.4  
\end{tabular}
\begin{tabular}{llllll}
\multicolumn{6}{c}{doubly twisted vortex ring in the 2+4 model}\\
\hline\hline
$a_{2,0,0,0}$ & $a_{4,0,0,0}$ & $a_{2,2,0,0}$ & $a_{2,0,2,0}$ &
$a_{4,2,0,0}$ & $a_{0,0,0,2}$ \\
14.86 & 0.8817 & 1.715 & 0.04187 & 0.1163 & 144.0 
\end{tabular}
\begin{tabular}{llllll}
\multicolumn{6}{c}{doubly twisted vortex ring in the 2+6 model}\\
\hline\hline
$a_{2,0,0,0}$ & $a_{4,0,0,0}$ & $a_{2,2,0,0}$ & $a_{2,0,2,0}$ &
$a_{4,2,0,0}$ & $a_{0,0,0,2}$ \\
12.79 & -- & -- & -- & 0.01319 & 326.4 
\end{tabular}
\end{center}
\end{table}

Calculating the oscillation amplitude numerically within the effective
theory, we can estimate the lowest value of the baryon charge density
on the ring; for simplicity we normalize the maximum baryon charge
density to unity. The result is shown in Tab.~\ref{tab:Bamps}.
We can see that the effective theory estimate is quite good in the
case of the 2+4 model, whereas in the 2+6 model the effective theory
overestimates the oscillation amplitudes by more than a factor of
two. 

\begin{table}[!htp]
\begin{center}
\caption{Oscillation amplitudes due to halfkinks on the worldsheet of
vortex rings measured in the baryon charge density; the PDE
calculations are compared to the effective field theory predictions. }
\label{tab:Bamps}
\begin{tabular}{l||ll}
\multicolumn{3}{c}{singly twisted vortex ring in the 2+6 model}\\
\hline\hline
$m_{\rm eff}$ &
$(\mathcal{B}_{\rm min}/\mathcal{B}_{\rm max})^{\rm PDE}$ &
$(\mathcal{B}_{\rm min}/\mathcal{B}_{\rm max})^{\rm EFT}$\\
0.560 & 0.920 & 0.732
\end{tabular}
\begin{tabular}{l||ll}
\multicolumn{3}{c}{doubly twisted vortex ring in the 2+4 model}\\
\hline\hline
$m_{\rm eff}$ &
$(\mathcal{B}_{\rm min}/\mathcal{B}_{\rm max})^{\rm PDE}$ &
$(\mathcal{B}_{\rm min}/\mathcal{B}_{\rm max})^{\rm EFT}$\\
0.367 & 0.955 & 0.967
\end{tabular}
\begin{tabular}{l||ll}
\multicolumn{3}{c}{doubly twisted vortex ring in the 2+6 model}\\
\hline\hline
$m_{\rm eff}$ &
$(\mathcal{B}_{\rm min}/\mathcal{B}_{\rm max})^{\rm PDE}$ &
$(\mathcal{B}_{\rm min}/\mathcal{B}_{\rm max})^{\rm EFT}$\\
0.862 & 0.915 & 0.830
\end{tabular}
\end{center}
\end{table}

Let us warn the reader that in the true calculation, the profile
function $f$ is not an axially symmetric function in the presence of
the kink. In principle all fields become functions of
$(r,\theta,\phi)$.
In the full PDE calculations, there is no Ansatz and so all fields
depend on all coordinates.
The effective theory is simply an approximation, valid when
$m_2\ll m$; however for practical calculations, we can barely observe
the oscillation in the PDE results if we take such small kink masses.
Nevertheless, even though we use a sizable kink mass, the effective
theory managed to reproduce the quantitative oscillation in the 2+4
model within the one-percent level.

Finally, and most important, the effective field theory captures the
qualitative behavior of the system quite well, although the
quantitative comparisons are not all very precise at this level of
leading order effective theory.

\section{Discussion and conclusion}\label{sec:discussion}

In this paper we have demonstrated that vortex strings can be
constructed in a simpler potential than that of BEC type used
previously. We constructed both straight vortices as well as vortex
rings with baryon numbers one and two. All these vortices were then
embedded with sine-Gordon-type halfkinks, both numerically and in the
effective field theory approach.
We find that only the 2+6 model has a ring-like structure for at
single vortex twisted once when the vortex potential is turned on,
whereas both the 2+4 model and the 2+6 model have ring-like structure
for doubly twisted vortices ($B=2$). 
An interesting observation is that in the 2+6 model, the energy
density of the vortex rings for both baryon number one and two
($B=1,2$) does not reveal the kink structure living on its worldline,
whereas the baryon charge density does show an oscillation due  
to said halfkinks.
However, the 2+4 model does not have this feature in common with the
2+6 model so both the energy density and the baryon charge density
oscillates on the vortex rings when kinks are embedded in the 2+4
model. 
Our effective theory approach to leading order has been compared to
full PDE calculations and at the qualitative level it works quite
well; however, at the quantitative level, it can predict numerical
quantities only within a factor of less than two.
There are two reasons for this, first we made some rough assumptions
about the dependencies of the fields when constructing the effective
theory and second we did not consider any backreaction from the kinks
onto its host soliton.

One future development could be to consider more precise effective
field theories on solitons, in particular taking backreaction and
massive modes into account.
An interesting question in this direction concerns the
(de-)stabilization at next-to-leading order of lumps on domain
walls \cite{Eto:2015uqa,Eto:2015wsa}.

In Refs.~\cite{Gudnason:2014nba,Gudnason:2014hsa} we considered the
linear kink potential $-\Re\phi_2$ on equal footing with the quadratic
kink potential $-(\Re\phi_2)^2$, considered in this paper.
In the latter references the linear kink potential was treated as a
small perturbation; however, mathematically, the linear kink potential
introduces a tiny shift in the vacuum, technically complicating the
topological structure of the vortex solution; in particular in
conjunction with the interpretation of the Skyrmion being absorbed
into the vortex string.
Although we have not studied the exact details, we expect the exact
solution to have two strings, one big and one very tiny, in the
situation where a unit-charge Skyrmion is absorbed into the string and
this effect is expected to break the axial symmetry of both vortices. 
Of course, in the approximation of a small perturbation ($m_2\ll m$),
the effect is so small that it is practically unobservable.
However, if one considers a sizable coefficient of the linear kink
potential, $m_2\lesssim m$, then this effect should become visible.
We will leave this interesting effect for future studies.

By embedding $U$ into an SL(2,$\mathbb{C}$) matrix $M$, one can
construct a supersymmetric Skyrme model \cite{Gudnason:2015ryh}. 
The bosonic part of the action contains the Skyrme term but no kinetic 
term as is the case of the supersymmetric BPS baby Skyrme
model \cite{Adam:2011hj,Adam:2013awa,Nitta:2014pwa}.
The potential $V_{\rm vortex}\propto\Tr[\sigma_3,U]^2$ considered in
this paper can be obtained by a twisted dimensional reduction along
one compactified spatial direction, implying that such a potential
term can be made supersymmetric. 
In such a theory, vortices may be BPS, preserving some fraction of
supersymmetry, which is possibly of a compacton type like BPS baby
Skyrmions preserving 1/4 of
supersymmetry \cite{Nitta:2014pwa,Nitta:2015uba}.

Recently, some progress has been achieved in deepening our
understanding of black hole Skyrme hair. In particular,
Refs.~\cite{Adam:2016vzf,Gudnason:2016kuu} show that the Skyrme term
is a necessity for having stable black hole scalar hair, whereas the
sixth-order derivative term is unable to stabilize the system; in our
terminology the 2+4 model allows for stable black hole hair, but the
2+6 model does not.
Axially symmetric black hole Skyrme hair configurations have been
studied in Refs.~\cite{Sawado:2003at,Sawado:2004yq}.
It would be interesting to study whether the vortex rings constructed
in this paper could be black hole hair with a fractional vortex
charge. This would have to be in the 2+4 or 2+4+6 model.
We will leave this problem for future studies.

\subsection*{Acknowledgments}

We thank Masayasu Harada and Bing-Ran He for useful discussions. 
S.~B.~G.~thanks the Recruitment Program of High-end Foreign
Experts for support.
The work of M.~N.~is supported in part by a Grant-in-Aid for
Scientific Research on Innovative Areas ``Topological Materials
Science'' (KAKENHI Grant No.~15H05855) and ``Nuclear Matter in Neutron
Stars Investigated by Experiments and Astronomical Observations''
(KAKENHI Grant No.~15H00841) from the the Ministry of Education,
Culture, Sports, Science (MEXT) of Japan. The work of M.~N.~is also
supported in part by the Japan Society for the Promotion of Science
(JSPS) Grant-in-Aid for Scientific Research (KAKENHI Grant
No.~16H03984) and by the MEXT-Supported Program for the Strategic
Research Foundation at Private Universities ``Topological Science''
(Grant No.~S1511006).

\appendix
\section{String splitting for higher vortex numbers}\label{app:stringsplitting}

In this appendix we consider a charge-two vortex ($n=2$) curled up
once to be a vortex ring of baryon charge $B=2$. This is in contrast
to the single vortex ($n=1$) twisted twice to give a \emph{stable} 
$B=2$ torus-like Skyrmion, as shown in Sec.~\ref{sec:B2rings}.
We expect, in view of the studies of Ref.~\cite{Gudnason:2014jga}
which uses instead the BEC-type potential to create vortices, that the
vortex rings with vortex charge larger than one are unstable.
To confirm this expectation in this model, i.e.~with the
potential \eqref{eq:Vvortex}, we performed various calculations for
different values of $m$.
\begin{figure}[!tp]
\begin{center}
\mbox{
\includegraphics[width=0.24\linewidth]{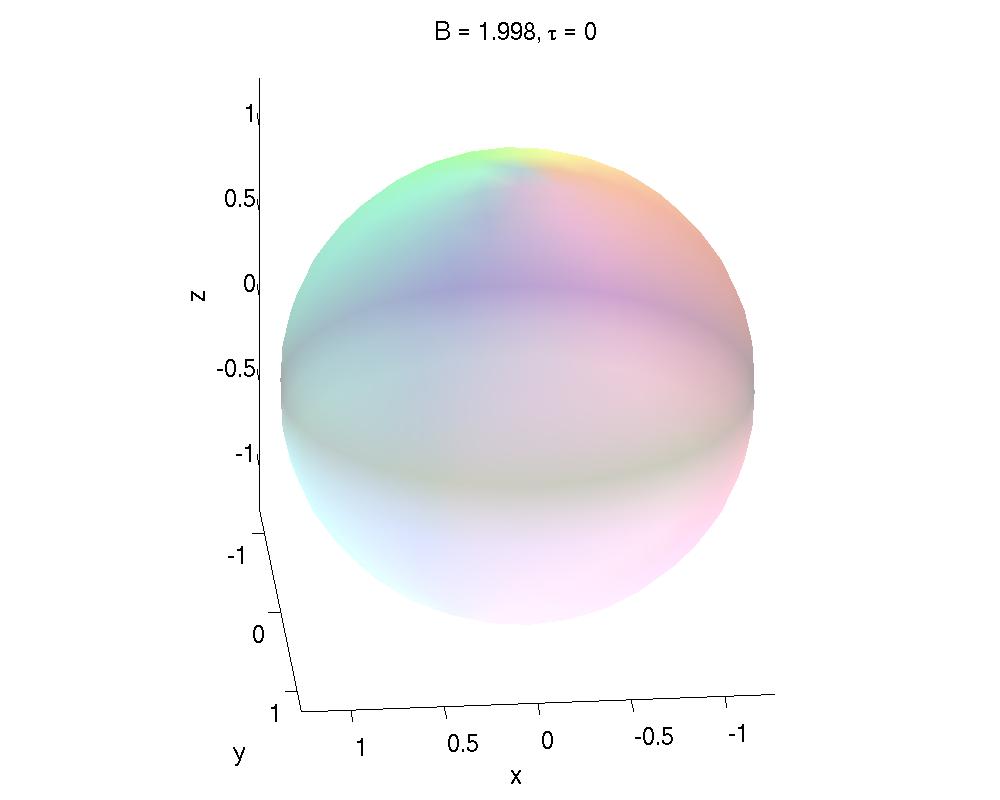}
\includegraphics[width=0.24\linewidth]{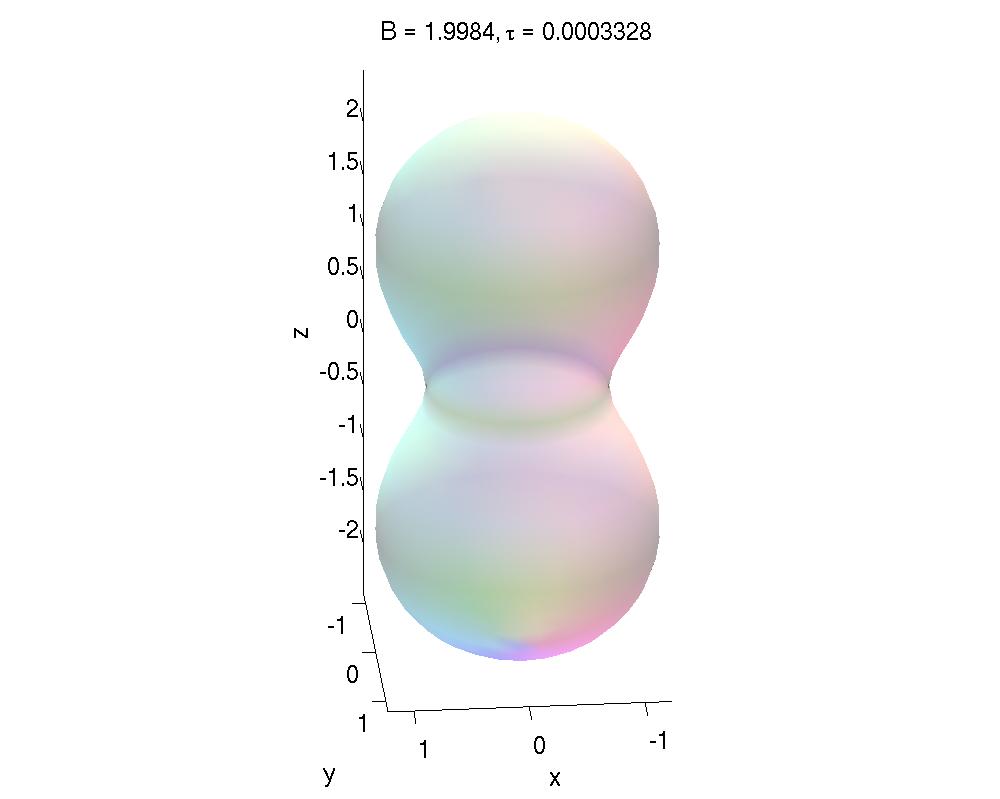}
\includegraphics[width=0.24\linewidth]{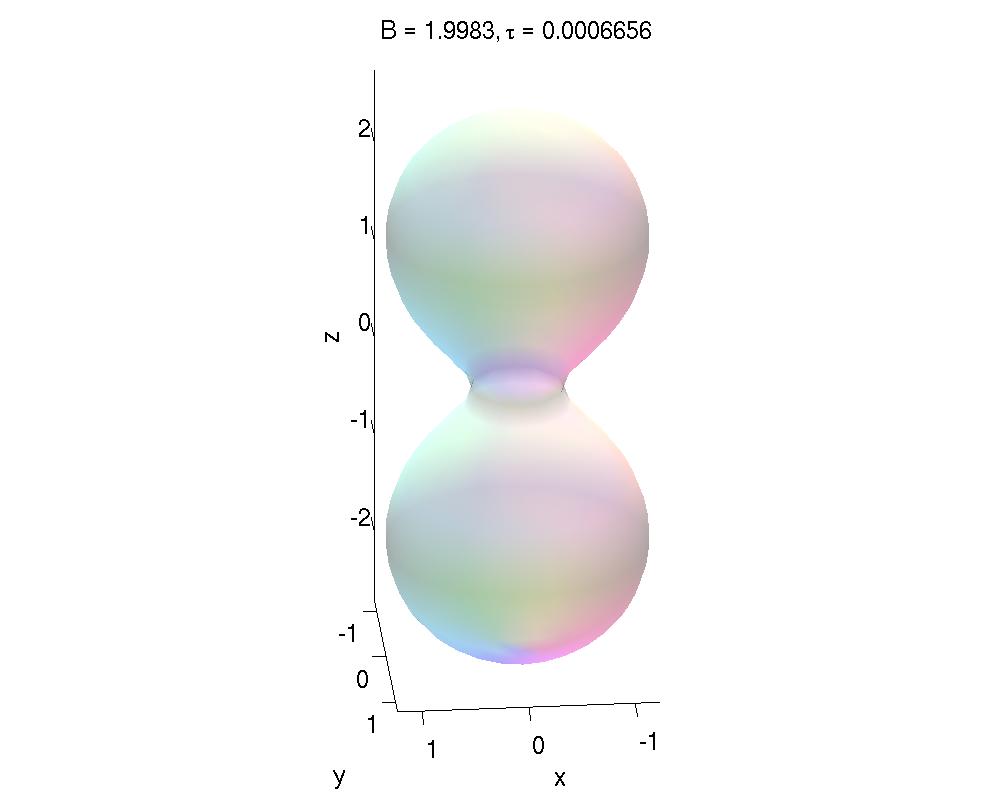}
\includegraphics[width=0.24\linewidth]{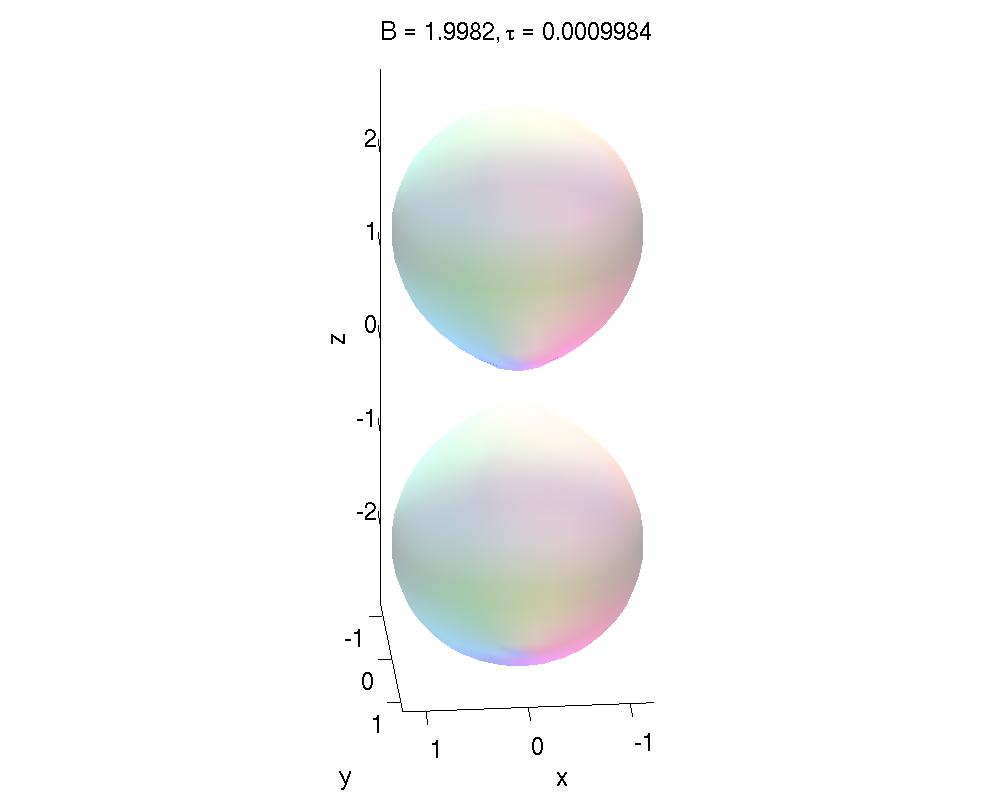}}
\caption{Isosurface of baryon charge density for the $B=2$ Skyrmion
constructed with the Ansatz \eqref{eq:B2k2initialcond}, i.e.~a
compactified charge-two vortex, as function of relaxation time
$\tau$. The vortex potential is turned off: $m=0$.
For the coloring scheme, see the text. }
\label{fig:v260ring12m0}
\vspace*{\floatsep}
\mbox{
\includegraphics[width=0.24\linewidth]{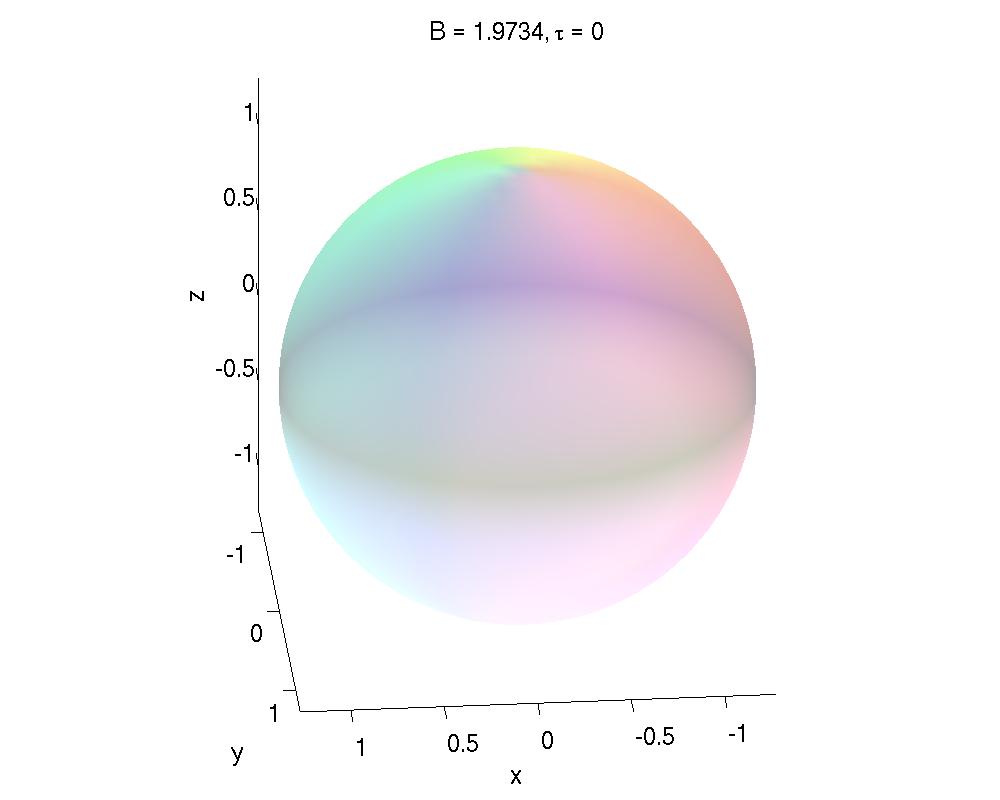}
\includegraphics[width=0.24\linewidth]{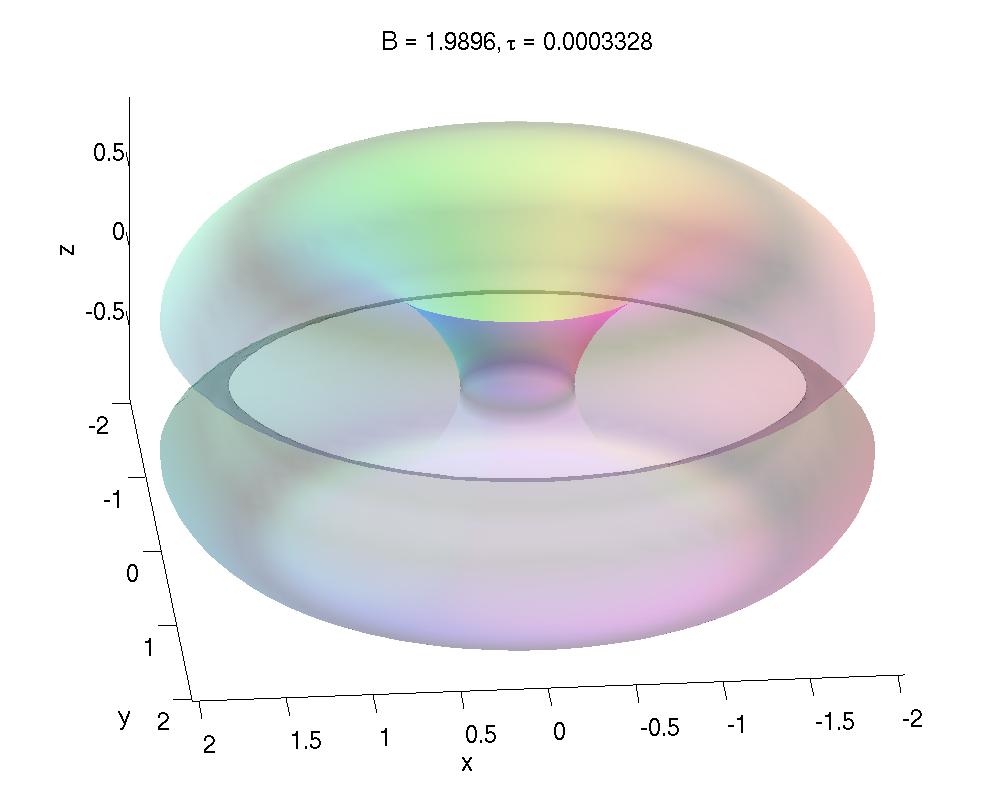}
\includegraphics[width=0.24\linewidth]{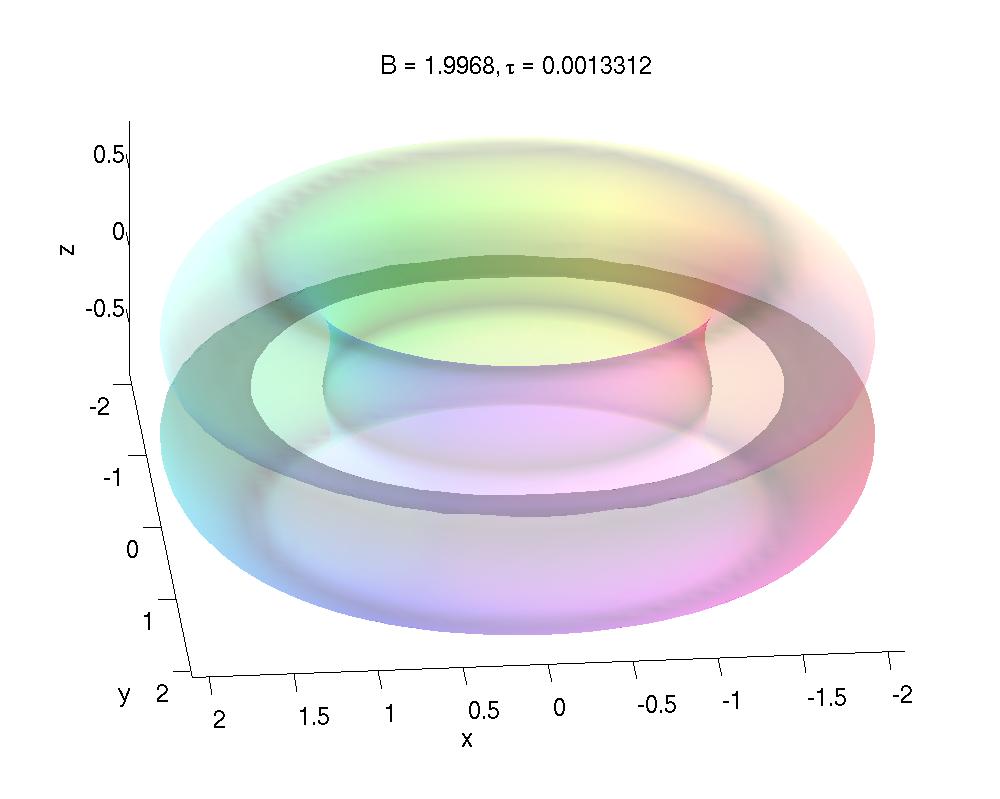}
\includegraphics[width=0.24\linewidth]{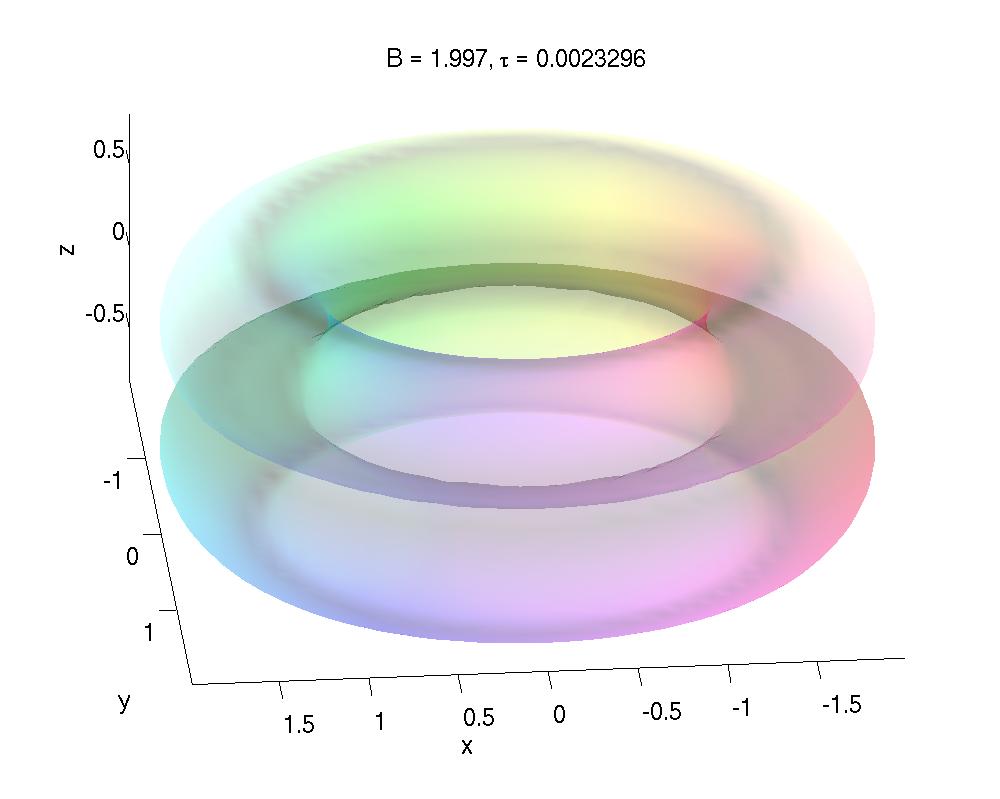}}
\mbox{
\includegraphics[width=0.24\linewidth]{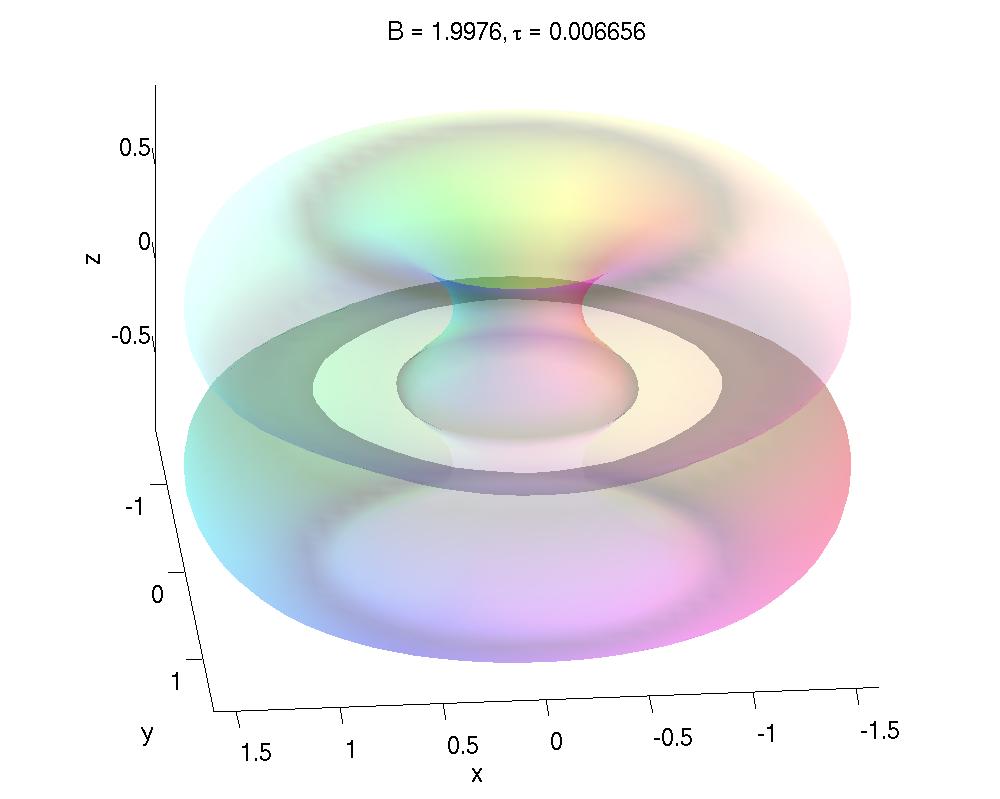}
\includegraphics[width=0.24\linewidth]{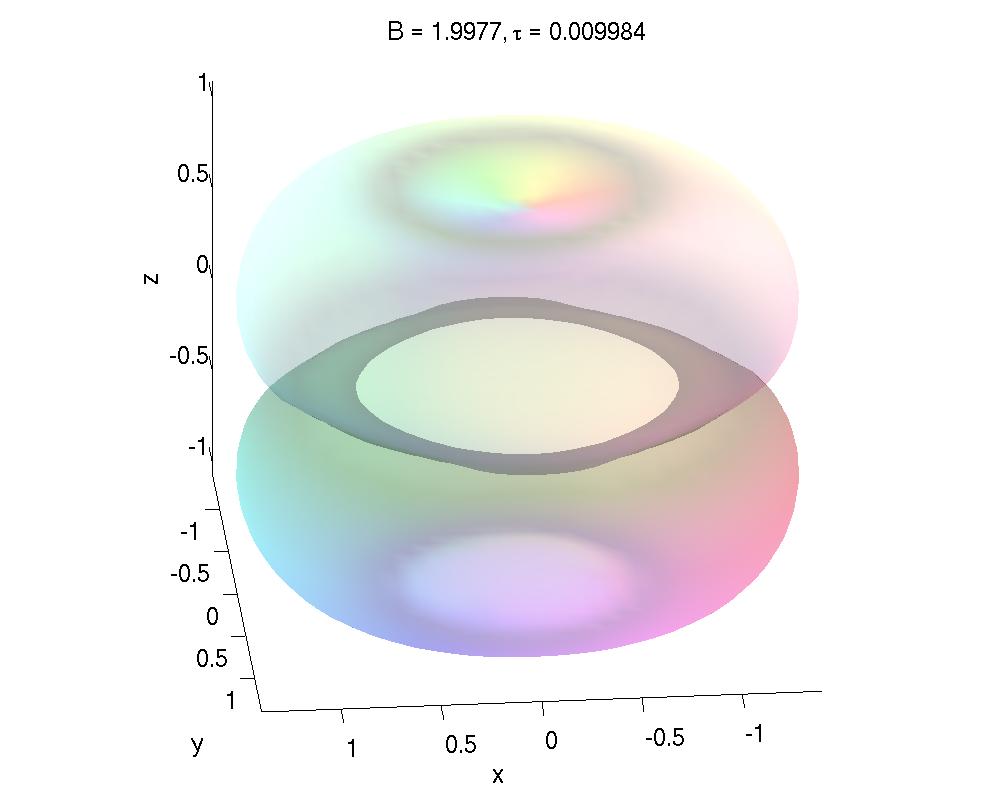}
\includegraphics[width=0.24\linewidth]{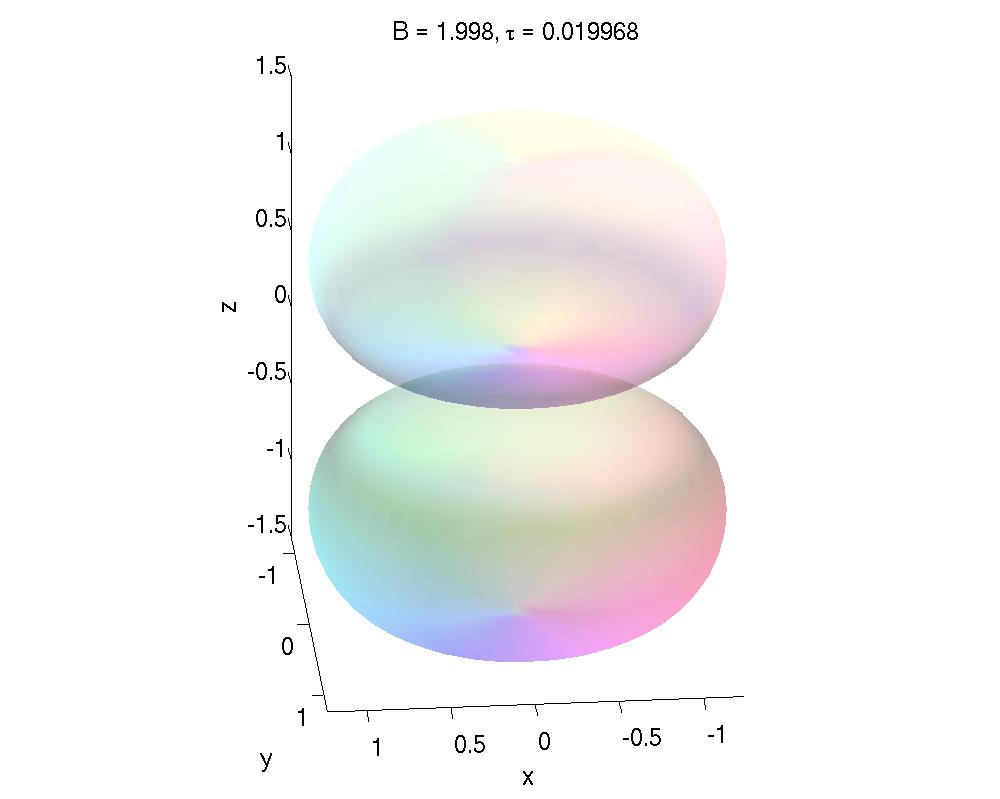}
\includegraphics[width=0.24\linewidth]{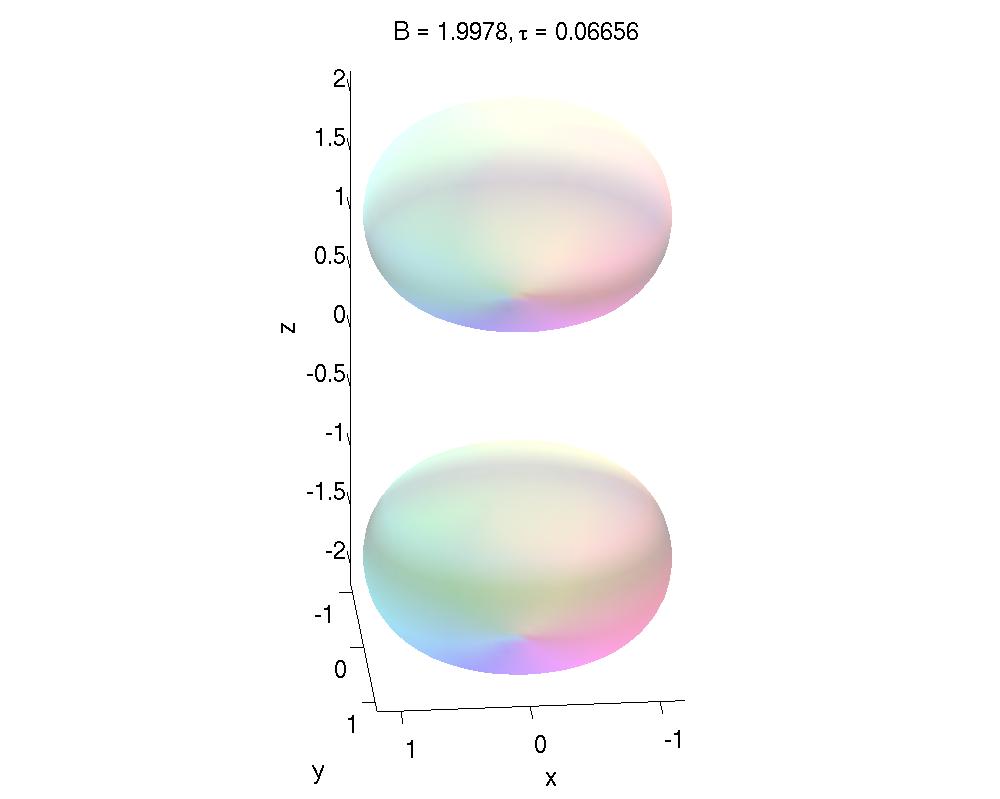}}
\caption{Isosurface of baryon charge density for the $B=2$ Skyrmion
constructed with the Ansatz \eqref{eq:B2k2initialcond}, i.e.~a
compactified charge-two vortex, as function of relaxation time
$\tau$. The vortex potential mass parameter is set as: $m=5$.
For the coloring scheme, see the text. }
\label{fig:v260ring12m5}
\end{center}
\end{figure}

Figs.~\ref{fig:v260ring12m0} and \ref{fig:v260ring12m5} show the
cooling-time evolution of the charge-two vortex compactified to a
$B=2$ Skyrmion, for $m=0$ and $m=5$, respectively.
Both configurations are created with the
Ansatz \eqref{eq:B2k2initialcond} and both configurations split up
into two disconnected single ($B=1$) Skyrmions.
Note that the $m=0$ configuration simply splits into two spherical
Skyrmions, whereas the $m=5$ one turns into a distinct torus-like
object before it splits up into single Skyrmions.
The two single Skyrmions are slightly torus-like for $m=5$, which
however look like squashed spheres at the level of isosurfaces at
half-maximum values of the baryon charge density, see
Fig.~\ref{fig:nout81_0_1_ms_baryonslice}.


\end{document}